\documentclass[twocolumn,superscriptaddress]{revtex4}

\usepackage{doi}
\usepackage{hyperref}
\hypersetup{
  colorlinks=true,        
  linkcolor=blue,         
  citecolor=cyan,         
}

\usepackage{graphicx}
\usepackage{dcolumn}
\usepackage{bm}
\usepackage{color}
\usepackage{enumitem}
\usepackage{amsmath}
\usepackage{subcaption}
\usepackage{amssymb}
\usepackage{orcidlink}
\usepackage{soul}
\usepackage{xcolor}
\usepackage{float}

\begin{document}

\title{Gravitational wave radiation from periodic orbits around a black hole in Einstein-Maxwell-scalar theory}

\author{Mirzabek Alloqulov}
\email{malloqulov@gmail.com}

\affiliation{School of Physics, Harbin Institute of Technology, Harbin 150001, People’s Republic of China}

\affiliation{University of Tashkent for Applied Sciences, Str. Gavhar 1, Tashkent 100149, Uzbekistan}

\affiliation{New Uzbekistan University, Movarounnahr str. 1, Tashkent 100000, Uzbekistan}

\author{Xojibonu Abdurashidova}
\email{abdurashidovaxojibonu@gmail.com}
\affiliation{Institute of Fundamental and Applied Research, National Research University TIIAME, Kori Niyoziy 39, Tashkent 100000, Uzbekistan}

\author{Sanjar Shaymatov}
\email{sanjar@astrin.uz}
\affiliation{Institute of Fundamental and Applied Research, National Research University TIIAME, Kori Niyoziy 39, Tashkent 100000, Uzbekistan}
\affiliation{National University of Uzbekistan, Tashkent, 100174, Uzbekistan}

\author{Raxmatillo Karimov}
\email{raxmatillo82@mail.ru}
\affiliation{Tashkent State Technical University, Tashkent 100095, Uzbekistan}

%
\date{\today}
\begin{abstract}
In this paper, we investigate the periodic orbits around a black hole (BH) in Einstein-Maxwell-scalar (EMS) theory and the corresponding gravitational waveforms. First, we study the test particle dynamics around the BH by using the Lagrangian formalism. Using the corresponding conditions, we explore the effect of the spacetime parameters on the marginally bound orbit (MBO)  and innermost stable circular orbit (ISCO). The results indicate that MBO and ISCO quantities decrease with the rise of the $\beta$ parameter and BH charge $Q$, and vice versa for the $\gamma$ parameter.  Subsequently, we plot the different zoom-whirl orbits around the BH in EMS theory by finding the test particle's corresponding energy and angular momentum. Finally, the gravitational waveforms of the extreme mass ratio inspirals, which contain a supermassive BH in EMS theory and a stellar mass object, are demonstrated. It is shown that the zoom-whirl orbits shrink as the BH charge increases, and, as a result, gravitational waveforms shift towards shorter times.      
\end{abstract}

\maketitle

\section{Introduction}

Gravitational wave (GW) astronomy has turned black holes (BHs) from theoretical possibilities into observable laboratories for strong-field gravity. The detections by the LIGO-Virgo collaboration~\cite{Abbott_2016}, alongside the Event Horizon Telescope imaging~\cite{Akiyama19L1, Akiyama19L6, Event}, have verified the existence of the event horizons while allowing their dynamics to be probed with high accuracy. These milestones raise a central question: does the geometry of astrophysical BHs strictly obey general relativity (GR), or do observations reveal imprints of additional fields and interactions?

Several theoretical arguments suggest that GR is an effective description of a more fundamental framework. One especially compelling extension arises in the low-energy regime of string theory, where additional scalar fields emerge and may develop non-minimal interactions with gauge fields~\cite{Green87book}. The Einstein-Maxwell-scalar (EMS) theory captures this idea that a massless scalar, i.e., the dilaton, interacts with the electromagnetic field through a coupling function~\cite{Gibbons88, Yu_2021}. The resulting charged BH solutions generalize the Reissner-Nordstr$\ddot{\mathrm{o}}$m (RN) spacetime~\cite{Reissner1916, Nordstrom1918} and contain dimensionless parameters, which are $\beta$ and $\gamma$, that control the scalar-electromagnetic coupling. In the limiting cases, Schwarzschild~\cite{1916SPAW.......189S,2015arXiv151202061B}, RN, or dilatonic BHs~\cite{Yu_2021} can be recovered. So far, various studies have investigated EMS BHs from different perspectives; see Refs.~\cite{Qiu:2020gox, Turimov:2020fme, Qiu:2021qrt, Herrera-Aguilar:2021qfq, Richarte:2021fbi, Feng:2022bst, Mazharimousavi:2022tnr, Zahid:2022zzt, Mazharimousavi:2022uhs, Kurbonov:2023uyr, Alloqulov:2024zln, Al-Badawi:2024dzc, Carrasco-H:2024fgc, Wu:2024sng, Alloqulov:2024hoj, Belmahi:2024wag, Yunusov:2024xzu, Zhang:2025jlb, Wu:2025hcu, Zhuang:2025eal, Wu:2025zwr}.  

It must be emphasized that electromagnetic observations probe only part of the spacetime and provide limited access to the dynamical strong-field regime near the event horizon. This happens because they are most sensitive to the region where photons propagate and are not ideally suited to mapping the dynamical strong-field region near the horizon. In contrast, GWs directly encode the dynamics of the gravitational field, providing a complementary and more direct probe of the strong-field regime and spacetime geometry near compact objects. In particular, extreme-mass-ratio inspirals (EMRIs), in which a stellar-mass compact object spirals into a supermassive BH (SMBH)~\cite{Hughes01EMRI, Amaro-Seoane18LRR, Babak17PRD}, are prime targets for future spaceborne detectors such as the Laser Interferometer Space Antenna (LISA)~\cite{Amaro-Seoane2017LISA}, Taiji~\cite{HuTaiji} and TianQin~\cite{Gong_2021}. Note that low-frequency GWs are emitted through these systems. Low-frequency radiation emitted from EMRIs offers a direct probe into near-horizon environments and the surrounding matter, making these systems valuable tools. The waveforms produced by an EMRI system depend directly on the orbital trajectory of the secondary celestial body. Previous works~\cite{Levin_2008, Grossman_2009, Misra_2010} establish that periodic geodesic trajectories occur in both Schwarzschild and Kerr metrics. Such trajectories serve as a foundational element of EMRI modeling by capturing the key mechanics of bound motion around BHs. By definition, a test particle on a periodic orbit completes a discrete number of radial and angular cycles before returning to its starting position, typically exhibiting zoom-whirl behaviour. Hence, the classification of these orbits depends on three integers: the zoom number $z$, the whirl number $w$, and the vertex number $v$~\cite{Levin_2008, Levin_2009}. Importantly, the properties of these periodic orbits remain closely linked to the underlying BH geometry~\cite{Levin2010, Babar17PRD, Tu23, Mustapha2020, wei2019, Deng20, yang2024}. Several studies investigate the periodic orbits and corresponding gravitational waveforms within different spacetimes~\cite{Liu:2018vea, Lin:2023rmo, Yao:2023ziq, Lin:2022llz, Chan:2025ocy, Wang:2022tfo, Lin:2023eyd, Haroon:2025rzx, Habibina:2022ztd, Zhang:2022psr, Lin:2022wda, Gao:2021arw, Lin:2021noq, Gao:2020wjz, Deng:2020hxw, Azreg-Ainou:2020bfl, Wei:2019zdf, Pugliese:2013xfa, Zhang:2022zox, Healy:2009zm, Wang:2025wob, Alloqulov:2025bxh, Wei:2025qlh, Yang:2024lmj, Shabbir:2025kqh, Junior:2024tmi,  Jiang:2024cpe, Yang:2024cnd, QiQi:2024dwc, Alloqulov:2025ucf, Wang:2025hla, Lu:2025cxx, Zare:2025aek, Gong:2025mne, Li:2025sfe, Choudhury:2025qsh, Chen:2025aqh, Deng:2025wzz, Li:2025eln,Sharipov:2026CPC, Ahmed:2025azu, Glampedakis2002, Alloqulovmod, Tan_2026, Shi_2026, Huang:2026tul, wang2026newtypemultibranchperiodic, Gogoi_2026, Zhang_2026, 2026arXiv260700812S,Lu_2026, Das_2026,Hua_2026,2026arXiv260724154A,bravogaete2026,2026arXiv260209453X}. 

Inspired by the aforementioned considerations, in the present work, we delve into the analysis of periodic orbits in the vicinity of the EMS BH and the gravitational waveforms emitted by the EMRI system as well, which comprises a stellar-mass object orbiting the supermassive EMS BH. To investigate the behavior of test particles around the BH within the framework of EMS theory, a Lagrangian formalism is utilized. Furthermore, under appropriate conditions, our analysis scrutinizes the characteristics of MBOs and ISCOs for the test particles. Subsequently, numerical methods are employed to solve the equations of motion, allowing the graphical representation of periodic orbits across a range of values for $z$, $w$ and $v$. As a final step, the gravitational waveforms associated with these periodic orbits are investigated using the numerical kludge formalism.

This paper is structured as follows: Section~\ref{section2} presents the EMS BH metric and the dynamics of the test particles, along with an investigation of MBOs and ISCOs. Section~\ref{section3} analyzes how BH charge affects the rational number and presents plots for the different periodic orbits, while Section~\ref{section4} evaluates the generated gravitational waveforms. Section~\ref{con} summarizes our key findings and discussions of their implications. Throughout the paper, we adopt geometric units $G=c=1$ and the metric signature $(–, +, +, +)$.

\section{Spacetime of the black hole in EMS theory}\label{section2}
\begin{figure*}
\includegraphics[scale=0.4]{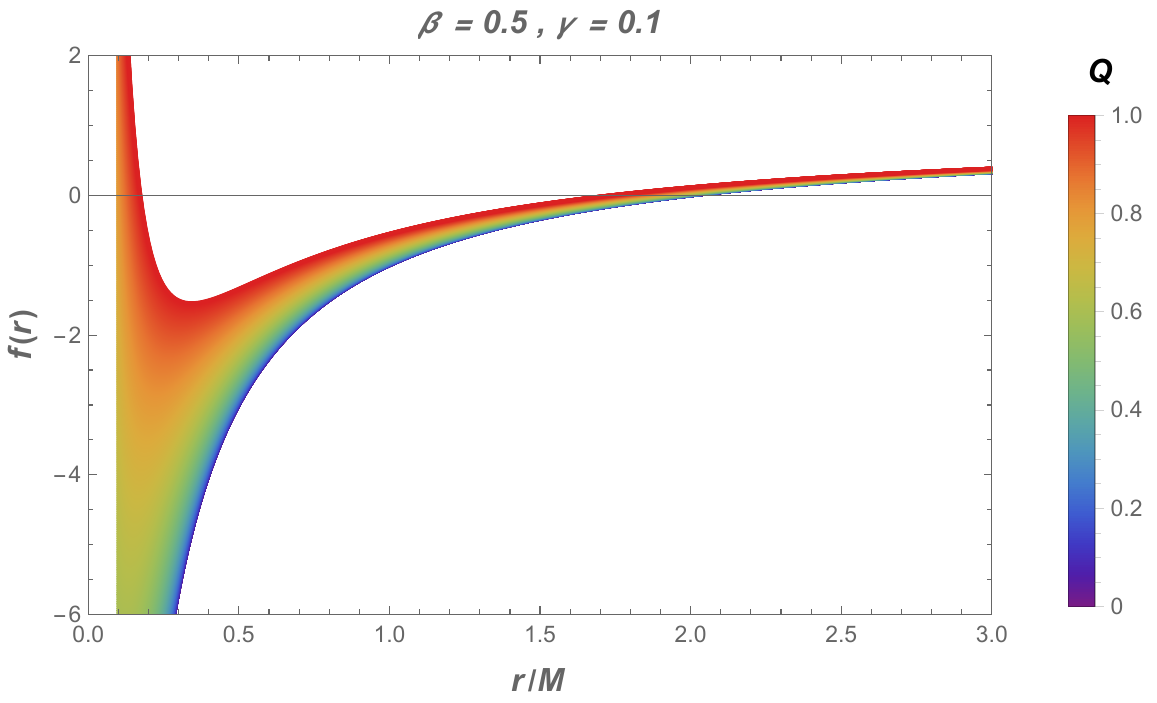}
\includegraphics[scale=0.4]{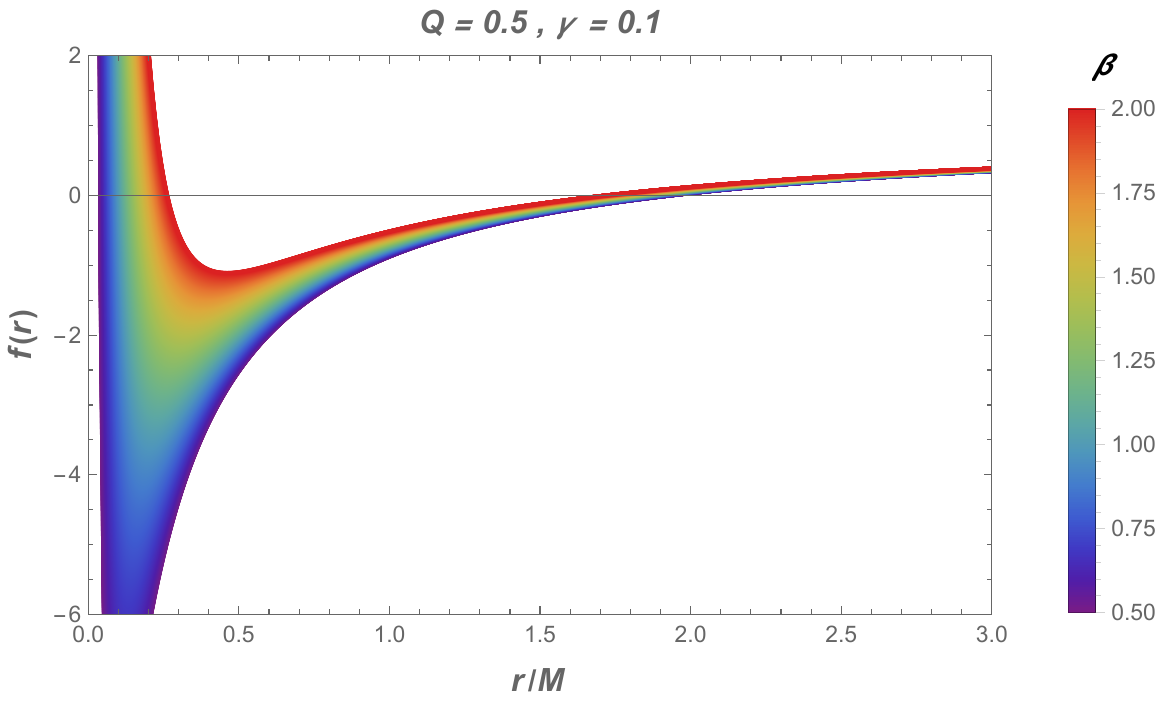}
\includegraphics[scale=0.4]{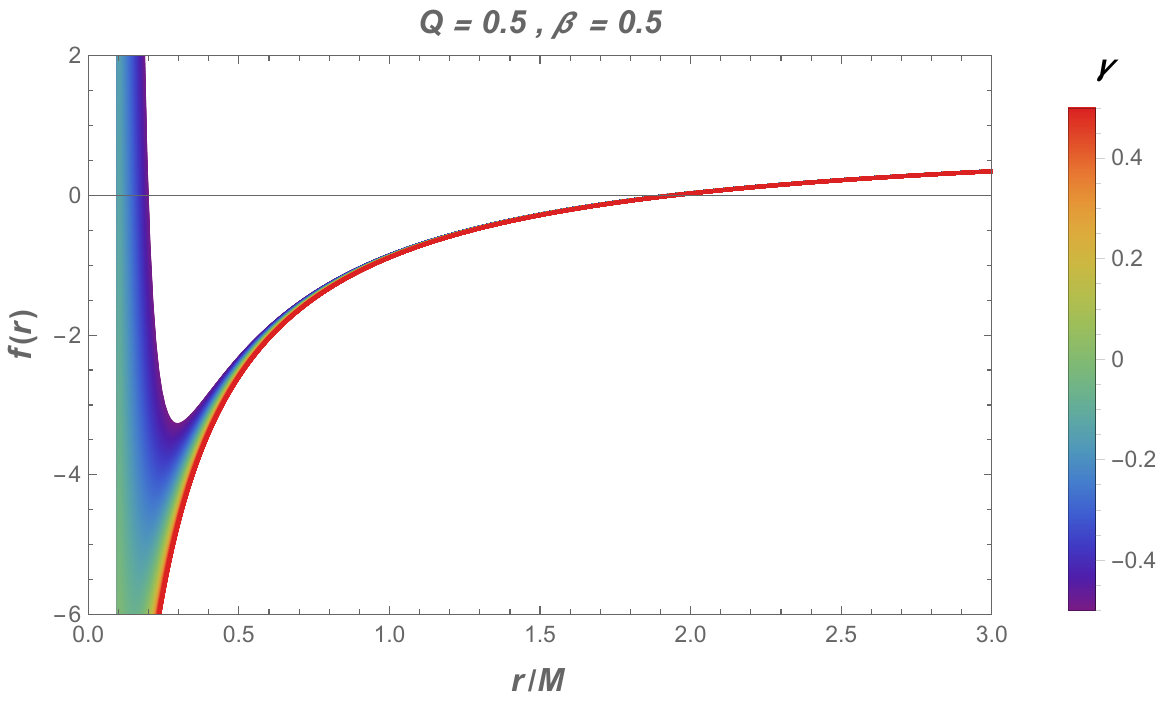}
\caption{The plot shows the radial profile of the metric function $f(r)$ for the different values of the spacetime parameters.}
\label{fig:metric}
\end{figure*}
\begin{figure*}
\includegraphics[scale=0.35]{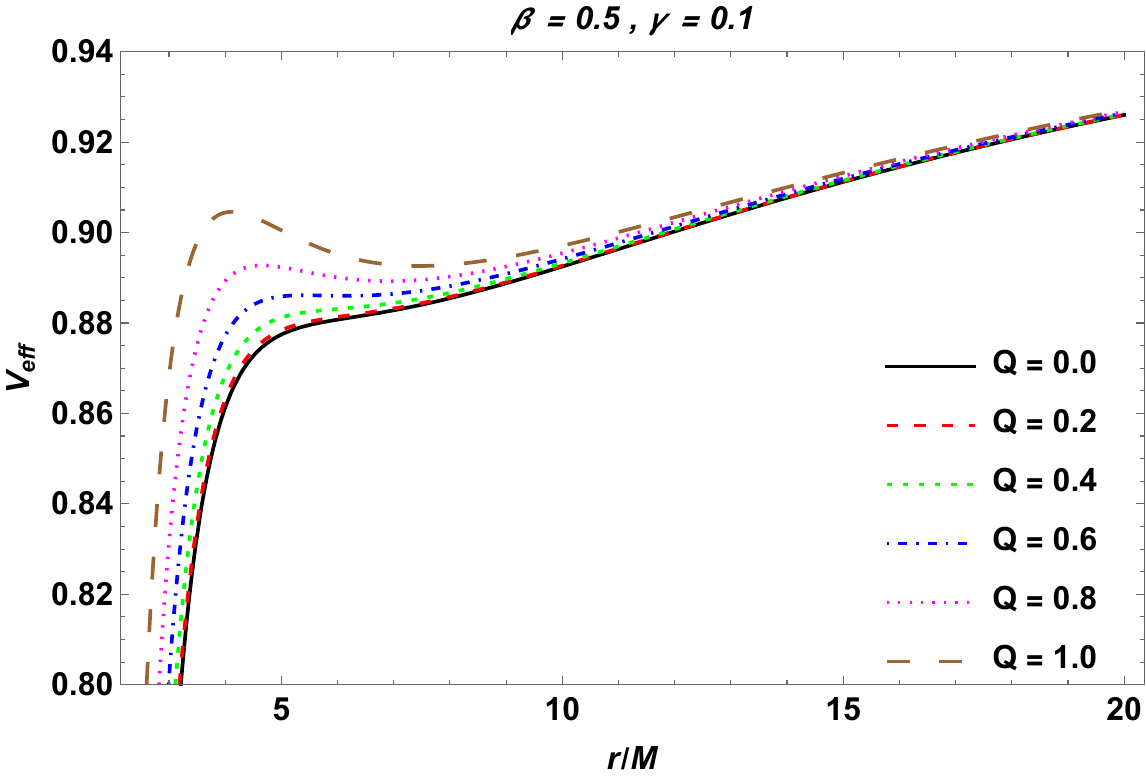}
\includegraphics[scale=0.35]{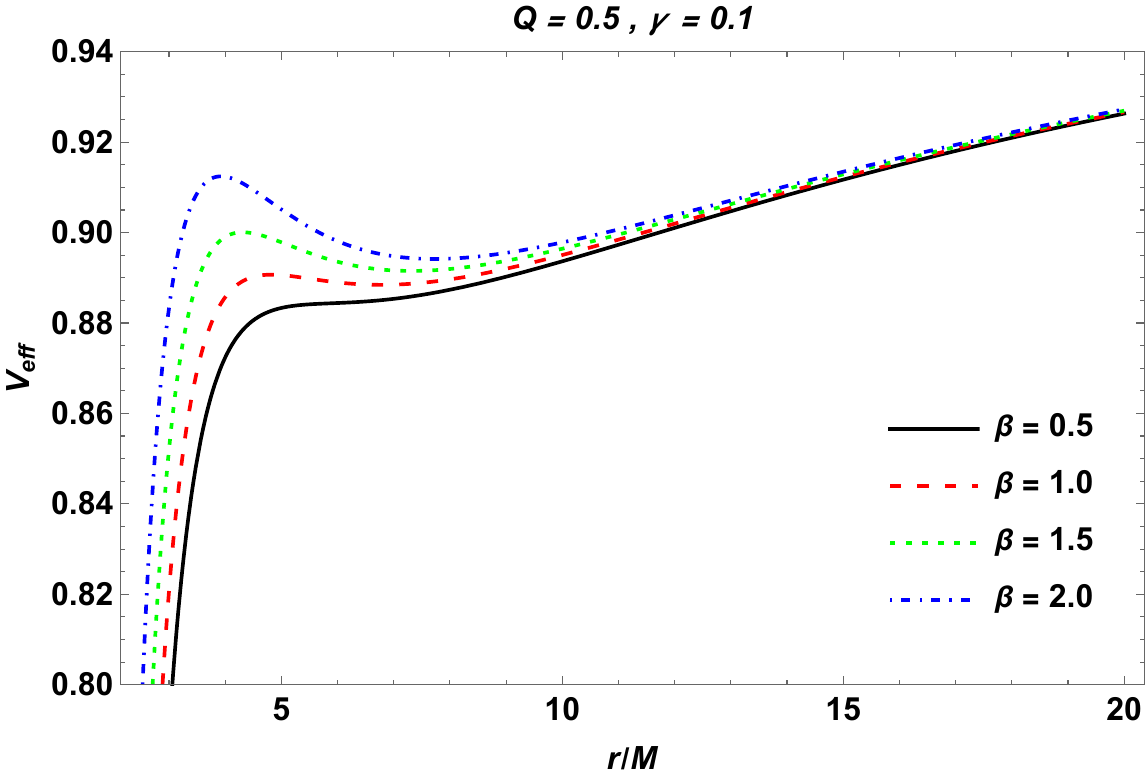}
\includegraphics[scale=0.35]{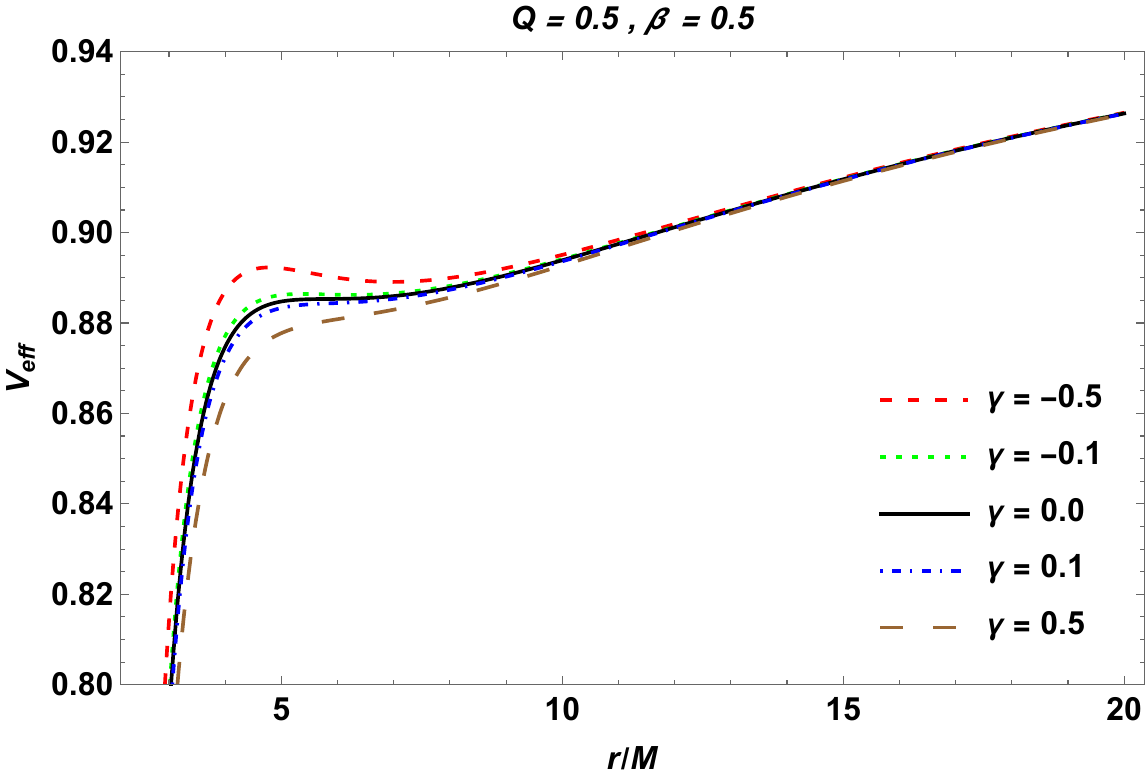}
\includegraphics[scale=0.35]{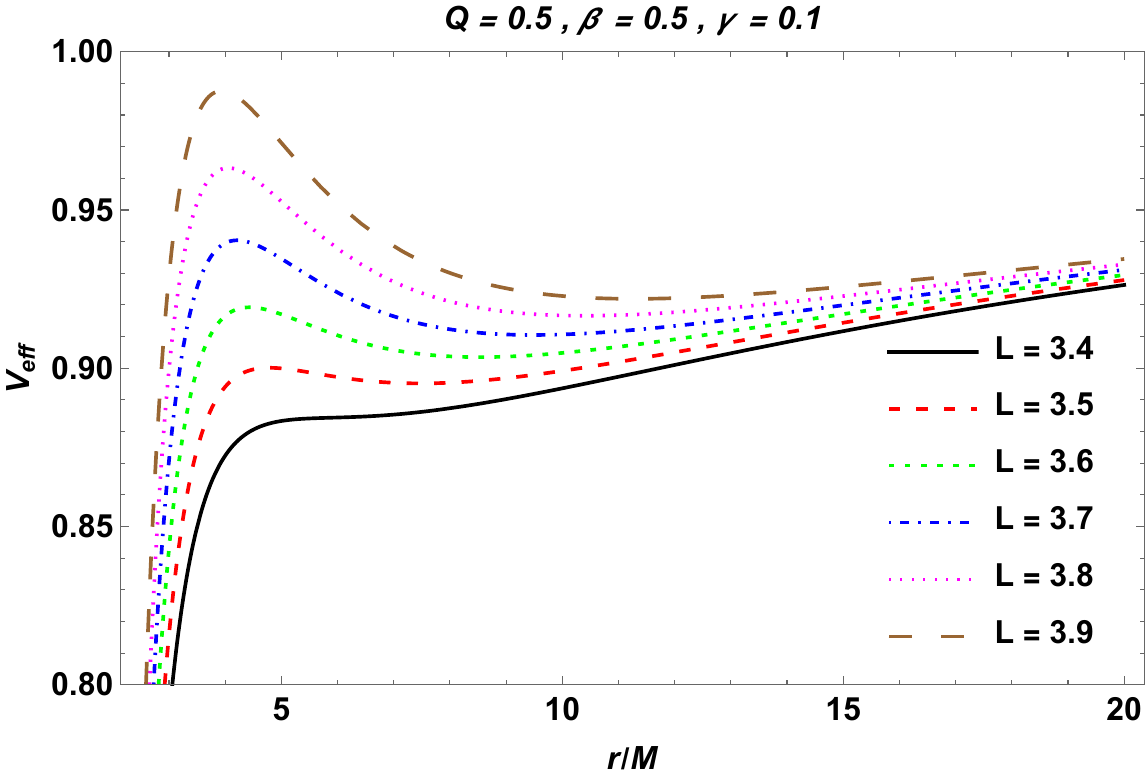}
\caption{The top panels illustrate the radial profile of the effective potential for the different values of the BH charge (left) and $\beta$ parameter (right). Here, $\beta=0.5$ \& $\gamma=0.1$ and $Q=0.5$ \& $\gamma=0.1$ for the left and right panels, respectively. Bottom panel: the radial dependence of the effective potential for the different values of the $\gamma$ parameter (left) and orbital angular momentum (right). We set $Q=\beta=0.5$ and $Q=\beta=0.5$ \& $\gamma=0.1$ for the left and right panels, respectively.}
\label{fig:eff}
\end{figure*}
\begin{figure*}
\includegraphics[scale=0.4]{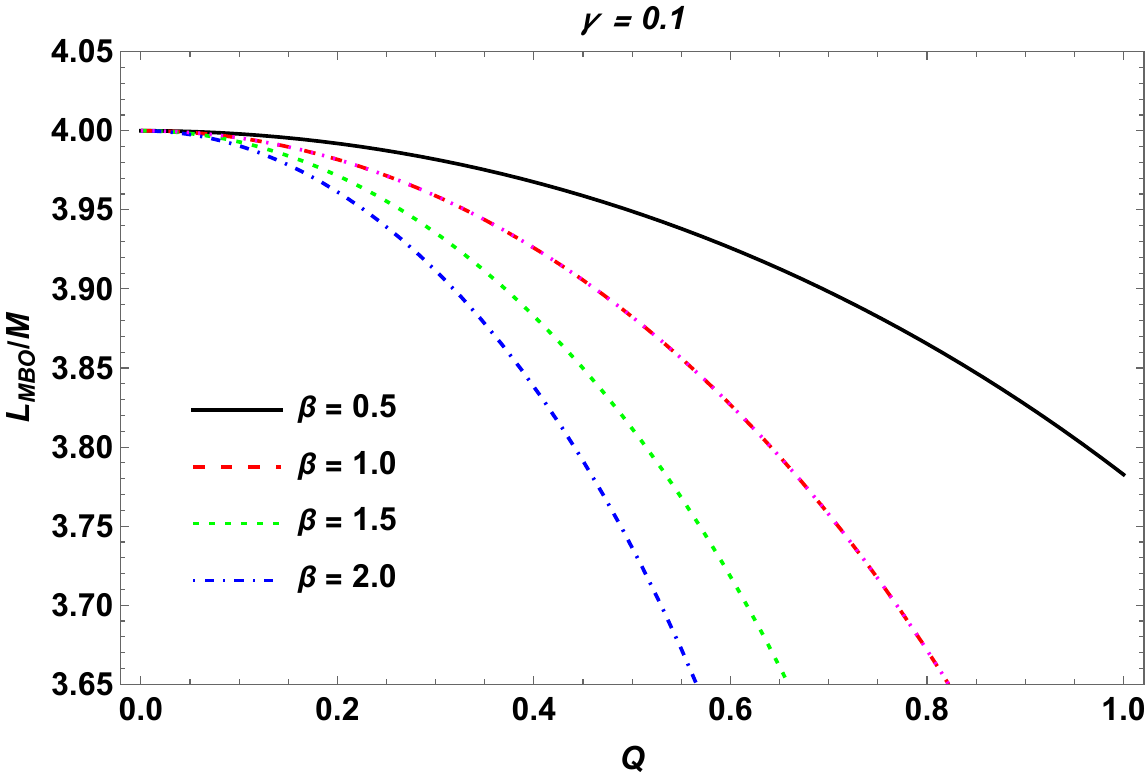}
\includegraphics[scale=0.4]{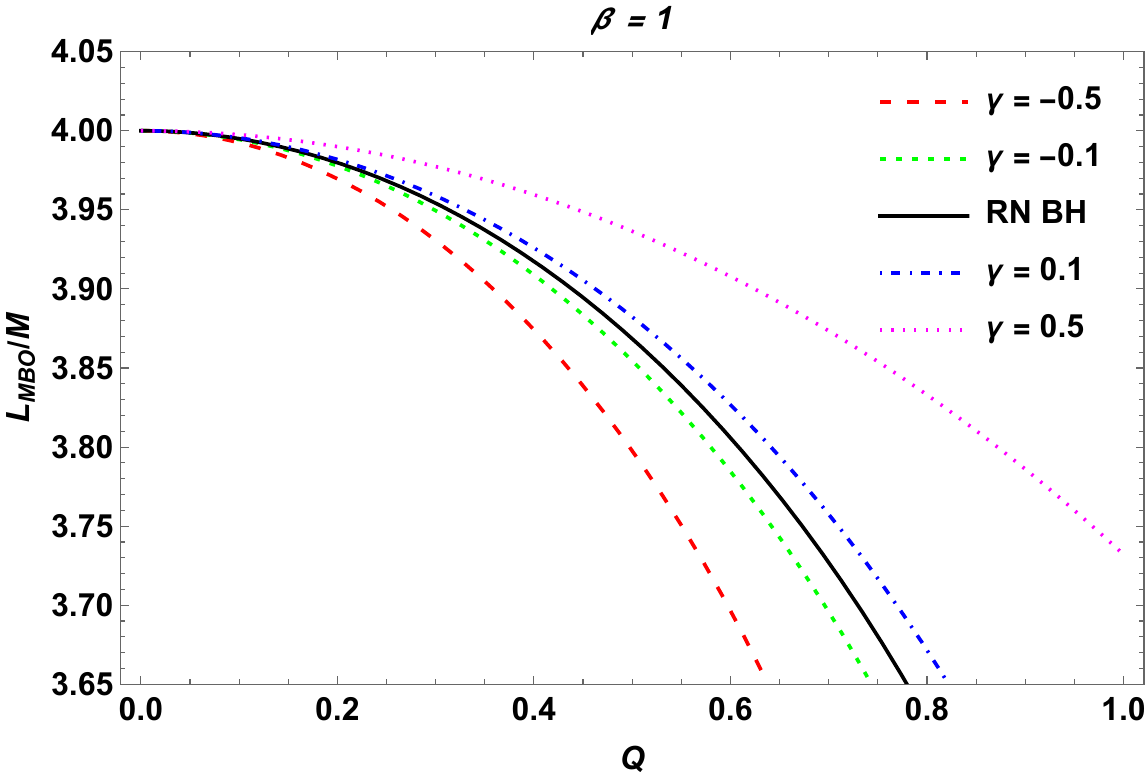}
\includegraphics[scale=0.4]{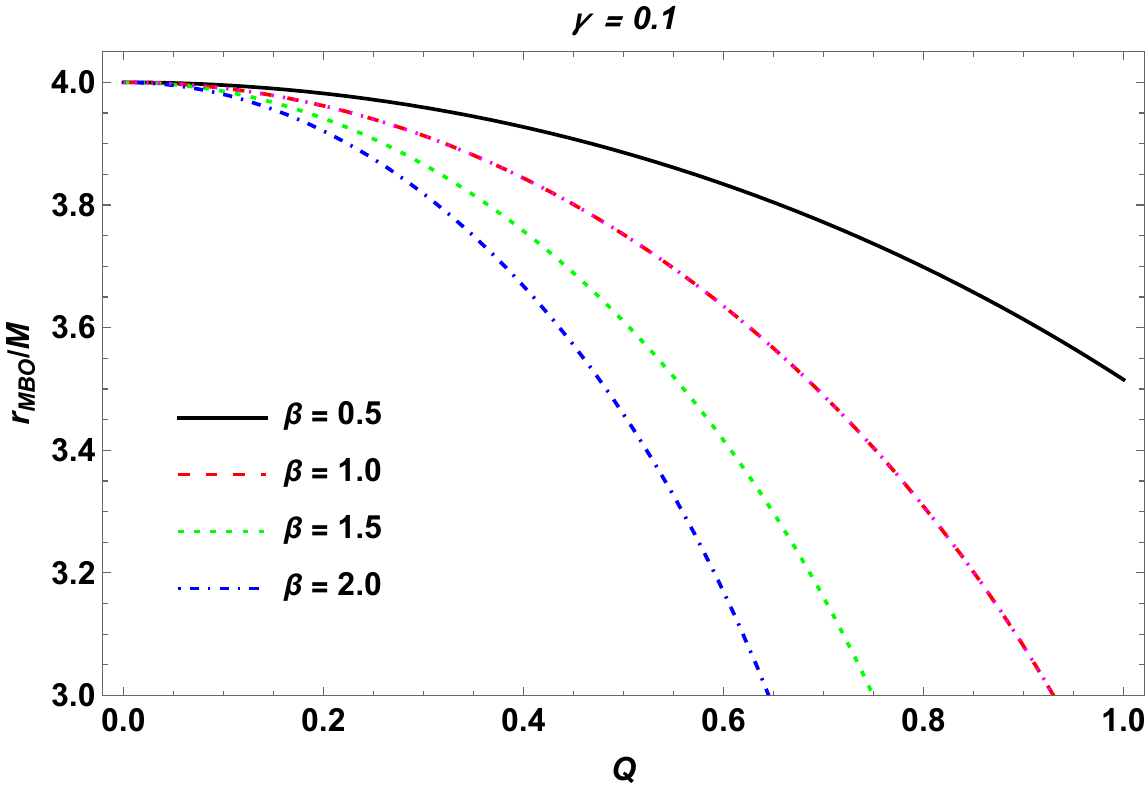}
\includegraphics[scale=0.4]{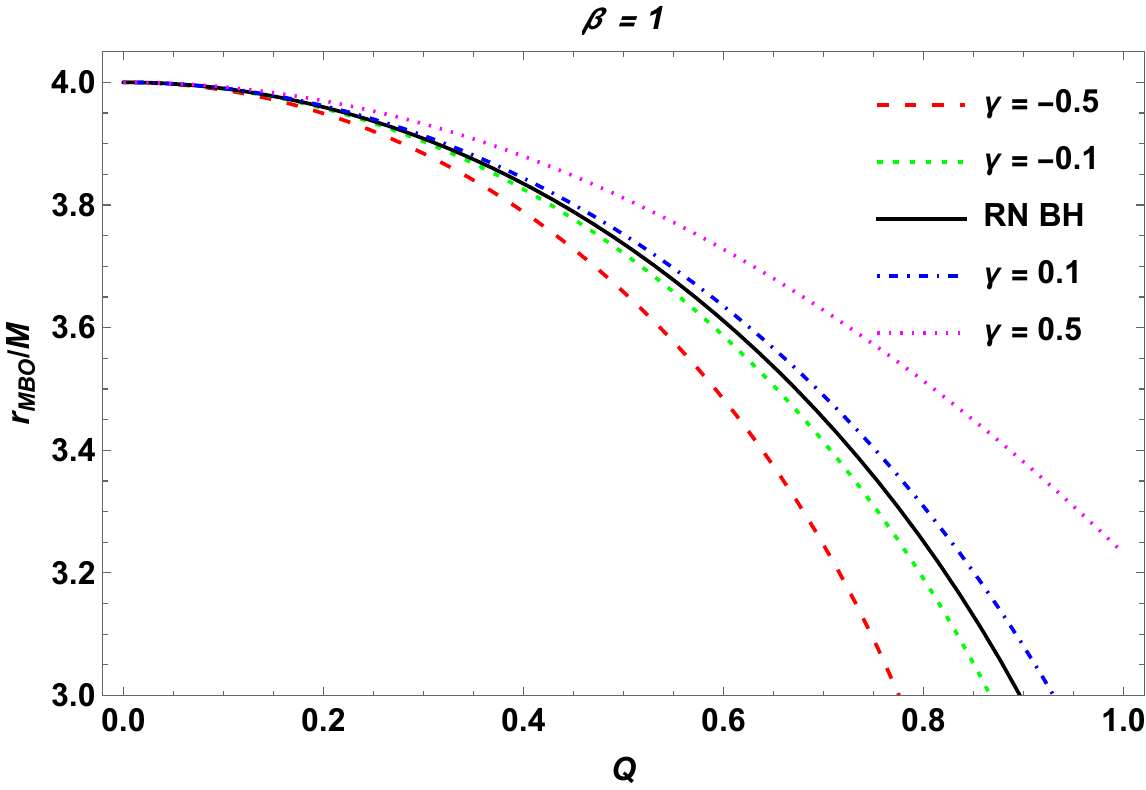}
\caption{Top panels: the dependence of the orbital angular momentum of a test particle moving along MBO on the BH charge for the different values of the $\beta$ (left) and $\gamma$ (right) parameters. Here, we set $\gamma=0.1$ and $\beta=1$ for the left and right panels, respectively.}
\label{fig:mbo}
\end{figure*}
\begin{figure*}
\includegraphics[scale=0.4]{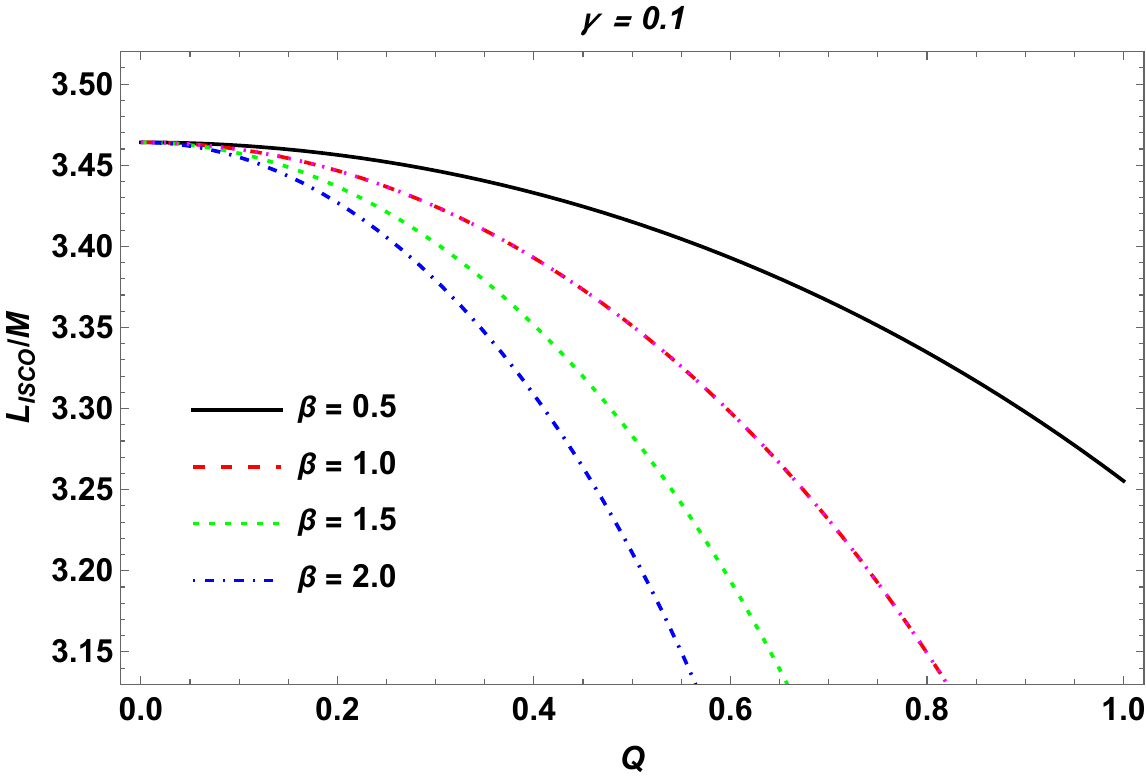}
\includegraphics[scale=0.4]{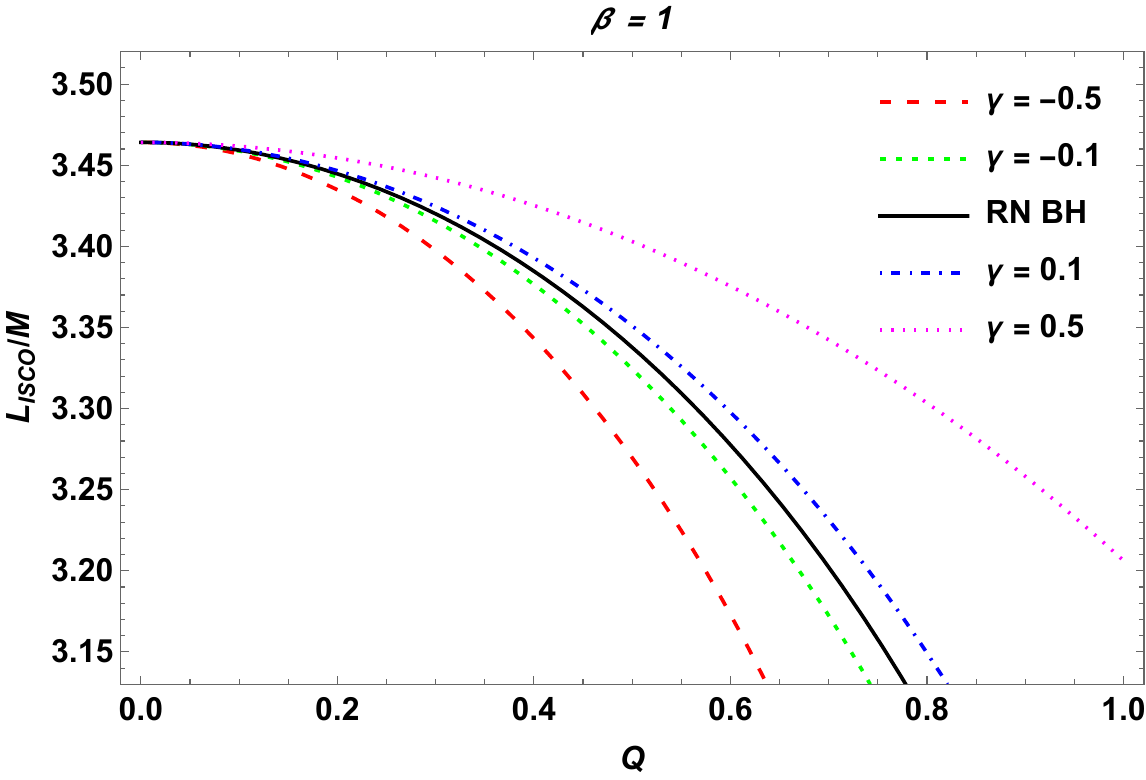}
\includegraphics[scale=0.4]{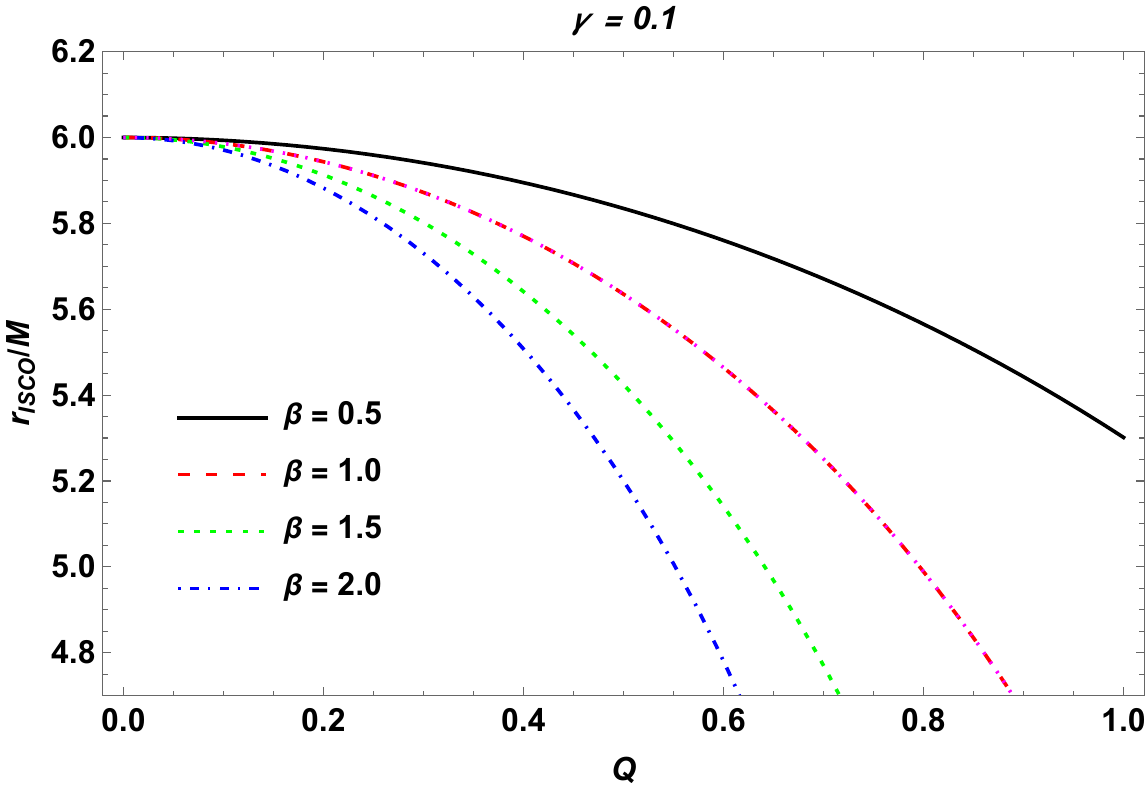}
\includegraphics[scale=0.4]{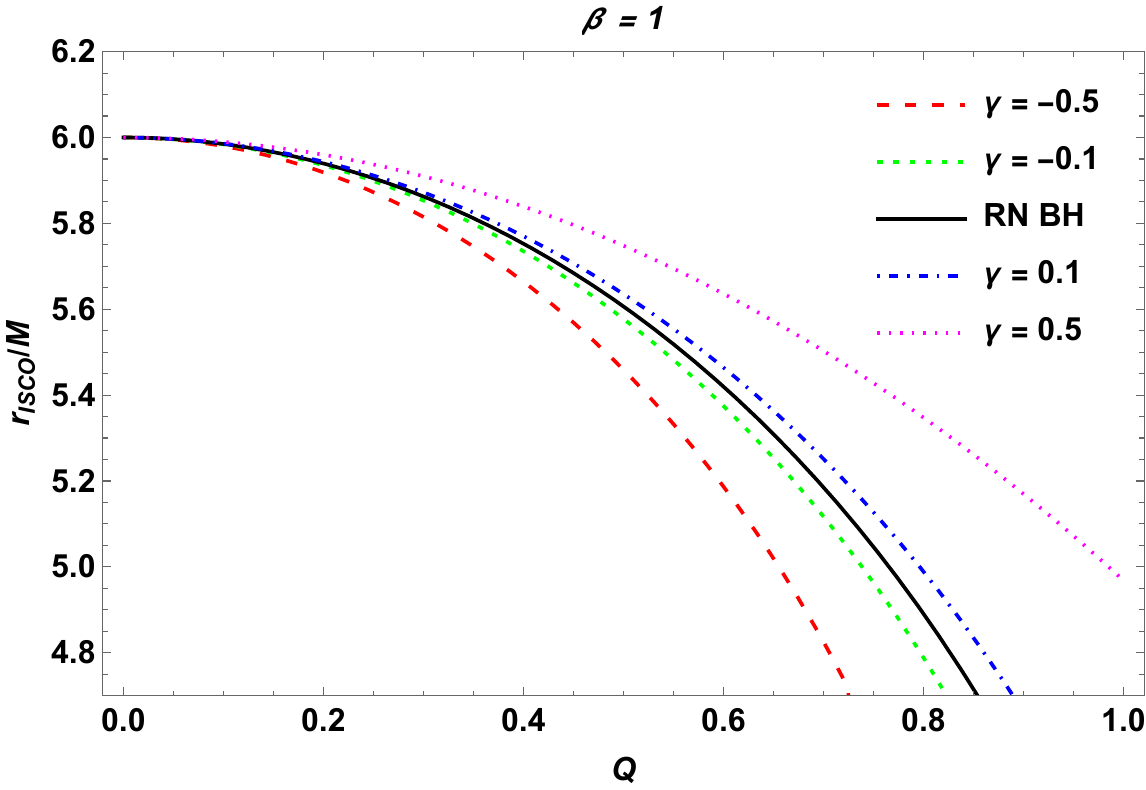}
\includegraphics[scale=0.4]{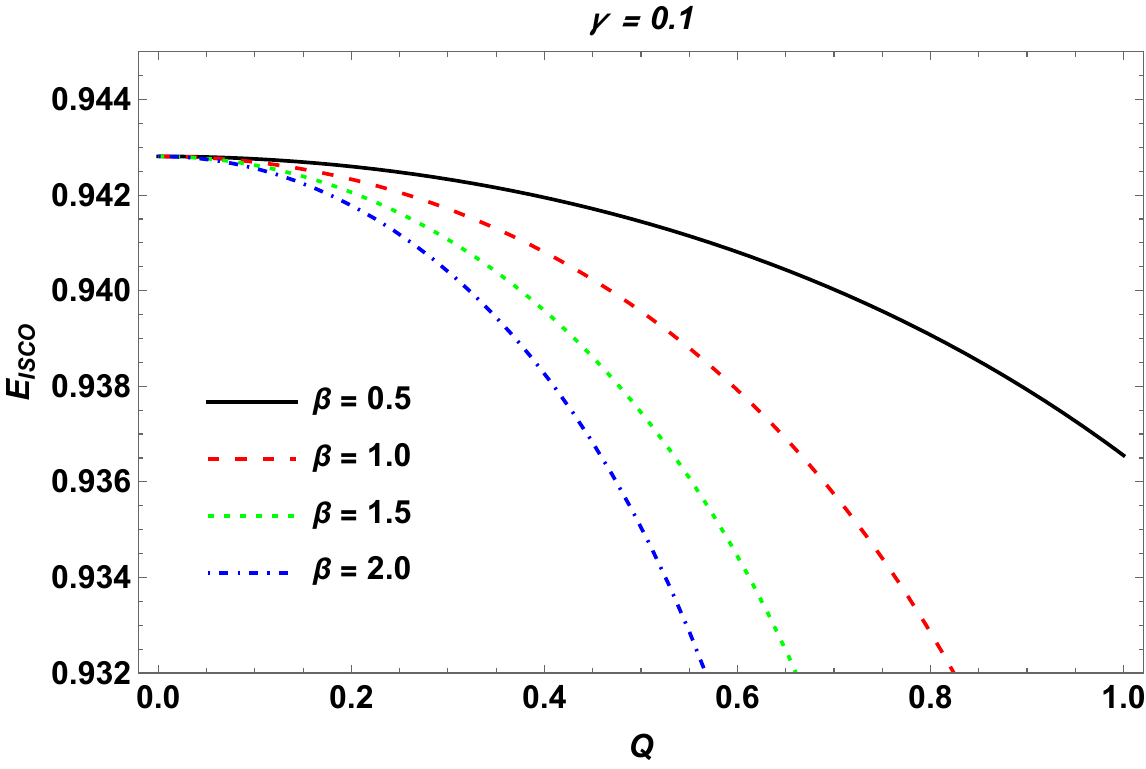}
\includegraphics[scale=0.4]{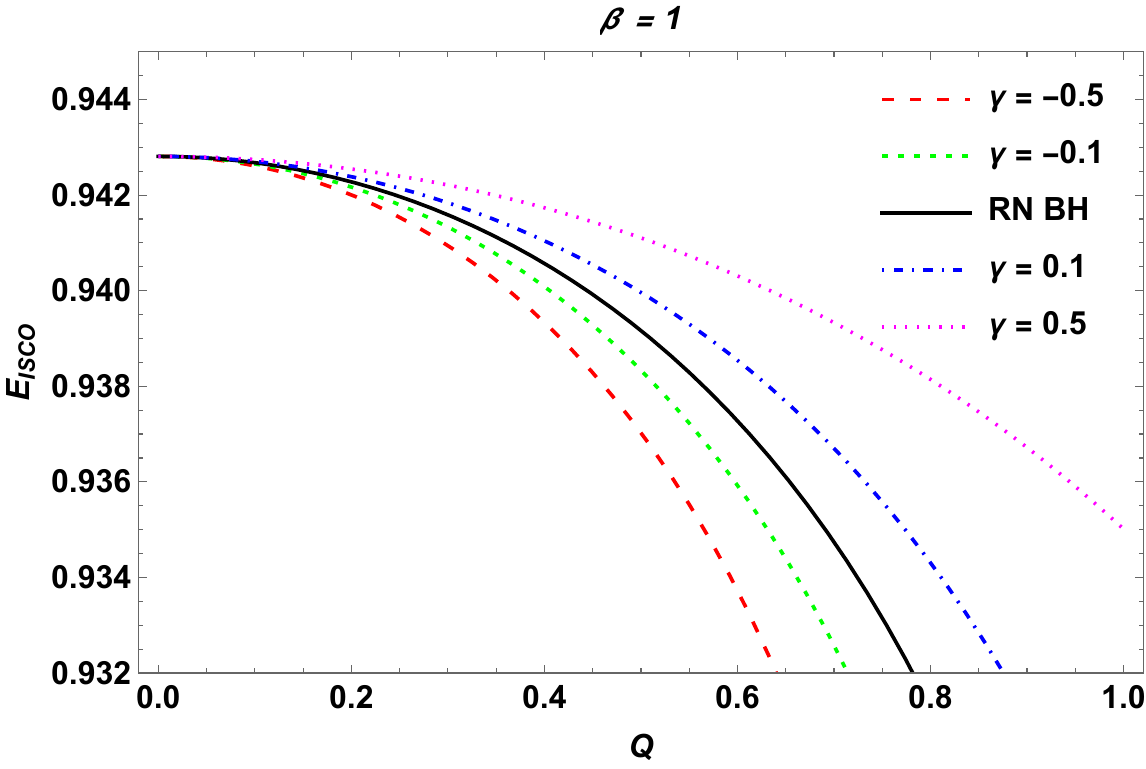}
\caption{The plot shows the dependence of ISCO parameters, i.e., the orbital angular momentum, radius and energy, on the BH charge for the different values of the $\beta$ (left) and $\gamma$ (right) parameters. We set $\gamma=0.1$ and $\beta=1$ for the left and right panels, respectively.}
\label{fig:isco}
\end{figure*}
\begin{figure*}
\includegraphics[scale=0.4]{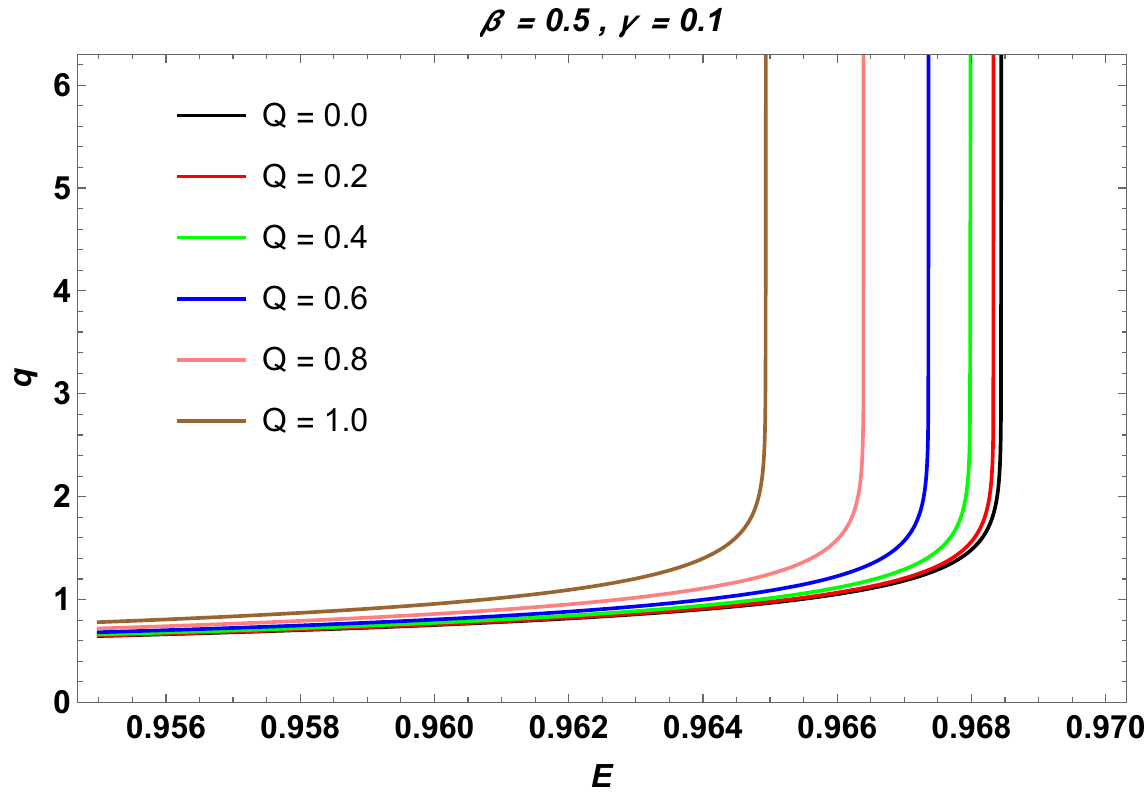}
\includegraphics[scale=0.4]{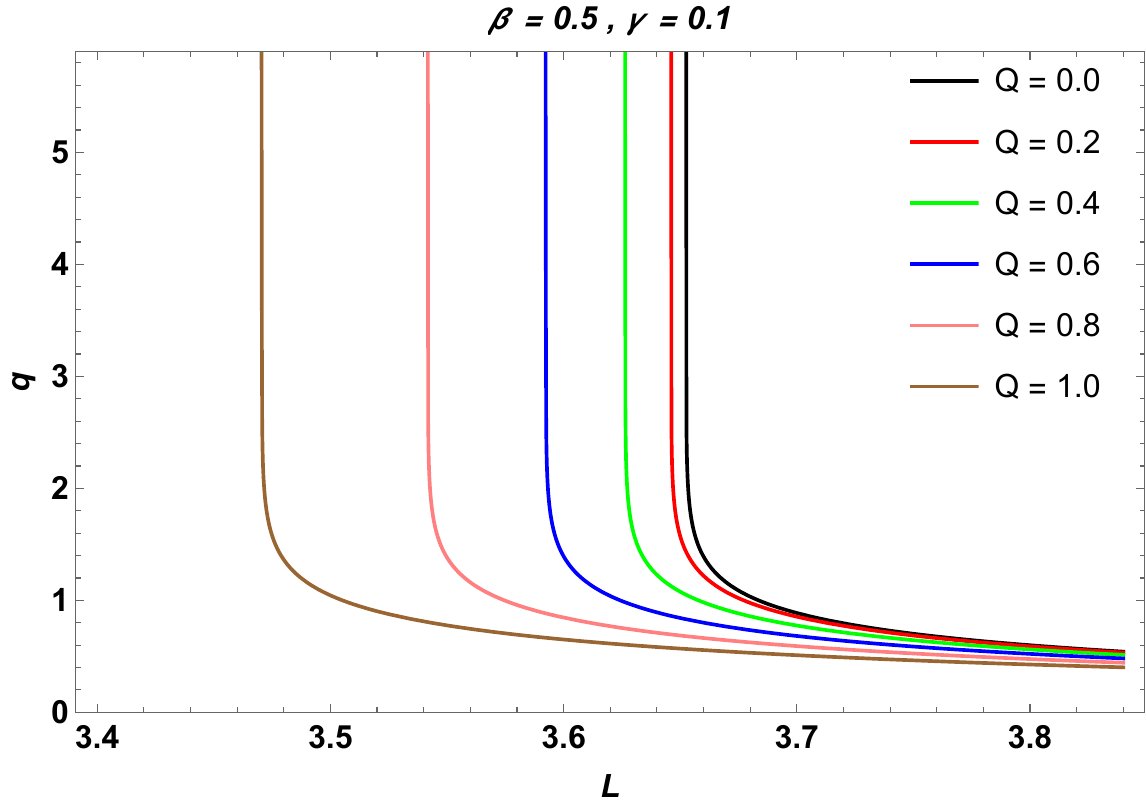}
\caption{The plot illustrates the rational number $q$ as a function of the energy $E$ (left panel) and orbital angular momentum $L$ (right panel). We fix $L=\frac{1}{2}(L_{MBO}+L_{ISCO})$ and $E=0.96$ for the left and right panels, respectively, and $\beta=0.5$ and $\gamma=0.1$ for both panels. }
\label{fig:q}
\end{figure*}
\begin{figure*}
    \centering
    \includegraphics[width=0.32\textwidth]{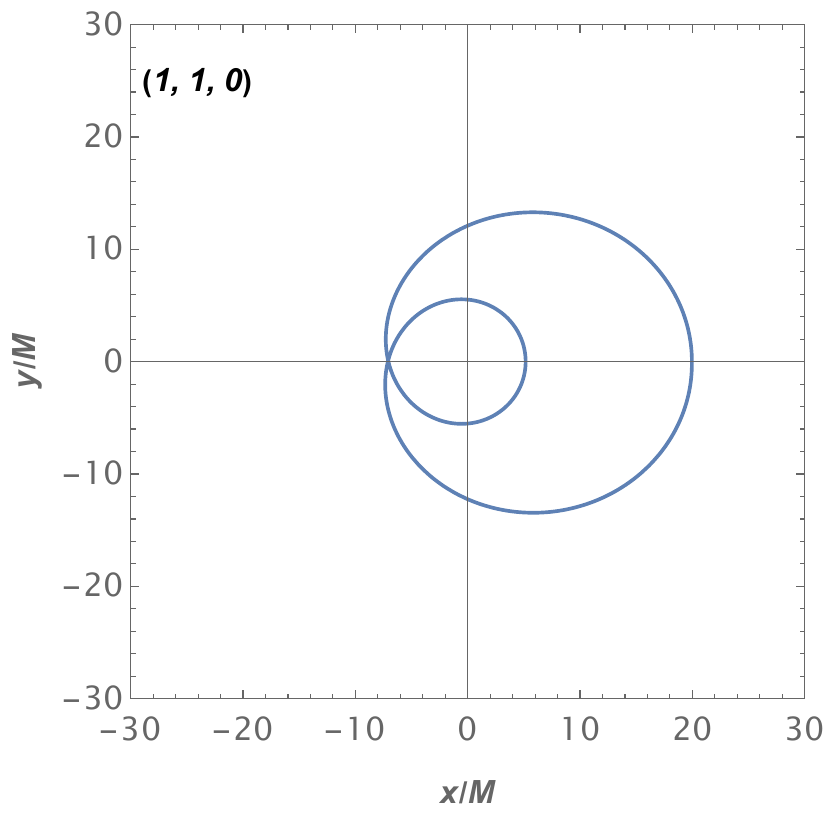} \hfill
    \includegraphics[width=0.32\textwidth]{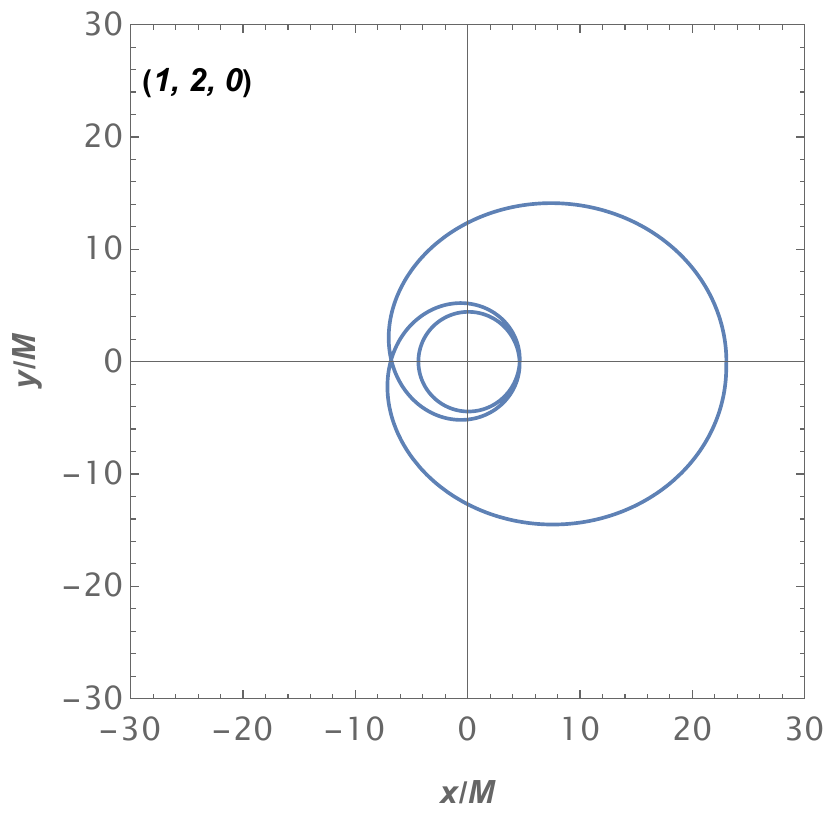} \hfill
    \includegraphics[width=0.32\textwidth]{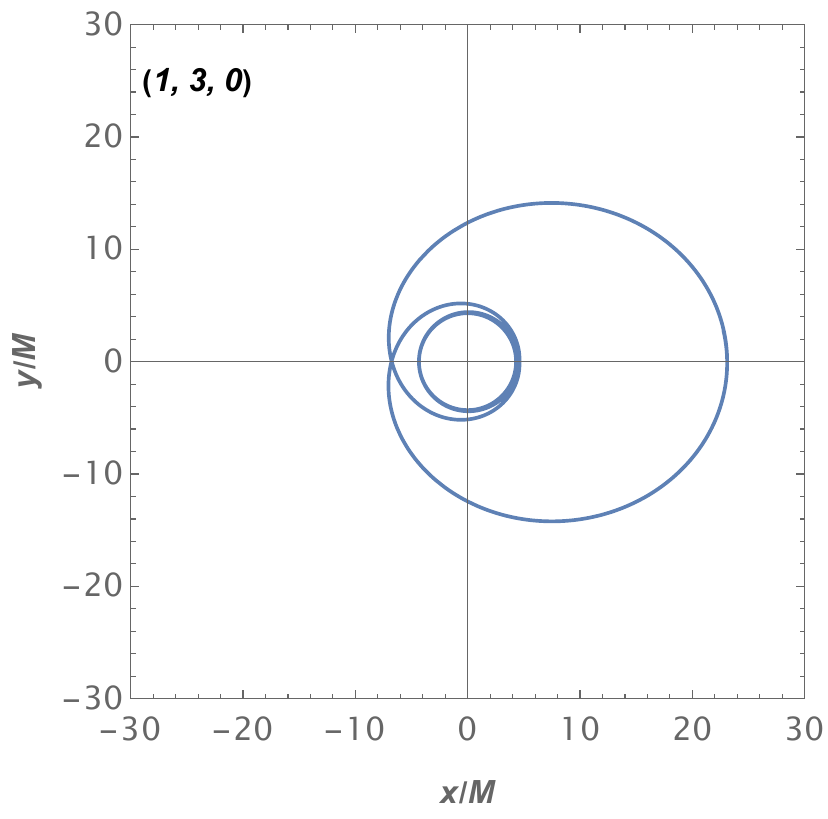} \\
    
    \vspace{0.2cm} 
    \includegraphics[width=0.32\textwidth]{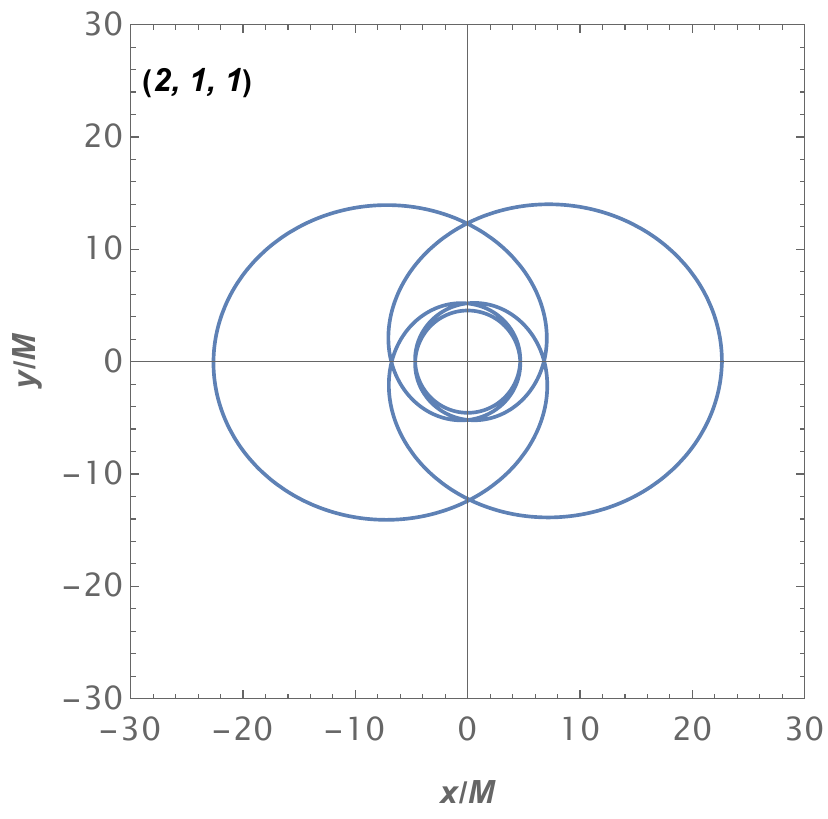} \hfill
    \includegraphics[width=0.32\textwidth]{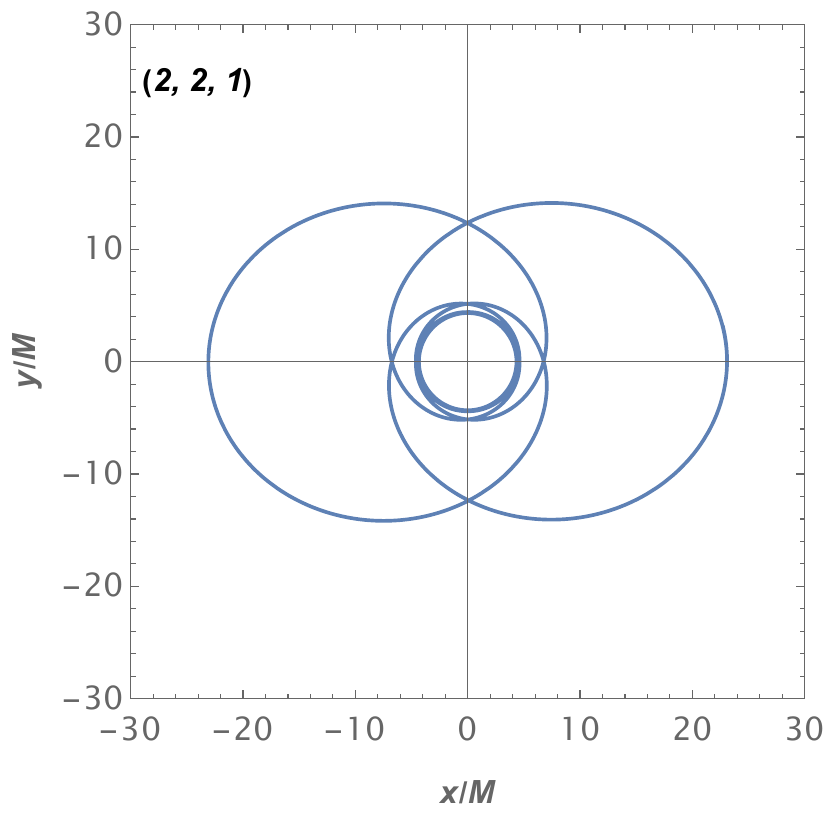} \hfill
    \includegraphics[width=0.32\textwidth]{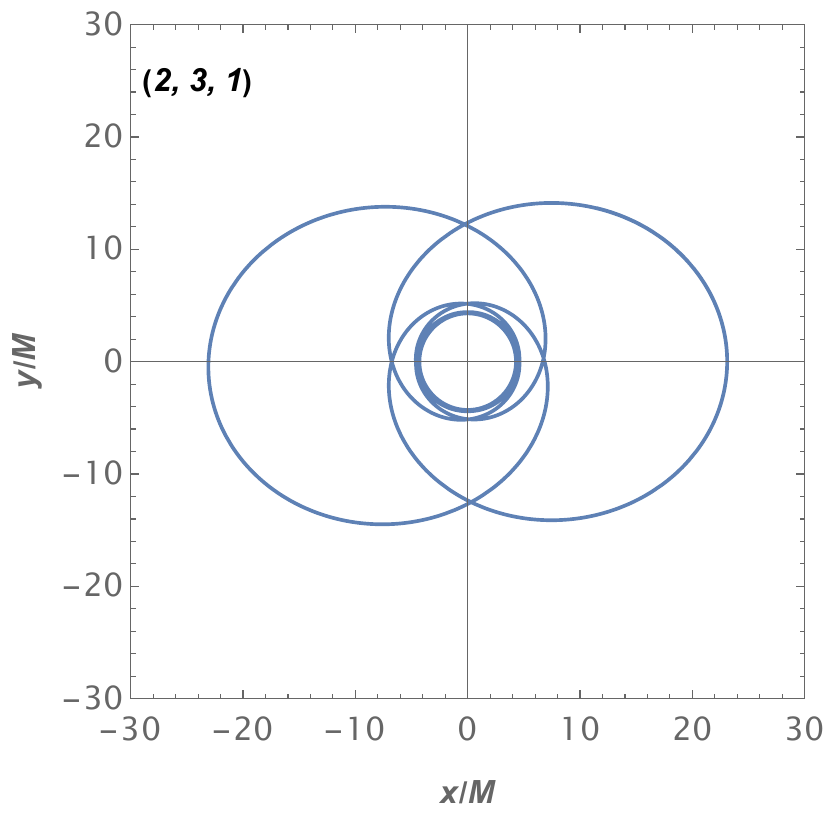} \\
    
    \vspace{0.2cm} 
    \includegraphics[width=0.32\textwidth]{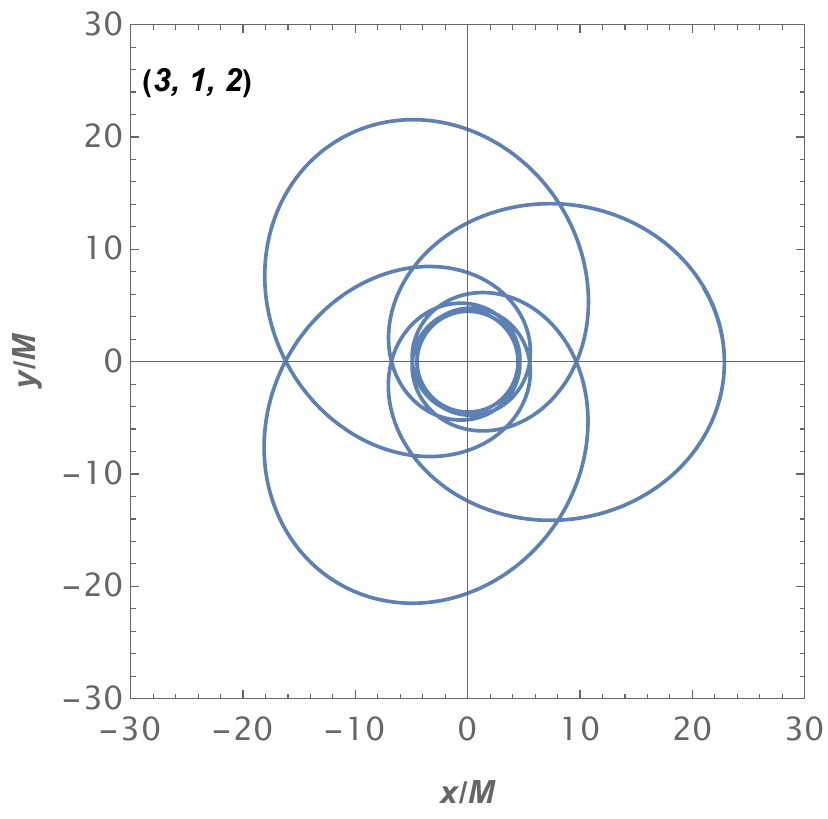} \hfill
    \includegraphics[width=0.32\textwidth]{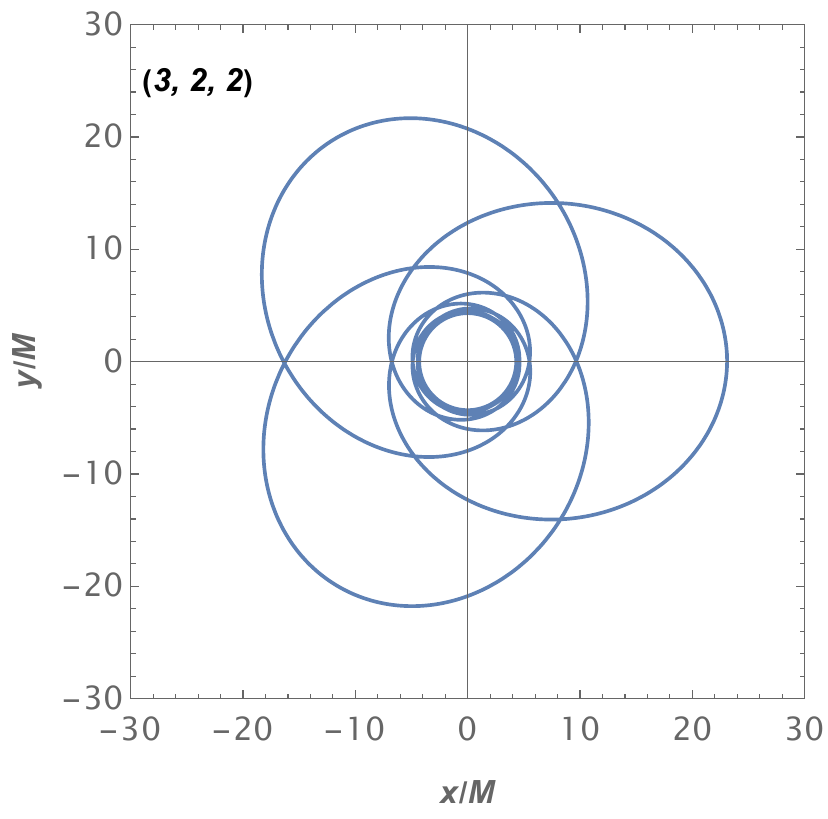} \hfill
    \includegraphics[width=0.32\textwidth]{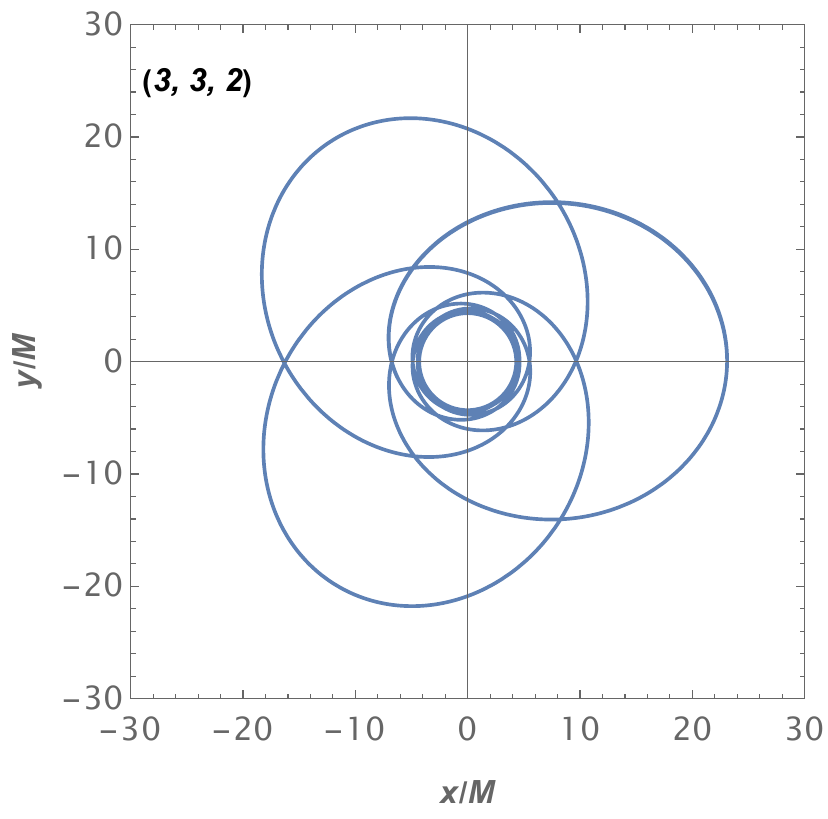} \\
    
    \vspace{0.2cm} 
    \includegraphics[width=0.32\textwidth]{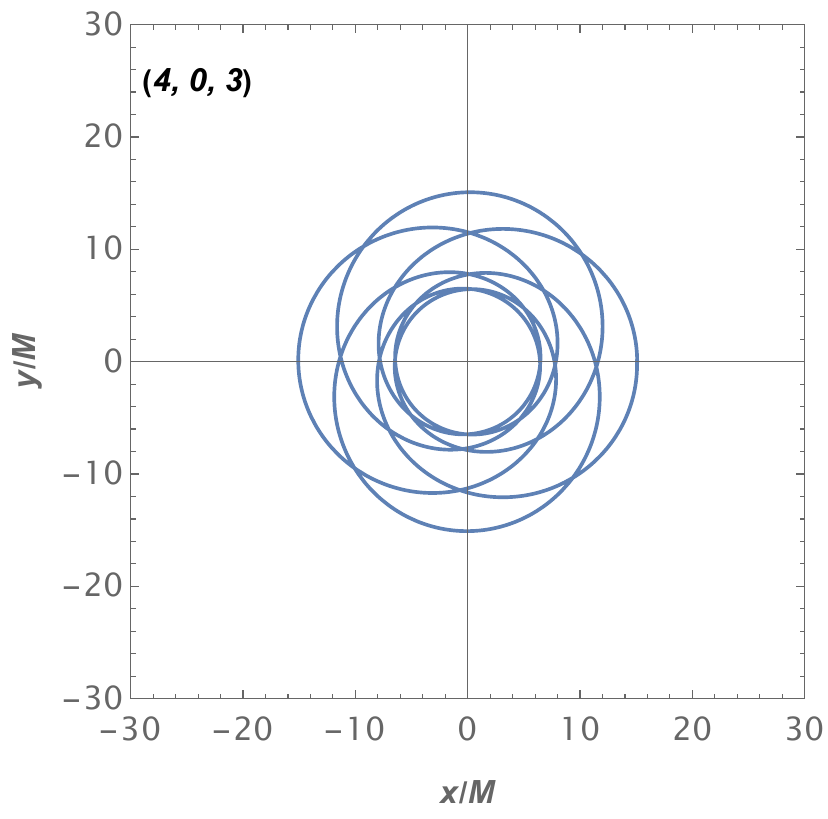} \hfill
    \includegraphics[width=0.32\textwidth]{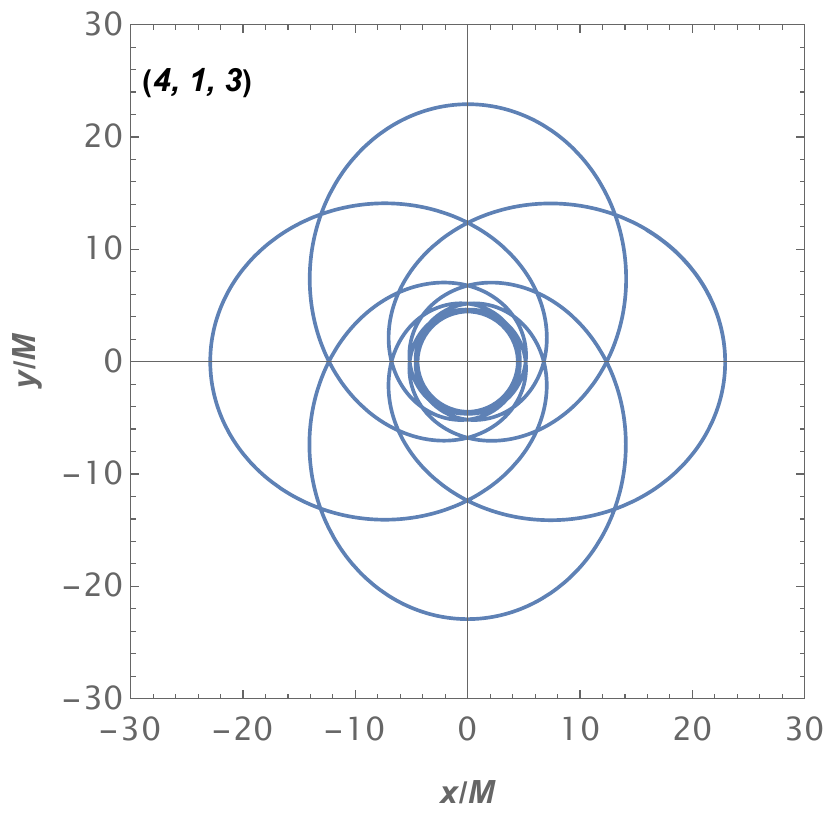} \hfill
    \includegraphics[width=0.32\textwidth]{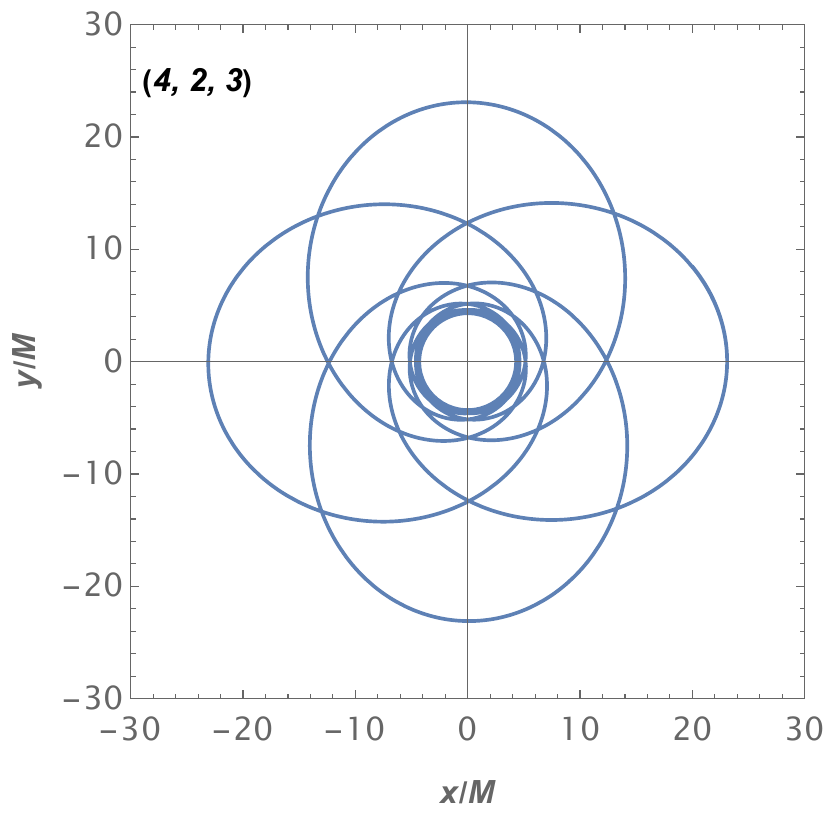}
    
    \caption{The plot illustrates the periodic orbits, which are characterized by different $z$, $w$ and $v$ integers. We set the spacetime parameters and orbital angular momentum as $Q=0.4$, $\beta=0.5$ \& $\gamma=0.1$ and $L=\frac{1}{2}(L_{MBO}+L_{ISCO})$.}
    \label{fig:orbits}
\end{figure*}
\begin{figure*}
\begin{subfigure}[b]{0.45\textwidth}
\includegraphics[width=\textwidth]{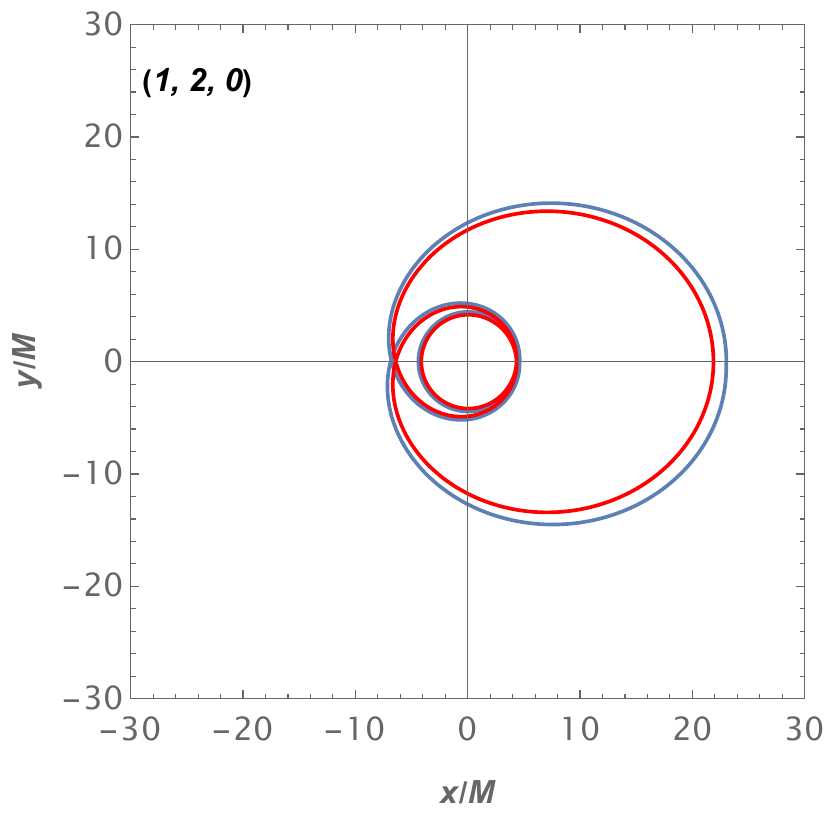}
\end{subfigure}
\begin{subfigure}[b]{0.52\textwidth}
\includegraphics[width=\textwidth]{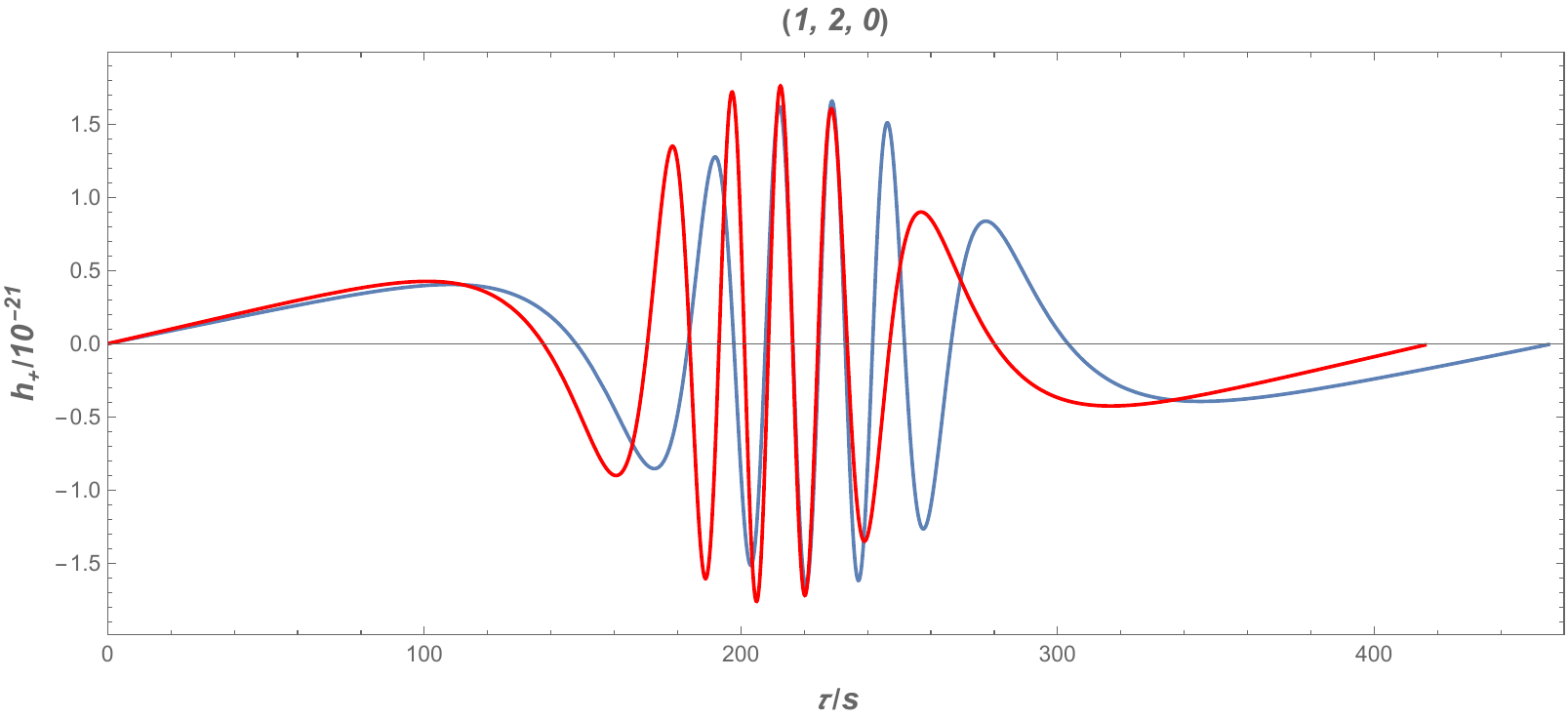}
\includegraphics[width=\textwidth]{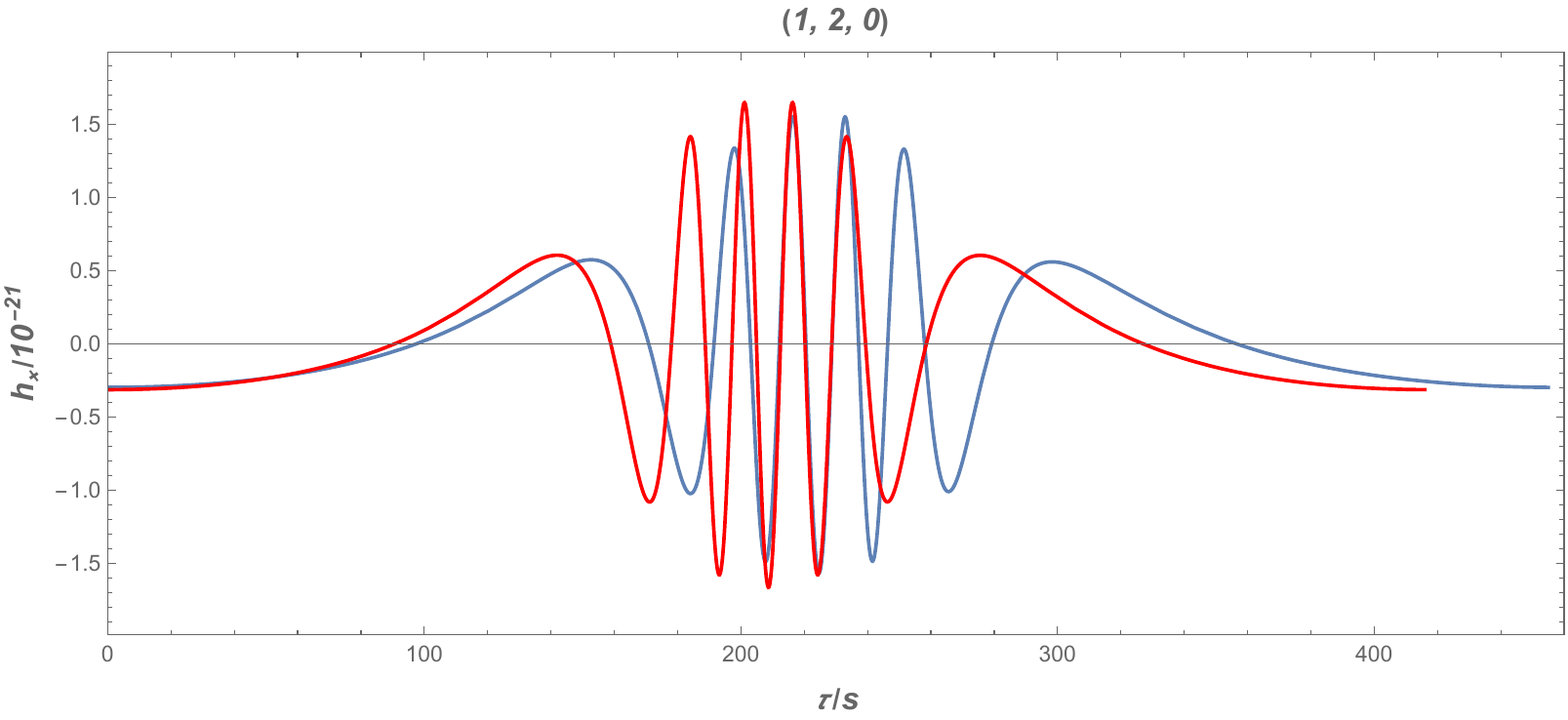}    
\end{subfigure} 
\caption{The plot represents the periodic orbit $(1,2,0)$  and the related gravitational waveforms of the EMRI system. The blue and red lines denote $Q=0.4$ and $Q=0.8$, respectively. The other spacetime parameters are fixed as $\beta=0.5$ and $\gamma=0.1$. }
\label{fig:gw1}
\end{figure*}
\begin{figure*}
\begin{subfigure}[b]{0.45\textwidth}
\includegraphics[width=\textwidth]{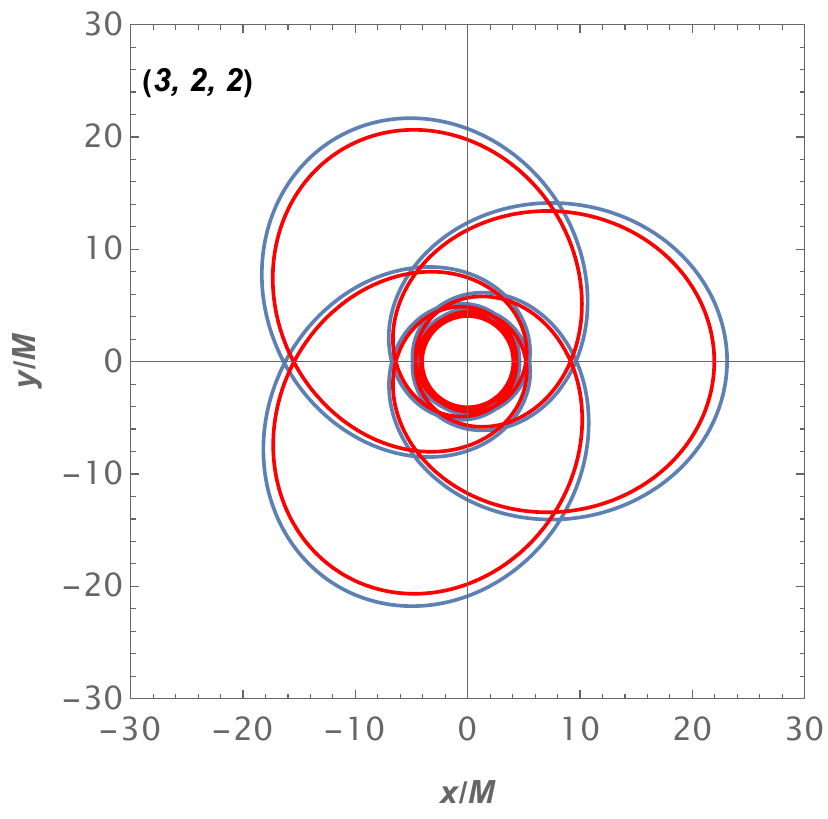}
\end{subfigure}
\begin{subfigure}[b]{0.52\textwidth}
\includegraphics[width=\textwidth]{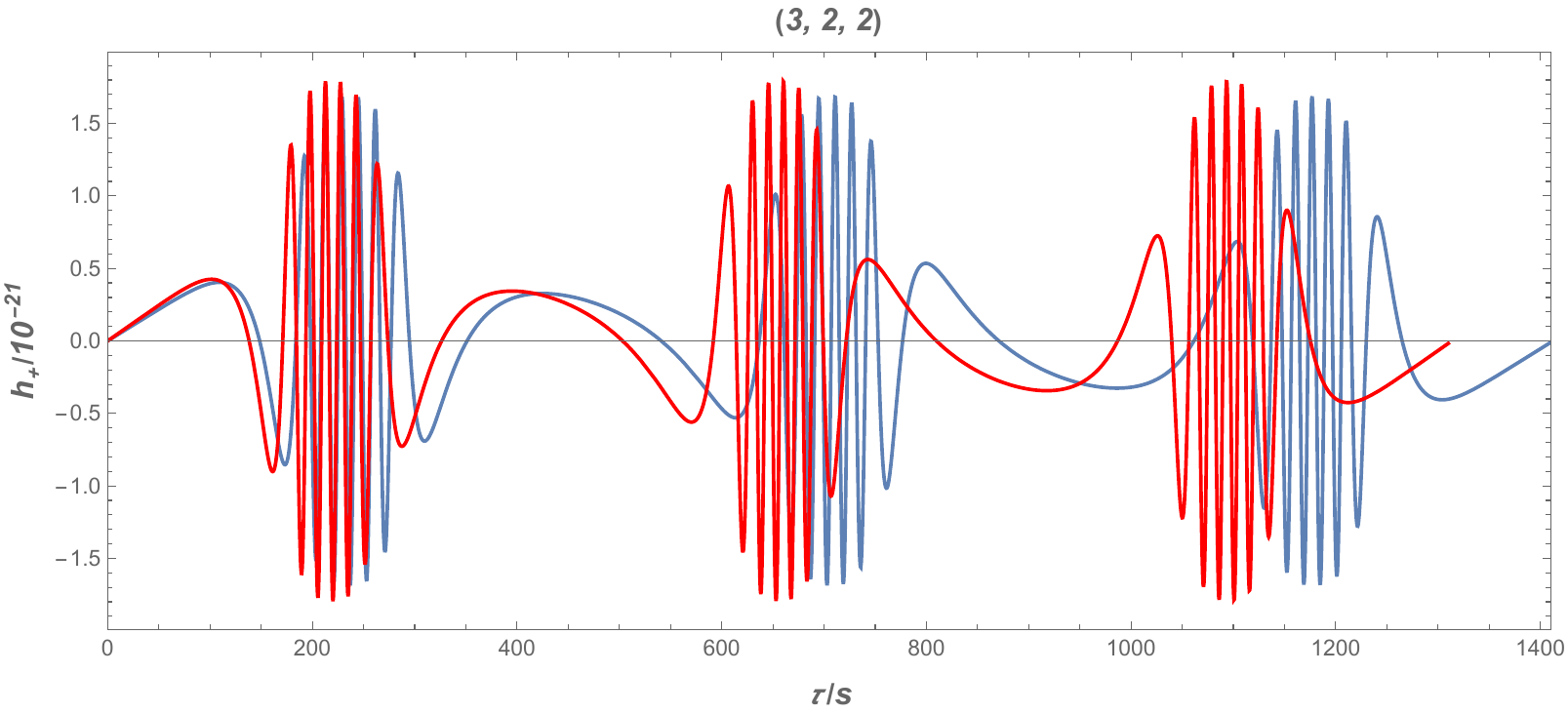}
\includegraphics[width=\textwidth]{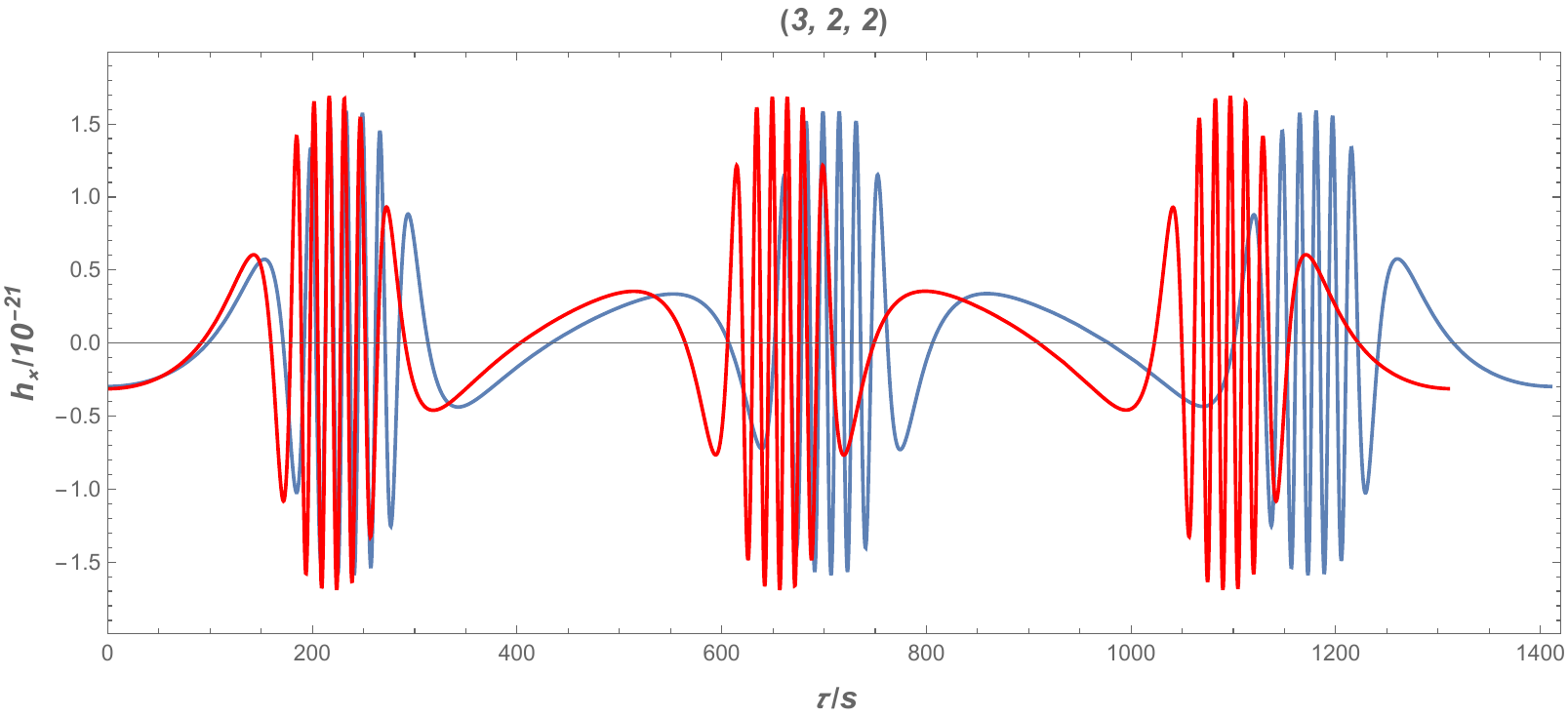}    
\end{subfigure} 
\caption{The plot represents the periodic orbit $(3,2,2)$ and the related gravitational waveforms of the EMRI system. The blue and red lines denote $Q=0.4$ and $Q=0.8$, respectively. The other spacetime parameters are fixed as $\beta=0.5$ and $\gamma=0.1$. }
\label{fig:gw2}
\end{figure*}
\begin{figure*}
\begin{subfigure}[b]{0.45\textwidth}
\includegraphics[width=\textwidth]{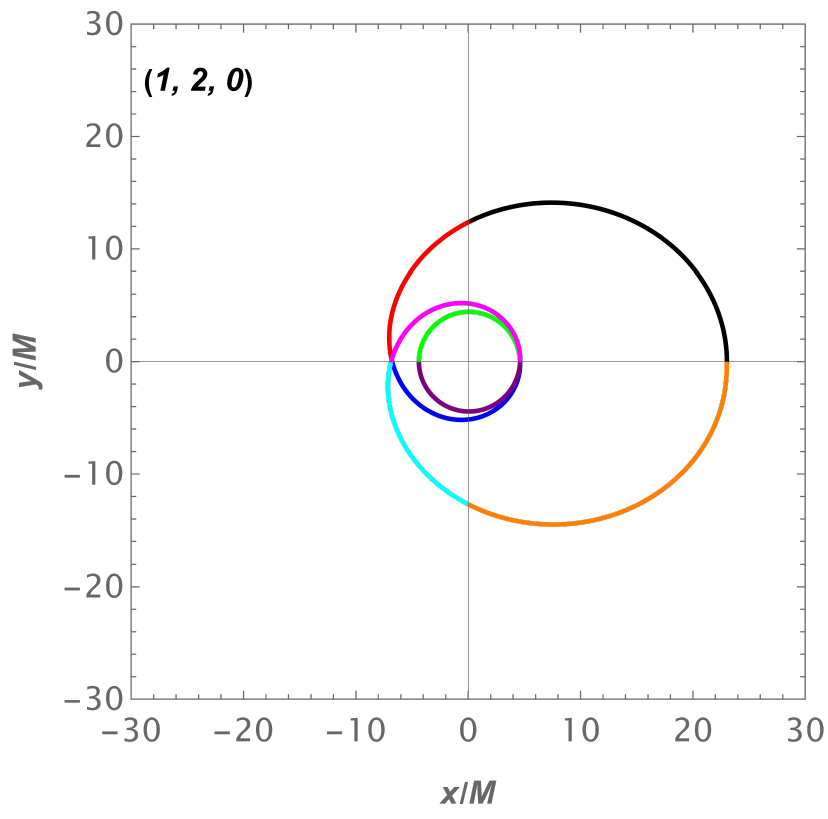}
\end{subfigure}
\begin{subfigure}[b]{0.52\textwidth}
\includegraphics[width=\textwidth]{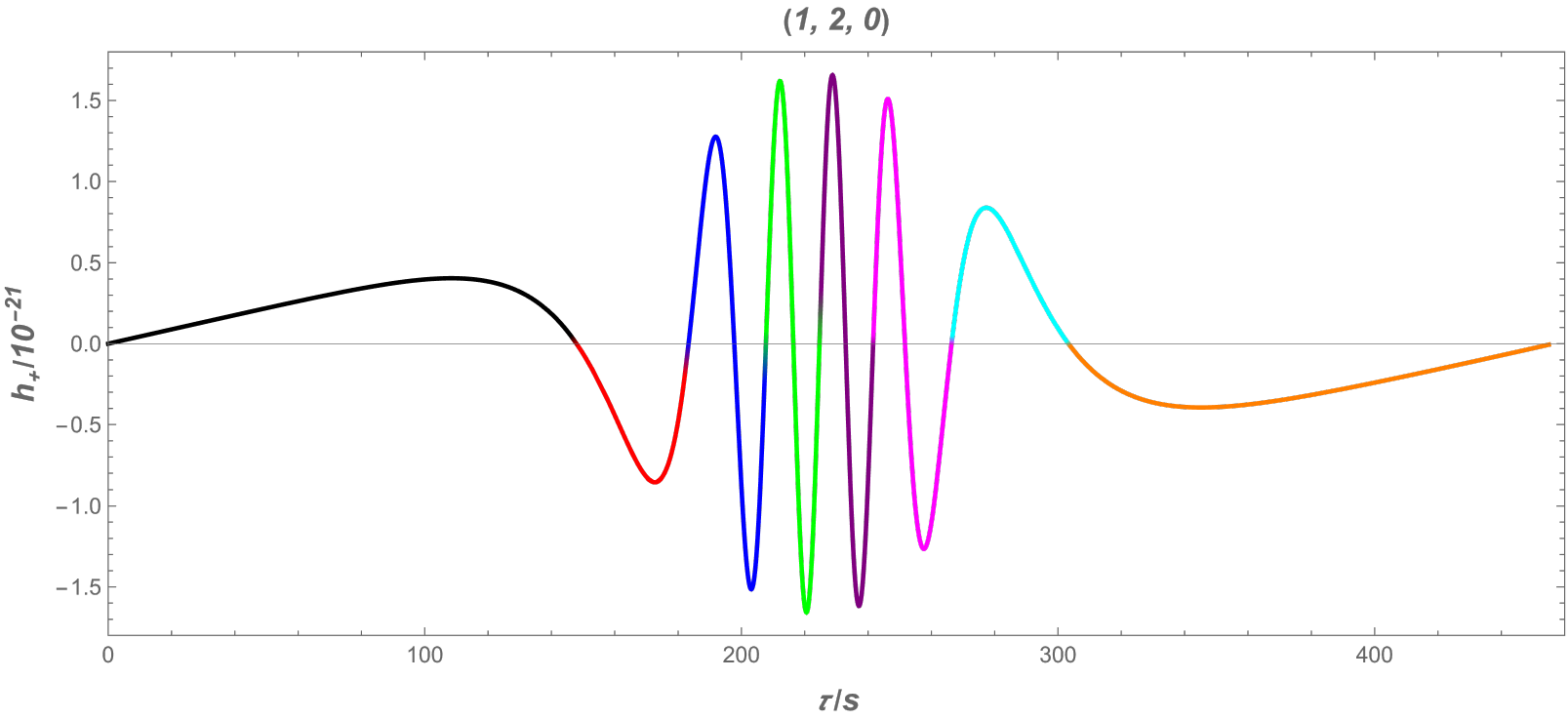}
\includegraphics[width=\textwidth]{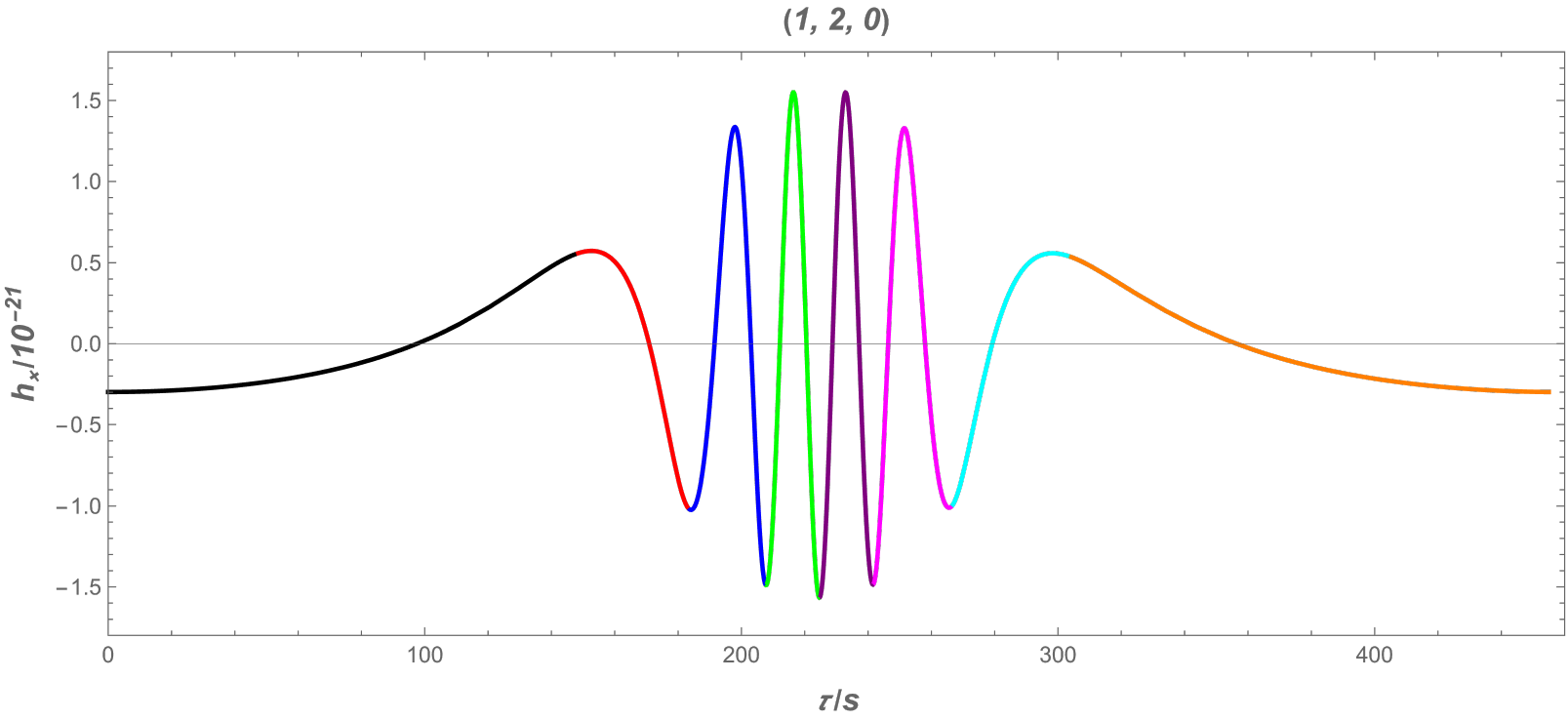}    
\end{subfigure} 
\caption{The figure demonstrates the $(1,2,0)$ periodic orbit and corresponding gravitational waveforms of the EMRI system. The different colors were used to indicate which part of the periodic orbit produces which part of the gravitational waveforms.}
\label{fig:gw3}
\end{figure*}
In this section, we provide a brief review of the spacetime metric in EMS theory together with the test particle dynamics around the BH. One can write the line element as follows~\cite{Gibbons88, Yu_2021}
\begin{equation}
ds^2=-f(r)dt^2+\frac{dr^2}{f(r)}+h(r)(d\theta^2+\sin^2{\theta}d\phi^2)\,,
\end{equation}
where $f(r)$ and $h(r)$ are the radial functions, and they can be written explicitly as
\begin{align}
    f(r)=1-\frac{2M}{r}+\frac{\beta Q^2}{h(r)}\,, \quad h(r)=r^2(1+\frac{\gamma Q^2}{M r})\,,
\end{align}
where $\beta$ and $\gamma$ refer to the dimensionless coupling constants, while $Q$ denotes the BH charge. One can recover the Schwarzschild BH when $Q=0$ or $\beta=\gamma=0$, while the RN BH when $\beta=1$ and $\gamma=0$. Fig.~\ref{fig:metric} illustrates the radial dependence of the metric function for the different values of the BH charge and coupling parameters. The event horizon radius can be found by solving $f(r)=0$. Therefore, one can see from this figure how the BH event horizon changes under the influence of the spacetime parameters.

Furthermore, we explore the dynamics of the test particles around the BH in EMS theory by using the Lagrangian formalism. The Lagrangian of the test particles with mass $m$ can be written as follows~\cite{1983mtbh.book.....C}
\begin{equation}
\mathcal{L}=\frac{1}{2}mg_{\mu\nu}\frac{dx^{\mu}}{d\tau}\frac{dx^{\nu}}{d\tau}\,,
\end{equation}
where $\tau$ represents the proper time. We set $m=1$ for simplicity, and the generalized momentum can be written as~\cite{1983mtbh.book.....C}
\begin{equation}
    p_{\mu}=\frac{\partial \mathcal{L}}{\partial \dot{x}^{\mu}}=g_{\mu\nu} \dot{x}^{\nu}\,.
\end{equation}
Using the above equation, one can obtain the equations of motion as
\begin{eqnarray}\label{eq:eqmotion}
p_{t} &=& -\left(1-\frac{2M}{r}+\frac{\beta Q^2}{h(r)}\right)\,\dot{t}=-E \, , \nonumber\\
p_{\phi} &=& h(r)\,\sin^{2}\theta\dot{\phi}=L \, , \nonumber \\
p_{r} &=& \left(1-\frac{2M}{r}+\frac{\beta Q^2}{h(r)}\right)^{-1}\dot{r} \, , \nonumber\\
p_{\theta} &=& h(r)\,\dot{\theta} \, ,
\end{eqnarray}
where $E$ and $L$ are, respectively, the energy and orbital angular momentum of the particle. The motion on the equatorial plane is considered in this work for simplicity, i.e., $\theta=\pi/2$. The following radial equation of motion can be written by using the normalization condition $g_{\mu \nu}\dot{x}^{\mu}\dot{x}^{\nu}=-1$
\begin{eqnarray}
\dot{r}^{2}+V_{eff}=E^{2}\, ,
\end{eqnarray}
where $V_{eff}$ refers to the effective potential, and it reads as
\begin{equation}
V_{eff}=\left(1-\frac{2M}{r}+\frac{\beta Q^2}{h(r)}\right)\left(1+\frac{L^{2}}{h(r)}\right)\, .
\end{equation}

Fig.~\ref{fig:eff} demonstrates the radial profile of the effective potential for the different values of the BH charge, coupling parameters and the orbital angular momentum. One can observe from the top panels of this figure that the values of the effective potential increase under the influence of the BH charge and the $\beta$ parameter. The bottom panel shows that there is a slight decrease with the increase of the $\gamma$ parameter, while the rise in orbital angular momentum causes an increase in the effective potential. 
It is worth noting that we focus on investigating periodic orbits around the BH in EMS theory. The following conditions must be satisfied for the periodic orbits~\cite{Dadhich22a}
\begin{equation}
    L_{ISCO}\leq L \quad\mbox{and}\quad E_{ISCO}\leq E \leq E_{MBO}=1\, ,
\end{equation}
where $E_{ISCO}$ and $L_{ISCO}$ are, respectively, the energy and the orbital angular momentum of the particle that is moving on the ISCO, while $E_{MBO}$ denotes the energy of the particle moving along the MBO. Therefore, examining the MBO and ISCO is essential to explore the characteristics of the periodic orbits. One can determine the MBO by using the following conditions
\begin{equation}\label{condition}
    V_{eff}=1\quad\mbox{and}\quad \frac{d V_{eff}}{dr}=0\,.
\end{equation}
The radius and orbital angular momentum of the MBO can be obtained numerically based on the above conditions. Fig.~\ref{fig:mbo} shows the orbital angular momentum (top panels) and radius (bottom panels) of the MBO as a function of the BH charge for different values of the $\beta$ and $\gamma$ parameters. We can see from these panels that there is a decrease in the values of the orbital angular momentum and radius of the MBO under the influence of the BH charge and $\beta$ parameter, and vice versa for the $\gamma$ parameter. The ISCO serves as another important bound orbit. We can define the ISCO parameters by setting the following conditions
\begin{equation}
    \dot{r}=0, \quad \frac{d V_{eff}}{dr}=0 \quad\mbox{and}\quad \frac{d^2V_{eff}}{dr^2}=0.
\end{equation}
We again choose the numerical way to investigate the ISCO parameters. We plot the dependence of the orbital angular momentum, radius and energy of the test particle moving along the ISCO on the BH charge for different values of the $\beta$ and $\gamma$ parameters. It can be seen from this figure that the values of the ISCO parameters decrease with the rise of the $\beta$ parameter and BH charge, and vice versa for the $\gamma$ parameter.

\section{Periodic orbits around the black hole in EMS theory}\label{section3}

This section is devoted to the periodic orbits of the test particles around the BH in EMS theory. It should be noted that every periodic orbit is characterized by the zoom, whirl and vertex numbers. The rational number, which is the ratio of the fundamental frequencies, can be defined as follows~\cite{Levin_2008}
\begin{equation}
q=\frac{\omega_{\phi}}{\omega_{r}}-1=w+\frac{v}{z}\, ,
\end{equation}
where $\omega_{\phi}$ and $\omega_r$ represent, respectively, the angular and radial frequencies. One can rewrite the above equation by considering the equations of motion as~\cite{Shabbir:2025kqh, Jiang:2024cpe, Alloqulov:2025ucf}
\begin{eqnarray}
q&=&\frac{1}{\pi} \int_{r_1}^{r_2}\frac{\dot{\phi}}{\dot{r}}-1 = \nonumber \\
&=&\frac{1}{\pi}\int_{r_1}^{r_2} \frac{1}{h(r)\sqrt{E^2-\left(1-\frac{2M}{r}+\frac{\beta Q^2}{h(r)}\right)\left(1+\frac{L^{2}}{h(r)}\right)}} dr
\end{eqnarray}
where $r_1$ and $r_2$ denote the periapsis and apoapsis radii of the periodic orbits, respectively. In order to get more information about the behaviour of the rational number, we plot it as a function of the energy and orbital angular momentum for the different values of the BH charge in Fig.~\ref{fig:q}. It is evident from this figure that $q$ grows as the energy $E$ increases, with a steep rise occurring as $E$ approaches its maximum value. On the other hand, $q$ decreases with increasing orbital angular momentum $L$, showing a sharp increase as $L$ nears its minimum. Additionally, the BH charge $Q$ causes the rational number $q$ to shift toward smaller values of both energy and orbital angular momentum. Next, we turn to demonstrate the periodic orbits by determining the associated values of the energy, while maintaining the orbital angular momentum as $L=\frac{1}{2}(L_{ISCO}+L_{MBO})$. Fig.~\ref{fig:orbits} illustrates the periodic orbits, which are characterized by different $(z,w,v)$ of the test particles around the BH in EMS theory. One can observe from this figure that the massive particles possessing a larger zoom number $z$ display periodic orbits with increasingly intricate structural configurations, while those with a greater whirl number $w$ complete extra loops around the BH before reaching the next apoapsis.

\section{Numerical kludge gravitational waveforms}\label{section4}

Here, we investigate the gravitational waveforms from the periodic orbits around the BH in EMS theory. To construct the gravitational waveforms, we consider the extreme mass-ratio inspiral (EMRI) system. It contains a stellar-mass object moving in a periodic orbit around a supermassive EMS BH. The GWs produced by this EMRI system could carry signatures of the periodic orbits and the supermassive EMS BH. The adiabatic approximation method, which is a widely recognized method, can be utilized to calculate the gravitational waveforms. Over the brief term, both energy and orbital angular momentum stay approximately unchanged, since the small body's orbital parameters shift over timescales far exceeding its orbital period. Consequently, across a limited number of cycles, the motion can be treated as a geodesic, allowing the gravitational waveforms generated by these periodic orbits to be characterized over a full cycle. Importantly, within such a short duration, the influence of gravitational radiation on the small object's trajectory can be neglected. To investigate the gravitational waveforms, we first solve the equations of motion numerically. Then, we generate the gravitational waveforms by applying the symmetric and trace-free (STF) mass quadrupole equation to gravitational radiation~\cite{Yang:2024lmj}. It can be written as follows~\cite{RevMod}
\begin{equation}
I^{ij}=\Big[\int d^3 xx^ix^jT^{tt}(t,x^i)\Big]^{(\text{STF})}  \,.  
\end{equation}
Here, $T^{tt}$ denotes the $tt$-component of the stress-energy tensor for the small celestial object on the trajectory $Z^i(t)$. One can define $T^{tt}$ as
\begin{equation}
T^{tt}(t,x^i)=m\delta^3(x^i-Z^i(t))\, .
\end{equation}
It is important to project the trajectory of the secondary celestial object into the Cartesian coordinate system by utilizing the Boyer-Lindquist coordinates as a fictitious spherical polar coordinate system
\begin{align}
x=r\sin{\theta}\cos{\phi}, \quad y=r\sin{\theta}\sin{\phi}, \quad z=r\cos{\theta}\, .
\end{align}
By considering the above equations, one can obtain the metric perturbations as
\begin{equation}
h_{ij}=\frac{2}{D_L}\frac{d^2I_{ij}}{dt^2}=\frac{2m}{D_L}(a_ix_j+a_jx_i+2v_iv_j)\,,
\end{equation}
where $v_i$ and $a_i$, respectively, denote the velocity and acceleration of the secondary celestial object. This waveform illustrates the fundamental propagation behavior of gravitational waves through space. To match what actual detectors would observe, the modeled gravitational wave signal must be transformed into the detector's reference frame. Consequently, it is necessary to project the gravitational wave onto a coordinate system that aligns with the detector's orientation~\cite{Poisson_Will_2014, 2024arXiv241101858M}. The coordinate basis can be written in the following form
\begin{equation}\label{coordsystem}
\begin{array}{l}
e_{X}=[\cos\zeta,-\sin\zeta,0]\,,\\
e_{Y}=[\cos\iota\sin\zeta,\cos\iota\cos\zeta,-\sin\iota]\,,\\
e_{Z}=[\sin\iota\sin\zeta,\sin\iota\cos\zeta,\cos\iota]\,.
\end{array}
\end{equation}
In the above equation, $\zeta$ denotes the longitude of the periastron, while $\iota$ represents the inclination angle of the test particle's orbit relative to the line of sight. The vectors $e_X$, $e_Y$, and $e_Z$ denote the orthogonal coordinate basis of the detector's reference frame, which serve to decompose the gravitational wave signal into its individual polarization components. After that, the corresponding GW polarizations take the following form~\cite{Poisson_Will_2014}
\begin{eqnarray}
 h_{+}&=&\frac{1}{2}(e^i_X e^j_X-e^i_Ye^j_Y)h_{ij}, \\
 h_{\times}&=&\frac{1}{2}(e^i_Xe^j_Y+e^i_Ye^j_X)h_{ij}.
\end{eqnarray}
The above equation can be written by considering Eq.~(\ref{coordsystem}) as
\begin{eqnarray}
h_{+}&=&\frac{1}{2}(h_{\zeta\zeta}-h_{\iota\iota}), \\
h_{\times}&=& h_{\iota\zeta}.
\end{eqnarray}
\begin{eqnarray}
h_{\zeta\zeta}&=&h_{xx}\cos^2{\zeta}-h_{xy}\sin{2\zeta}+h_{yy}\sin^2{\zeta}, \\
h_{\iota\iota}&=&\cos{\iota}[h_{xx}\sin^2{\zeta}+h_{xy}\sin{2\zeta}+h_{yy}\cos^2{\zeta}]\nonumber\\&+&h_{zz}\sin^2{\iota}-\sin{2\iota}[h_{xz}\sin{\zeta}+h_{yz}\cos{\zeta}], \\
h_{\iota\zeta}&=&\cos{\iota}\Big[\frac{1}{2}h_{xx}\sin{2\zeta}+h_{xy}\cos{2\zeta}-\frac{1}{2}h_{yy}\sin{2\zeta}\Big]\nonumber\\&+&\sin{\iota}[h_{yz}\sin{\zeta}-h_{xz}\cos{\zeta}].
\end{eqnarray}

In order to vizualize the GWs, we consider the EMRI system. We assume the masses of the primary and secondary objects in the EMRI system are $M\sim 10^7 M_{\odot}$ and $m\sim10M_{\odot}$, respectively. Also, we set the luminosity distance as $D_{L}=200 \text{Mpc}$, and the inclination angle and latitude as $\iota=\zeta=\pi/4$. In Fig.~\ref{fig:gw1}, we plot the $(1,2,0)$ periodic orbit for $Q=0.4$ (blue line) and $Q=0.8$ (red line) and the corresponding gravitational waveforms by fixing other spacetime parameters as $\beta=0.5$ and $\gamma=0.1$. To be more informative, we plot a similar figure for the $(3,2,2)$ periodic orbit in Fig.~\ref{fig:gw2}. It can be seen from these figures that the gravitational waveforms comprise two distinct stages, which are the zoom and whirl stages. The zoom stage corresponds to the calm portion of the waveform, during which GWs are emitted as the small object follows an elliptical orbit far from the SMBH. In contrast, the whirl stage is represented by the rapidly oscillating portion, which arises when the small object begins a whirl motion in close proximity to the SMBH. Notably, the sharp increase in GW frequency during the whirl stage directly gives rise to these intense oscillations. These figures also present a comparison between the $Q=0.4$ and $Q=0.8$ cases, represented by the blue and red lines, respectively. As the charge increases, both the zoom and whirl regions shrink, causing the gravitational wave signal to arrive earlier. In addition, in Fig.~\ref{fig:gw3}, we plot the $(1,2,0)$ periodic orbit and its corresponding gravitational waveforms, using different colors to highlight which part of the periodic orbit produces which part of the gravitational waveforms. One can also see from this figure that the zoom and whirl stages correspond to the calm and rapidly oscillating portions, respectively.

\section{Conclusions}\label{con}

In this paper, we comprehensively examined the dynamics of test particles around the BH in EMS theory. First, we provided a brief analysis of the spacetime metric in EMS theory. We then turned to analyze the motion of the test particles around the BH in EMS theory by using the Lagrangian formalism. From the equation of motion, we derived the effective potential analytically and examined its dependence on the orbital and spacetime parameters. We showed that the potential barrier of the effective potential increases with increasing the BH charge, the parameter $\beta$ and the orbital angular momentum $L$. In contrast, increasing the parameter $\gamma$ decreases the shape of the effective potential (see Fig.~\ref{fig:eff}). These results demonstrated that the EMS spacetime parameters significantly modify the orbital dynamics of test particles in the vicinity of the black hole.

We further investigated the MBO and ISCO by imposing the relevant conditions on the effective potential and examined the MBO and ISCO parameters on the BH charge for different values of the coupling parameters. Our results showed that both the MBO and ISCO parameters decrease with increasing BH charge and the parameter $\beta$, while they increase with increasing the parameter $\gamma$ (see Figs.~\ref{fig:mbo} and~\ref{fig:isco}). 

Next, we investigated the influence the BH charge $Q$ on the rational number $q$ numerically. We found that increasing the BH charge $Q$ shifts $q$-curves toward lower values of both the energy $E$ and orbital angular momentum $L$ (see Fig.~\ref{fig:q}). This behavior shows that the BH charge $Q$ significantly modifies the energy and angular momentum conditions required for periodic motion. We then numerically determined the energy of periodic orbits characterized by different configurations $(z,w,v)$, fixing the orbital angular momentum at $L=\frac{1}{2}(L_{ISCO}+L_{MBO})$. Using these results, we constructed the corresponding periodic trajectories and the orbital structure in EMS spacetime (see Fig.~\ref{fig:orbits}).  

Finally, we examined the gravitational wave signatures of periodic motion in an EMRI system consisting of the SMBH and stellar-mass compact object with masses $M\sim 10^{7}M_{\odot}$ and $m\sim 10M_{\odot}$, respectively. Analyzing periodic orbits with configurations $(1,2,0)$ and $(3,2,2)$ and their corresponding gravitational waveforms for different values of the BH charge, we found that the charge modifies the spacetime geometry and therefore changes the orbital frequencies and phase evolution of the system (see Figs.~\ref{fig:gw1} and~\ref{fig:gw2}). These modifications are directly imprinted on the emitted gravitational-wave signal. 

To further illustrate the connection between the orbital dynamics and the waveform structure, we considered different segments of the periodic orbit $(1,2,0)$ to the corresponding features of the gravitational waveform using different colors (see Fig.~\ref{fig:gw3}). In addition, the resulting correlation shows that changes in the orbital dynamics induced by the BH charge are encoded in the structure of the gravitational waveforms. These results suggest that EMRI waveforms can, in principle, serve as a sensitive probe of the underlying EMS BH geometry and help constrain deviations from the corresponding BH spacetime.

\section*{ACKNOWLEDGEMENT}

This research was funded by the National Natural Science Foundation of China (NSFC) under Grant No. U2541210.

\bibliographystyle{apsrev4-1}  

\bibliography{ref}

\begin{thebibliography}{110}%
\makeatletter
\providecommand \@ifxundefined [1]{%
 \@ifx{#1\undefined}
}%
\providecommand \@ifnum [1]{%
 \ifnum #1\expandafter \@firstoftwo
 \else \expandafter \@secondoftwo
 \fi
}%
\providecommand \@ifx [1]{%
 \ifx #1\expandafter \@firstoftwo
 \else \expandafter \@secondoftwo
 \fi
}%
\providecommand \natexlab [1]{#1}%
\providecommand \enquote  [1]{``#1''}%
\providecommand \bibnamefont  [1]{#1}%
\providecommand \bibfnamefont [1]{#1}%
\providecommand \citenamefont [1]{#1}%
\providecommand \href@noop [0]{\@secondoftwo}%
\providecommand \href [0]{\begingroup \@sanitize@url \@href}%
\providecommand \@href[1]{\@@startlink{#1}\@@href}%
\providecommand \@@href[1]{\endgroup#1\@@endlink}%
\providecommand \@sanitize@url [0]{\catcode `\\12\catcode `\$12\catcode `\&12\catcode `\#12\catcode `\^12\catcode `\_12\catcode `\%12\relax}%
\providecommand \@@startlink[1]{}%
\providecommand \@@endlink[0]{}%
\providecommand \url  [0]{\begingroup\@sanitize@url \@url }%
\providecommand \@url [1]{\endgroup\@href {#1}{\urlprefix }}%
\providecommand \urlprefix  [0]{URL }%
\providecommand \Eprint [0]{\href }%
\providecommand \doibase [0]{http://dx.doi.org/}%
\providecommand \selectlanguage [0]{\@gobble}%
\providecommand \bibinfo  [0]{\@secondoftwo}%
\providecommand \bibfield  [0]{\@secondoftwo}%
\providecommand \translation [1]{[#1]}%
\providecommand \BibitemOpen [0]{}%
\providecommand \bibitemStop [0]{}%
\providecommand \bibitemNoStop [0]{.\EOS\space}%
\providecommand \EOS [0]{\spacefactor3000\relax}%
\providecommand \BibitemShut  [1]{\csname bibitem#1\endcsname}%
\let\auto@bib@innerbib\@empty
\bibitem [{\citenamefont {Abbott}\ and\ \citenamefont {et~al.}(2016)}]{Abbott_2016}%
  \BibitemOpen
  \bibfield  {author} {\bibinfo {author} {\bibfnamefont {B.}~\bibnamefont {Abbott}}\ and\ \bibinfo {author} {\bibnamefont {et~al.}},\ }\href {\doibase 10.1103/physrevlett.116.061102} {\bibfield  {journal} {\bibinfo  {journal} {Physical Review Letters}\ }\textbf {\bibinfo {volume} {116}} (\bibinfo {year} {2016}),\ 10.1103/physrevlett.116.061102}\BibitemShut {NoStop}%
\bibitem [{\citenamefont {Akiyama}\ and\ \citenamefont {et~al.}(2019)}]{Akiyama19L1}%
  \BibitemOpen
  \bibfield  {author} {\bibinfo {author} {\bibfnamefont {K.}~\bibnamefont {Akiyama}}\ and\ \bibinfo {author} {\bibnamefont {et~al.}},\ }\href {\doibase 10.3847/2041-8213/ab0ec7} {\bibfield  {journal} {\bibinfo  {journal} {ApJ.}\ }\textbf {\bibinfo {volume} {875}},\ \bibinfo {eid} {L1} (\bibinfo {year} {2019})},\ \Eprint {http://arxiv.org/abs/1906.11238} {arXiv:1906.11238 [astro-ph.GA]} \BibitemShut {NoStop}%
\bibitem [{\citenamefont {{Akiyama}}\ and\ \citenamefont {et~al.}(2019)}]{Akiyama19L6}%
  \BibitemOpen
  \bibfield  {author} {\bibinfo {author} {\bibfnamefont {K.}~\bibnamefont {{Akiyama}}}\ and\ \bibinfo {author} {\bibnamefont {et~al.}},\ }\href {\doibase 10.3847/2041-8213/ab1141} {\bibfield  {journal} {\bibinfo  {journal} {Astrophys. J.}\ }\textbf {\bibinfo {volume} {875}},\ \bibinfo {eid} {L6} (\bibinfo {year} {2019})},\ \Eprint {http://arxiv.org/abs/1906.11243} {arXiv:1906.11243 [astro-ph.GA]} \BibitemShut {NoStop}%
\bibitem [{\citenamefont {Collaboration}\ and\ \citenamefont {et~al.}(2022)}]{Event}%
  \BibitemOpen
  \bibfield  {author} {\bibinfo {author} {\bibfnamefont {E.~H.~T.}\ \bibnamefont {Collaboration}}\ and\ \bibinfo {author} {\bibnamefont {et~al.}},\ }\href {\doibase 10.3847/2041-8213/ac6674} {\bibfield  {journal} {\bibinfo  {journal} {The Astrophysical Journal Letters}\ }\textbf {\bibinfo {volume} {930}},\ \bibinfo {pages} {L12} (\bibinfo {year} {2022})}\BibitemShut {NoStop}%
\bibitem [{\citenamefont {{Green}}\ \emph {et~al.}(1987)\citenamefont {{Green}}, \citenamefont {{Schwarz}},\ and\ \citenamefont {{Witten}}}]{Green87book}%
  \BibitemOpen
  \bibfield  {author} {\bibinfo {author} {\bibfnamefont {M.~B.}\ \bibnamefont {{Green}}}, \bibinfo {author} {\bibfnamefont {J.~H.}\ \bibnamefont {{Schwarz}}}, \ and\ \bibinfo {author} {\bibfnamefont {E.}~\bibnamefont {{Witten}}},\ }\href@noop {} {\emph {\bibinfo {title} {{Superstring theory. Volume 1 - Introduction}}}}\ (\bibinfo {year} {1987})\BibitemShut {NoStop}%
\bibitem [{\citenamefont {Gibbons}\ and\ \citenamefont {ichi Maeda}(1988)}]{Gibbons88}%
  \BibitemOpen
  \bibfield  {author} {\bibinfo {author} {\bibfnamefont {G.}~\bibnamefont {Gibbons}}\ and\ \bibinfo {author} {\bibfnamefont {K.}~\bibnamefont {ichi Maeda}},\ }\href {\doibase https://doi.org/10.1016/0550-3213(88)90006-5} {\bibfield  {journal} {\bibinfo  {journal} {Nuclear Physics B}\ }\textbf {\bibinfo {volume} {298}},\ \bibinfo {pages} {741} (\bibinfo {year} {1988})}\BibitemShut {NoStop}%
\bibitem [{\citenamefont {Yu}\ \emph {et~al.}(2021)\citenamefont {Yu}, \citenamefont {Qiu},\ and\ \citenamefont {Gao}}]{Yu_2021}%
  \BibitemOpen
  \bibfield  {author} {\bibinfo {author} {\bibfnamefont {S.}~\bibnamefont {Yu}}, \bibinfo {author} {\bibfnamefont {J.}~\bibnamefont {Qiu}}, \ and\ \bibinfo {author} {\bibfnamefont {C.}~\bibnamefont {Gao}},\ }\href {\doibase 10.1088/1361-6382/abf2f5} {\bibfield  {journal} {\bibinfo  {journal} {Classical and Quantum Gravity}\ }\textbf {\bibinfo {volume} {38}},\ \bibinfo {pages} {105006} (\bibinfo {year} {2021})}\BibitemShut {NoStop}%
\bibitem [{\citenamefont {Reissner}(1916)}]{Reissner1916}%
  \BibitemOpen
  \bibfield  {author} {\bibinfo {author} {\bibfnamefont {H.}~\bibnamefont {Reissner}},\ }\href {\doibase 10.1002/andp.19163550905} {\bibfield  {journal} {\bibinfo  {journal} {Annalen der Physik}\ }\textbf {\bibinfo {volume} {355}},\ \bibinfo {pages} {106} (\bibinfo {year} {1916})}\BibitemShut {NoStop}%
\bibitem [{\citenamefont {Nordstr{\"o}m}(1918)}]{Nordstrom1918}%
  \BibitemOpen
  \bibfield  {author} {\bibinfo {author} {\bibfnamefont {G.}~\bibnamefont {Nordstr{\"o}m}},\ }\href@noop {} {\bibfield  {journal} {\bibinfo  {journal} {Koninklijke Nederlandse Akademie van Wetenschappen Proceedings Series B Physical Sciences}\ }\textbf {\bibinfo {volume} {20}},\ \bibinfo {pages} {1238} (\bibinfo {year} {1918})}\BibitemShut {NoStop}%
\bibitem [{\citenamefont {{Schwarzschild}}(1916)}]{1916SPAW.......189S}%
  \BibitemOpen
  \bibfield  {author} {\bibinfo {author} {\bibfnamefont {K.}~\bibnamefont {{Schwarzschild}}},\ }\href@noop {} {\bibfield  {journal} {\bibinfo  {journal} {Sitzungsberichte der Königlich Preussischen Akademie der Wissenschaften}\ ,\ \bibinfo {pages} {189}} (\bibinfo {year} {1916})}\BibitemShut {NoStop}%
\bibitem [{\citenamefont {{Blinder}}(2015)}]{2015arXiv151202061B}%
  \BibitemOpen
  \bibfield  {author} {\bibinfo {author} {\bibfnamefont {S.~M.}\ \bibnamefont {{Blinder}}},\ }\href {\doibase 10.48550/arXiv.1512.02061} {\bibfield  {journal} {\bibinfo  {journal} {arXiv e-prints}\ ,\ \bibinfo {eid} {arXiv:1512.02061}} (\bibinfo {year} {2015})},\ \Eprint {http://arxiv.org/abs/1512.02061} {arXiv:1512.02061 [physics.pop-ph]} \BibitemShut {NoStop}%
\bibitem [{\citenamefont {Qiu}\ and\ \citenamefont {Gao}(2020)}]{Qiu:2020gox}%
  \BibitemOpen
  \bibfield  {author} {\bibinfo {author} {\bibfnamefont {J.}~\bibnamefont {Qiu}}\ and\ \bibinfo {author} {\bibfnamefont {C.}~\bibnamefont {Gao}},\ }\href {\doibase 10.3390/universe6090148} {\bibfield  {journal} {\bibinfo  {journal} {Universe}\ }\textbf {\bibinfo {volume} {6}},\ \bibinfo {pages} {148} (\bibinfo {year} {2020})},\ \Eprint {http://arxiv.org/abs/2006.09764} {arXiv:2006.09764 [gr-qc]} \BibitemShut {NoStop}%
\bibitem [{\citenamefont {Turimov}\ \emph {et~al.}(2020)\citenamefont {Turimov}, \citenamefont {Rayimbaev}, \citenamefont {Abdujabbarov}, \citenamefont {Ahmedov},\ and\ \citenamefont {Stuchl{\'\i}k}}]{Turimov:2020fme}%
  \BibitemOpen
  \bibfield  {author} {\bibinfo {author} {\bibfnamefont {B.}~\bibnamefont {Turimov}}, \bibinfo {author} {\bibfnamefont {J.}~\bibnamefont {Rayimbaev}}, \bibinfo {author} {\bibfnamefont {A.}~\bibnamefont {Abdujabbarov}}, \bibinfo {author} {\bibfnamefont {B.}~\bibnamefont {Ahmedov}}, \ and\ \bibinfo {author} {\bibfnamefont {Z.}~\bibnamefont {Stuchl{\'\i}k}},\ }\href {\doibase 10.1103/PhysRevD.102.064052} {\bibfield  {journal} {\bibinfo  {journal} {Phys. Rev. D}\ }\textbf {\bibinfo {volume} {102}},\ \bibinfo {pages} {064052} (\bibinfo {year} {2020})},\ \Eprint {http://arxiv.org/abs/2008.08613} {arXiv:2008.08613 [gr-qc]} \BibitemShut {NoStop}%
\bibitem [{\citenamefont {Qiu}(2021)}]{Qiu:2021qrt}%
  \BibitemOpen
  \bibfield  {author} {\bibinfo {author} {\bibfnamefont {J.}~\bibnamefont {Qiu}},\ }\href {\doibase 10.1140/epjc/s10052-021-09890-3} {\bibfield  {journal} {\bibinfo  {journal} {Eur. Phys. J. C}\ }\textbf {\bibinfo {volume} {81}},\ \bibinfo {pages} {1094} (\bibinfo {year} {2021})},\ \Eprint {http://arxiv.org/abs/2101.03034} {arXiv:2101.03034 [gr-qc]} \BibitemShut {NoStop}%
\bibitem [{\citenamefont {Herrera-Aguilar}\ \emph {et~al.}(2021)\citenamefont {Herrera-Aguilar}, \citenamefont {Paschalis},\ and\ \citenamefont {Romero-Figueroa}}]{Herrera-Aguilar:2021qfq}%
  \BibitemOpen
  \bibfield  {author} {\bibinfo {author} {\bibfnamefont {A.}~\bibnamefont {Herrera-Aguilar}}, \bibinfo {author} {\bibfnamefont {J.~E.}\ \bibnamefont {Paschalis}}, \ and\ \bibinfo {author} {\bibfnamefont {C.~E.}\ \bibnamefont {Romero-Figueroa}},\ }\href@noop {} {\  (\bibinfo {year} {2021})},\ \Eprint {http://arxiv.org/abs/2110.04445} {arXiv:2110.04445 [hep-th]} \BibitemShut {NoStop}%
\bibitem [{\citenamefont {Richarte}\ \emph {et~al.}(2022)\citenamefont {Richarte}, \citenamefont {Martins},\ and\ \citenamefont {Fabris}}]{Richarte:2021fbi}%
  \BibitemOpen
  \bibfield  {author} {\bibinfo {author} {\bibfnamefont {M.~G.}\ \bibnamefont {Richarte}}, \bibinfo {author} {\bibfnamefont {{\'E}.~L.}\ \bibnamefont {Martins}}, \ and\ \bibinfo {author} {\bibfnamefont {J.~C.}\ \bibnamefont {Fabris}},\ }\href {\doibase 10.1103/PhysRevD.105.064043} {\bibfield  {journal} {\bibinfo  {journal} {Phys. Rev. D}\ }\textbf {\bibinfo {volume} {105}},\ \bibinfo {pages} {064043} (\bibinfo {year} {2022})},\ \Eprint {http://arxiv.org/abs/2111.01595} {arXiv:2111.01595 [gr-qc]} \BibitemShut {NoStop}%
\bibitem [{\citenamefont {Feng}\ \emph {et~al.}(2022)\citenamefont {Feng}, \citenamefont {Li}, \citenamefont {Liang},\ and\ \citenamefont {Yang}}]{Feng:2022bst}%
  \BibitemOpen
  \bibfield  {author} {\bibinfo {author} {\bibfnamefont {H.}~\bibnamefont {Feng}}, \bibinfo {author} {\bibfnamefont {M.}~\bibnamefont {Li}}, \bibinfo {author} {\bibfnamefont {G.-R.}\ \bibnamefont {Liang}}, \ and\ \bibinfo {author} {\bibfnamefont {R.-J.}\ \bibnamefont {Yang}},\ }\href {\doibase 10.1088/1475-7516/2022/04/027} {\bibfield  {journal} {\bibinfo  {journal} {JCAP}\ }\textbf {\bibinfo {volume} {04}},\ \bibinfo {pages} {027} (\bibinfo {year} {2022})},\ \Eprint {http://arxiv.org/abs/2203.02924} {arXiv:2203.02924 [gr-qc]} \BibitemShut {NoStop}%
\bibitem [{\citenamefont {Mazharimousavi}(2022{\natexlab{a}})}]{Mazharimousavi:2022tnr}%
  \BibitemOpen
  \bibfield  {author} {\bibinfo {author} {\bibfnamefont {S.~H.}\ \bibnamefont {Mazharimousavi}},\ }\href {\doibase 10.1140/epjc/s10052-022-10198-z} {\bibfield  {journal} {\bibinfo  {journal} {Eur. Phys. J. C}\ }\textbf {\bibinfo {volume} {82}},\ \bibinfo {pages} {238} (\bibinfo {year} {2022}{\natexlab{a}})}\BibitemShut {NoStop}%
\bibitem [{\citenamefont {Zahid}\ \emph {et~al.}(2022)\citenamefont {Zahid}, \citenamefont {Rayimbaev}, \citenamefont {Khan}, \citenamefont {Ren}, \citenamefont {Ahmedov},\ and\ \citenamefont {Ibragimov}}]{Zahid:2022zzt}%
  \BibitemOpen
  \bibfield  {author} {\bibinfo {author} {\bibfnamefont {M.}~\bibnamefont {Zahid}}, \bibinfo {author} {\bibfnamefont {J.}~\bibnamefont {Rayimbaev}}, \bibinfo {author} {\bibfnamefont {S.~U.}\ \bibnamefont {Khan}}, \bibinfo {author} {\bibfnamefont {J.}~\bibnamefont {Ren}}, \bibinfo {author} {\bibfnamefont {S.}~\bibnamefont {Ahmedov}}, \ and\ \bibinfo {author} {\bibfnamefont {I.}~\bibnamefont {Ibragimov}},\ }\href {\doibase 10.1140/epjc/s10052-022-10432-8} {\bibfield  {journal} {\bibinfo  {journal} {Eur. Phys. J. C}\ }\textbf {\bibinfo {volume} {82}},\ \bibinfo {pages} {494} (\bibinfo {year} {2022})}\BibitemShut {NoStop}%
\bibitem [{\citenamefont {Mazharimousavi}(2022{\natexlab{b}})}]{Mazharimousavi:2022uhs}%
  \BibitemOpen
  \bibfield  {author} {\bibinfo {author} {\bibfnamefont {S.~H.}\ \bibnamefont {Mazharimousavi}},\ }\href {\doibase 10.1088/1361-6382/ac8141} {\bibfield  {journal} {\bibinfo  {journal} {Class. Quant. Grav.}\ }\textbf {\bibinfo {volume} {39}},\ \bibinfo {pages} {167001} (\bibinfo {year} {2022}{\natexlab{b}})}\BibitemShut {NoStop}%
\bibitem [{\citenamefont {Kurbonov}\ \emph {et~al.}(2023)\citenamefont {Kurbonov}, \citenamefont {Rayimbaev}, \citenamefont {Alloqulov}, \citenamefont {Zahid}, \citenamefont {Abdulxamidov}, \citenamefont {Abdujabbarov},\ and\ \citenamefont {Kurbanova}}]{Kurbonov:2023uyr}%
  \BibitemOpen
  \bibfield  {author} {\bibinfo {author} {\bibfnamefont {N.}~\bibnamefont {Kurbonov}}, \bibinfo {author} {\bibfnamefont {J.}~\bibnamefont {Rayimbaev}}, \bibinfo {author} {\bibfnamefont {M.}~\bibnamefont {Alloqulov}}, \bibinfo {author} {\bibfnamefont {M.}~\bibnamefont {Zahid}}, \bibinfo {author} {\bibfnamefont {F.}~\bibnamefont {Abdulxamidov}}, \bibinfo {author} {\bibfnamefont {A.}~\bibnamefont {Abdujabbarov}}, \ and\ \bibinfo {author} {\bibfnamefont {M.}~\bibnamefont {Kurbanova}},\ }\href {\doibase 10.1140/epjc/s10052-023-11691-9} {\bibfield  {journal} {\bibinfo  {journal} {Eur. Phys. J. C}\ }\textbf {\bibinfo {volume} {83}},\ \bibinfo {pages} {506} (\bibinfo {year} {2023})}\BibitemShut {NoStop}%
\bibitem [{\citenamefont {Alloqulov}\ \emph {et~al.}(2024)\citenamefont {Alloqulov}, \citenamefont {Shaymatov}, \citenamefont {Ahmedov},\ and\ \citenamefont {Jawad}}]{Alloqulov:2024zln}%
  \BibitemOpen
  \bibfield  {author} {\bibinfo {author} {\bibfnamefont {M.}~\bibnamefont {Alloqulov}}, \bibinfo {author} {\bibfnamefont {S.}~\bibnamefont {Shaymatov}}, \bibinfo {author} {\bibfnamefont {B.}~\bibnamefont {Ahmedov}}, \ and\ \bibinfo {author} {\bibfnamefont {A.}~\bibnamefont {Jawad}},\ }\href {\doibase 10.1088/1674-1137/ad137f} {\bibfield  {journal} {\bibinfo  {journal} {Chin. Phys. C}\ }\textbf {\bibinfo {volume} {48}},\ \bibinfo {pages} {025101} (\bibinfo {year} {2024})},\ \Eprint {http://arxiv.org/abs/2401.03184} {arXiv:2401.03184 [gr-qc]} \BibitemShut {NoStop}%
\bibitem [{\citenamefont {Al-Badawi}\ \emph {et~al.}(2024)\citenamefont {Al-Badawi}, \citenamefont {Alloqulov}, \citenamefont {Shaymatov},\ and\ \citenamefont {Ahmedov}}]{Al-Badawi:2024dzc}%
  \BibitemOpen
  \bibfield  {author} {\bibinfo {author} {\bibfnamefont {A.}~\bibnamefont {Al-Badawi}}, \bibinfo {author} {\bibfnamefont {M.}~\bibnamefont {Alloqulov}}, \bibinfo {author} {\bibfnamefont {S.}~\bibnamefont {Shaymatov}}, \ and\ \bibinfo {author} {\bibfnamefont {B.}~\bibnamefont {Ahmedov}},\ }\href {\doibase 10.1088/1674-1137/ad5a70} {\bibfield  {journal} {\bibinfo  {journal} {Chin. Phys. C}\ }\textbf {\bibinfo {volume} {48}},\ \bibinfo {pages} {095105} (\bibinfo {year} {2024})},\ \Eprint {http://arxiv.org/abs/2401.04584} {arXiv:2401.04584 [gr-qc]} \BibitemShut {NoStop}%
\bibitem [{\citenamefont {Carrasco-H.}\ \emph {et~al.}(2024)\citenamefont {Carrasco-H.}, \citenamefont {Santos},\ and\ \citenamefont {Contreras}}]{Carrasco-H:2024fgc}%
  \BibitemOpen
  \bibfield  {author} {\bibinfo {author} {\bibfnamefont {M.}~\bibnamefont {Carrasco-H.}}, \bibinfo {author} {\bibfnamefont {N.~M.}\ \bibnamefont {Santos}}, \ and\ \bibinfo {author} {\bibfnamefont {E.}~\bibnamefont {Contreras}},\ }\href {\doibase 10.1016/j.dark.2024.101529} {\bibfield  {journal} {\bibinfo  {journal} {Phys. Dark Univ.}\ }\textbf {\bibinfo {volume} {45}},\ \bibinfo {pages} {101529} (\bibinfo {year} {2024})},\ \Eprint {http://arxiv.org/abs/2405.20442} {arXiv:2405.20442 [gr-qc]} \BibitemShut {NoStop}%
\bibitem [{\citenamefont {Wu}\ \emph {et~al.}(2024)\citenamefont {Wu}, \citenamefont {Feng},\ and\ \citenamefont {Chen}}]{Wu:2024sng}%
  \BibitemOpen
  \bibfield  {author} {\bibinfo {author} {\bibfnamefont {Y.}~\bibnamefont {Wu}}, \bibinfo {author} {\bibfnamefont {H.}~\bibnamefont {Feng}}, \ and\ \bibinfo {author} {\bibfnamefont {W.-Q.}\ \bibnamefont {Chen}},\ }\href {\doibase 10.1140/epjc/s10052-024-13454-6} {\bibfield  {journal} {\bibinfo  {journal} {Eur. Phys. J. C}\ }\textbf {\bibinfo {volume} {84}},\ \bibinfo {pages} {1075} (\bibinfo {year} {2024})},\ \Eprint {http://arxiv.org/abs/2410.14113} {arXiv:2410.14113 [gr-qc]} \BibitemShut {NoStop}%
\bibitem [{\citenamefont {Alloqulov}\ \emph {et~al.}(2025{\natexlab{a}})\citenamefont {Alloqulov}, \citenamefont {Shaymatov}, \citenamefont {Jawad},\ and\ \citenamefont {Zaripov}}]{Alloqulov:2024hoj}%
  \BibitemOpen
  \bibfield  {author} {\bibinfo {author} {\bibfnamefont {M.}~\bibnamefont {Alloqulov}}, \bibinfo {author} {\bibfnamefont {S.}~\bibnamefont {Shaymatov}}, \bibinfo {author} {\bibfnamefont {A.}~\bibnamefont {Jawad}}, \ and\ \bibinfo {author} {\bibfnamefont {O.}~\bibnamefont {Zaripov}},\ }\href {\doibase 10.1088/1572-9494/ad7831} {\bibfield  {journal} {\bibinfo  {journal} {Commun. Theor. Phys.}\ }\textbf {\bibinfo {volume} {77}},\ \bibinfo {pages} {015402} (\bibinfo {year} {2025}{\natexlab{a}})}\BibitemShut {NoStop}%
\bibitem [{\citenamefont {Belmahi}(2026)}]{Belmahi:2024wag}%
  \BibitemOpen
  \bibfield  {author} {\bibinfo {author} {\bibfnamefont {H.}~\bibnamefont {Belmahi}},\ }\href {\doibase 10.1142/s0219887825502482} {\bibfield  {journal} {\bibinfo  {journal} {Int. J. Geom. Meth. Mod. Phys.}\ }\textbf {\bibinfo {volume} {23}},\ \bibinfo {pages} {2550248} (\bibinfo {year} {2026})},\ \Eprint {http://arxiv.org/abs/2411.11622} {arXiv:2411.11622 [hep-th]} \BibitemShut {NoStop}%
\bibitem [{\citenamefont {Yunusov}\ \emph {et~al.}(2024)\citenamefont {Yunusov}, \citenamefont {Rayimbaev}, \citenamefont {Sarikulov}, \citenamefont {Zahid}, \citenamefont {Abdujabbarov},\ and\ \citenamefont {Stuchl{\'\i}k}}]{Yunusov:2024xzu}%
  \BibitemOpen
  \bibfield  {author} {\bibinfo {author} {\bibfnamefont {O.}~\bibnamefont {Yunusov}}, \bibinfo {author} {\bibfnamefont {J.}~\bibnamefont {Rayimbaev}}, \bibinfo {author} {\bibfnamefont {F.}~\bibnamefont {Sarikulov}}, \bibinfo {author} {\bibfnamefont {M.}~\bibnamefont {Zahid}}, \bibinfo {author} {\bibfnamefont {A.}~\bibnamefont {Abdujabbarov}}, \ and\ \bibinfo {author} {\bibfnamefont {Z.}~\bibnamefont {Stuchl{\'\i}k}},\ }\href {\doibase 10.1140/epjc/s10052-024-13500-3} {\bibfield  {journal} {\bibinfo  {journal} {Eur. Phys. J. C}\ }\textbf {\bibinfo {volume} {84}},\ \bibinfo {pages} {1240} (\bibinfo {year} {2024})}\BibitemShut {NoStop}%
\bibitem [{\citenamefont {Zhang}\ \emph {et~al.}(2025)\citenamefont {Zhang}, \citenamefont {Zhang},\ and\ \citenamefont {Zheng}}]{Zhang:2025jlb}%
  \BibitemOpen
  \bibfield  {author} {\bibinfo {author} {\bibfnamefont {C.-Y.}\ \bibnamefont {Zhang}}, \bibinfo {author} {\bibfnamefont {Z.}~\bibnamefont {Zhang}}, \ and\ \bibinfo {author} {\bibfnamefont {R.}~\bibnamefont {Zheng}},\ }\href {\doibase 10.1007/s11433-024-2607-1} {\bibfield  {journal} {\bibinfo  {journal} {Sci. China Phys. Mech. Astron.}\ }\textbf {\bibinfo {volume} {68}},\ \bibinfo {pages} {250411} (\bibinfo {year} {2025})},\ \Eprint {http://arxiv.org/abs/2503.08315} {arXiv:2503.08315 [gr-qc]} \BibitemShut {NoStop}%
\bibitem [{\citenamefont {Wu}\ \emph {et~al.}(2025)\citenamefont {Wu}, \citenamefont {Cai}, \citenamefont {Ban}, \citenamefont {Feng},\ and\ \citenamefont {Chen}}]{Wu:2025hcu}%
  \BibitemOpen
  \bibfield  {author} {\bibinfo {author} {\bibfnamefont {Y.}~\bibnamefont {Wu}}, \bibinfo {author} {\bibfnamefont {Z.}~\bibnamefont {Cai}}, \bibinfo {author} {\bibfnamefont {Z.}~\bibnamefont {Ban}}, \bibinfo {author} {\bibfnamefont {H.}~\bibnamefont {Feng}}, \ and\ \bibinfo {author} {\bibfnamefont {W.-Q.}\ \bibnamefont {Chen}},\ }\href {\doibase 10.1140/epjc/s10052-025-14831-5} {\bibfield  {journal} {\bibinfo  {journal} {Eur. Phys. J. C}\ }\textbf {\bibinfo {volume} {85}},\ \bibinfo {pages} {1085} (\bibinfo {year} {2025})},\ \Eprint {http://arxiv.org/abs/2504.10327} {arXiv:2504.10327 [gr-qc]} \BibitemShut {NoStop}%
\bibitem [{\citenamefont {Zhuang}\ \emph {et~al.}(2026)\citenamefont {Zhuang}, \citenamefont {Meng},\ and\ \citenamefont {Zhang}}]{Zhuang:2025eal}%
  \BibitemOpen
  \bibfield  {author} {\bibinfo {author} {\bibfnamefont {Z.}~\bibnamefont {Zhuang}}, \bibinfo {author} {\bibfnamefont {K.}~\bibnamefont {Meng}}, \ and\ \bibinfo {author} {\bibfnamefont {H.}~\bibnamefont {Zhang}},\ }\href {\doibase 10.1140/epjc/s10052-026-15674-4} {\bibfield  {journal} {\bibinfo  {journal} {Eur. Phys. J. C}\ }\textbf {\bibinfo {volume} {86}},\ \bibinfo {pages} {459} (\bibinfo {year} {2026})},\ \Eprint {http://arxiv.org/abs/2505.22033} {arXiv:2505.22033 [gr-qc]} \BibitemShut {NoStop}%
\bibitem [{\citenamefont {Wu}\ and\ \citenamefont {Chen}(2026)}]{Wu:2025zwr}%
  \BibitemOpen
  \bibfield  {author} {\bibinfo {author} {\bibfnamefont {Y.}~\bibnamefont {Wu}}\ and\ \bibinfo {author} {\bibfnamefont {W.-Q.}\ \bibnamefont {Chen}},\ }\href {\doibase 10.1088/1475-7516/2026/03/071} {\bibfield  {journal} {\bibinfo  {journal} {JCAP}\ }\textbf {\bibinfo {volume} {03}},\ \bibinfo {pages} {071} (\bibinfo {year} {2026})},\ \bibinfo {note} {[Erratum: JCAP 05, E03 (2026)]},\ \Eprint {http://arxiv.org/abs/2509.19381} {arXiv:2509.19381 [gr-qc]} \BibitemShut {NoStop}%
\bibitem [{\citenamefont {Hughes}(2001)}]{Hughes01EMRI}%
  \BibitemOpen
  \bibfield  {author} {\bibinfo {author} {\bibfnamefont {S.~A.}\ \bibnamefont {Hughes}},\ }\href {\doibase 10.1088/0264-9381/18/19/314} {\bibfield  {journal} {\bibinfo  {journal} {Class. Quantum Gravity}\ }\textbf {\bibinfo {volume} {18}},\ \bibinfo {pages} {4067–4073} (\bibinfo {year} {2001})}\BibitemShut {NoStop}%
\bibitem [{\citenamefont {{Amaro-Seoane}}(2018)}]{Amaro-Seoane18LRR}%
  \BibitemOpen
  \bibfield  {author} {\bibinfo {author} {\bibfnamefont {P.}~\bibnamefont {{Amaro-Seoane}}},\ }\href {\doibase 10.1007/s41114-018-0013-8} {\bibfield  {journal} {\bibinfo  {journal} {Living Rev. Rel.}\ }\textbf {\bibinfo {volume} {21}},\ \bibinfo {eid} {4} (\bibinfo {year} {2018})},\ \Eprint {http://arxiv.org/abs/1205.5240} {arXiv:1205.5240 [astro-ph.CO]} \BibitemShut {NoStop}%
\bibitem [{\citenamefont {{Babak}}\ \emph {et~al.}(2017)\citenamefont {{Babak}}, \citenamefont {{Gair}}, \citenamefont {{Sesana}}, \citenamefont {{Barausse}}, \citenamefont {{Sopuerta}}, \citenamefont {{Berry}}, \citenamefont {{Berti}}, \citenamefont {{Amaro-Seoane}}, \citenamefont {{Petiteau}},\ and\ \citenamefont {{Klein}}}]{Babak17PRD}%
  \BibitemOpen
  \bibfield  {author} {\bibinfo {author} {\bibfnamefont {S.}~\bibnamefont {{Babak}}}, \bibinfo {author} {\bibfnamefont {J.}~\bibnamefont {{Gair}}}, \bibinfo {author} {\bibfnamefont {A.}~\bibnamefont {{Sesana}}}, \bibinfo {author} {\bibfnamefont {E.}~\bibnamefont {{Barausse}}}, \bibinfo {author} {\bibfnamefont {C.~F.}\ \bibnamefont {{Sopuerta}}}, \bibinfo {author} {\bibfnamefont {C.~P.~L.}\ \bibnamefont {{Berry}}}, \bibinfo {author} {\bibfnamefont {E.}~\bibnamefont {{Berti}}}, \bibinfo {author} {\bibfnamefont {P.}~\bibnamefont {{Amaro-Seoane}}}, \bibinfo {author} {\bibfnamefont {A.}~\bibnamefont {{Petiteau}}}, \ and\ \bibinfo {author} {\bibfnamefont {A.}~\bibnamefont {{Klein}}},\ }\href {\doibase 10.1103/PhysRevD.95.103012} {\bibfield  {journal} {\bibinfo  {journal} {Phys. Rev. D}\ }\textbf {\bibinfo {volume} {95}},\ \bibinfo {eid} {103012} (\bibinfo {year} {2017})},\ \Eprint {http://arxiv.org/abs/1703.09722} {arXiv:1703.09722 [gr-qc]} \BibitemShut {NoStop}%
\bibitem [{\citenamefont {{Amaro-Seoane}}\ and\ \citenamefont {et~al. {(Laser Interferometer Space Antenna)}}(2017)}]{Amaro-Seoane2017LISA}%
  \BibitemOpen
  \bibfield  {author} {\bibinfo {author} {\bibfnamefont {P.}~\bibnamefont {{Amaro-Seoane}}}\ and\ \bibinfo {author} {\bibnamefont {et~al. {(Laser Interferometer Space Antenna)}}},\ }\href {https://arxiv.org/abs/1702.00786} {\enquote {\bibinfo {title} {Laser interferometer space antenna},}\ } (\bibinfo {year} {2017}),\ \Eprint {http://arxiv.org/abs/1702.00786} {arXiv:1702.00786 [astro-ph.IM]} \BibitemShut {NoStop}%
\bibitem [{\citenamefont {Hu}\ and\ \citenamefont {Wu}(2017)}]{HuTaiji}%
  \BibitemOpen
  \bibfield  {author} {\bibinfo {author} {\bibfnamefont {W.-R.}\ \bibnamefont {Hu}}\ and\ \bibinfo {author} {\bibfnamefont {Y.-L.}\ \bibnamefont {Wu}},\ }\href {\doibase 10.1093/nsr/nwx116} {\bibfield  {journal} {\bibinfo  {journal} {National Science Review}\ }\textbf {\bibinfo {volume} {4}},\ \bibinfo {pages} {685} (\bibinfo {year} {2017})},\ \Eprint {http://arxiv.org/abs/https://academic.oup.com/nsr/article-pdf/4/5/685/31566708/nwx116.pdf} {https://academic.oup.com/nsr/article-pdf/4/5/685/31566708/nwx116.pdf} \BibitemShut {NoStop}%
\bibitem [{\citenamefont {Gong}\ \emph {et~al.}(2021)\citenamefont {Gong}, \citenamefont {Luo},\ and\ \citenamefont {Wang}}]{Gong_2021}%
  \BibitemOpen
  \bibfield  {author} {\bibinfo {author} {\bibfnamefont {Y.}~\bibnamefont {Gong}}, \bibinfo {author} {\bibfnamefont {J.}~\bibnamefont {Luo}}, \ and\ \bibinfo {author} {\bibfnamefont {B.}~\bibnamefont {Wang}},\ }\href {\doibase 10.1038/s41550-021-01480-3} {\bibfield  {journal} {\bibinfo  {journal} {Nature Astronomy}\ }\textbf {\bibinfo {volume} {5}},\ \bibinfo {pages} {881–889} (\bibinfo {year} {2021})}\BibitemShut {NoStop}%
\bibitem [{\citenamefont {Levin}\ and\ \citenamefont {Perez-Giz}(2008)}]{Levin_2008}%
  \BibitemOpen
  \bibfield  {author} {\bibinfo {author} {\bibfnamefont {J.}~\bibnamefont {Levin}}\ and\ \bibinfo {author} {\bibfnamefont {G.}~\bibnamefont {Perez-Giz}},\ }\href {\doibase 10.1103/physrevd.77.103005} {\bibfield  {journal} {\bibinfo  {journal} {Phys. Rev. D}\ }\textbf {\bibinfo {volume} {77}} (\bibinfo {year} {2008}),\ 10.1103/physrevd.77.103005}\BibitemShut {NoStop}%
\bibitem [{\citenamefont {Grossman}\ and\ \citenamefont {Levin}(2009)}]{Grossman_2009}%
  \BibitemOpen
  \bibfield  {author} {\bibinfo {author} {\bibfnamefont {R.}~\bibnamefont {Grossman}}\ and\ \bibinfo {author} {\bibfnamefont {J.}~\bibnamefont {Levin}},\ }\href {\doibase 10.1103/physrevd.79.043017} {\bibfield  {journal} {\bibinfo  {journal} {Phys. Rev. D}\ }\textbf {\bibinfo {volume} {79}} (\bibinfo {year} {2009}),\ 10.1103/physrevd.79.043017}\BibitemShut {NoStop}%
\bibitem [{\citenamefont {Misra}\ and\ \citenamefont {Levin}(2010{\natexlab{a}})}]{Misra_2010}%
  \BibitemOpen
  \bibfield  {author} {\bibinfo {author} {\bibfnamefont {V.}~\bibnamefont {Misra}}\ and\ \bibinfo {author} {\bibfnamefont {J.}~\bibnamefont {Levin}},\ }\href {\doibase 10.1103/physrevd.82.083001} {\bibfield  {journal} {\bibinfo  {journal} {Phys. Rev. D}\ }\textbf {\bibinfo {volume} {82}} (\bibinfo {year} {2010}{\natexlab{a}}),\ 10.1103/physrevd.82.083001}\BibitemShut {NoStop}%
\bibitem [{\citenamefont {Levin}(2009)}]{Levin_2009}%
  \BibitemOpen
  \bibfield  {author} {\bibinfo {author} {\bibfnamefont {J.}~\bibnamefont {Levin}},\ }\href {\doibase 10.1088/0264-9381/26/23/235010} {\bibfield  {journal} {\bibinfo  {journal} {Class. Quantum Gravity}\ }\textbf {\bibinfo {volume} {26}},\ \bibinfo {pages} {235010} (\bibinfo {year} {2009})}\BibitemShut {NoStop}%
\bibitem [{\citenamefont {Misra}\ and\ \citenamefont {Levin}(2010{\natexlab{b}})}]{Levin2010}%
  \BibitemOpen
  \bibfield  {author} {\bibinfo {author} {\bibfnamefont {V.}~\bibnamefont {Misra}}\ and\ \bibinfo {author} {\bibfnamefont {J.}~\bibnamefont {Levin}},\ }\href {\doibase 10.1103/PhysRevD.82.083001} {\bibfield  {journal} {\bibinfo  {journal} {Phys. Rev. D}\ }\textbf {\bibinfo {volume} {82}},\ \bibinfo {pages} {083001} (\bibinfo {year} {2010}{\natexlab{b}})}\BibitemShut {NoStop}%
\bibitem [{\citenamefont {{Babar}}\ \emph {et~al.}(2017)\citenamefont {{Babar}}, \citenamefont {{Babar}},\ and\ \citenamefont {{Lim}}}]{Babar17PRD}%
  \BibitemOpen
  \bibfield  {author} {\bibinfo {author} {\bibfnamefont {G.~Z.}\ \bibnamefont {{Babar}}}, \bibinfo {author} {\bibfnamefont {A.~Z.}\ \bibnamefont {{Babar}}}, \ and\ \bibinfo {author} {\bibfnamefont {Y.-K.}\ \bibnamefont {{Lim}}},\ }\href {\doibase 10.1103/PhysRevD.96.084052} {\bibfield  {journal} {\bibinfo  {journal} {Phys. Rev. D}\ }\textbf {\bibinfo {volume} {96}},\ \bibinfo {eid} {084052} (\bibinfo {year} {2017})},\ \Eprint {http://arxiv.org/abs/1710.09581} {arXiv:1710.09581 [gr-qc]} \BibitemShut {NoStop}%
\bibitem [{\citenamefont {Tu}\ \emph {et~al.}(2023)\citenamefont {Tu}, \citenamefont {Zhu},\ and\ \citenamefont {Wang}}]{Tu23}%
  \BibitemOpen
  \bibfield  {author} {\bibinfo {author} {\bibfnamefont {Z.-Y.}\ \bibnamefont {Tu}}, \bibinfo {author} {\bibfnamefont {T.}~\bibnamefont {Zhu}}, \ and\ \bibinfo {author} {\bibfnamefont {A.}~\bibnamefont {Wang}},\ }\href {\doibase 10.1103/PhysRevD.108.024035} {\bibfield  {journal} {\bibinfo  {journal} {Phys. Rev. D}\ }\textbf {\bibinfo {volume} {108}},\ \bibinfo {pages} {024035} (\bibinfo {year} {2023})}\BibitemShut {NoStop}%
\bibitem [{\citenamefont {Azreg-A\"{\i}nou}\ \emph {et~al.}(2020)\citenamefont {Azreg-A\"{\i}nou}, \citenamefont {Chen}, \citenamefont {Deng}, \citenamefont {Jamil}, \citenamefont {Zhu}, \citenamefont {Wu},\ and\ \citenamefont {Lim}}]{Mustapha2020}%
  \BibitemOpen
  \bibfield  {author} {\bibinfo {author} {\bibfnamefont {M.}~\bibnamefont {Azreg-A\"{\i}nou}}, \bibinfo {author} {\bibfnamefont {Z.}~\bibnamefont {Chen}}, \bibinfo {author} {\bibfnamefont {B.}~\bibnamefont {Deng}}, \bibinfo {author} {\bibfnamefont {M.}~\bibnamefont {Jamil}}, \bibinfo {author} {\bibfnamefont {T.}~\bibnamefont {Zhu}}, \bibinfo {author} {\bibfnamefont {Q.}~\bibnamefont {Wu}}, \ and\ \bibinfo {author} {\bibfnamefont {Y.-K.}\ \bibnamefont {Lim}},\ }\href {\doibase 10.1103/PhysRevD.102.044028} {\bibfield  {journal} {\bibinfo  {journal} {Phys. Rev. D}\ }\textbf {\bibinfo {volume} {102}},\ \bibinfo {pages} {044028} (\bibinfo {year} {2020})}\BibitemShut {NoStop}%
\bibitem [{\citenamefont {Wei}\ \emph {et~al.}(2019{\natexlab{a}})\citenamefont {Wei}, \citenamefont {Yang},\ and\ \citenamefont {Liu}}]{wei2019}%
  \BibitemOpen
  \bibfield  {author} {\bibinfo {author} {\bibfnamefont {S.-W.}\ \bibnamefont {Wei}}, \bibinfo {author} {\bibfnamefont {J.}~\bibnamefont {Yang}}, \ and\ \bibinfo {author} {\bibfnamefont {Y.-X.}\ \bibnamefont {Liu}},\ }\href {\doibase 10.1103/PhysRevD.99.104016} {\bibfield  {journal} {\bibinfo  {journal} {Phys. Rev. D}\ }\textbf {\bibinfo {volume} {99}},\ \bibinfo {pages} {104016} (\bibinfo {year} {2019}{\natexlab{a}})}\BibitemShut {NoStop}%
\bibitem [{\citenamefont {{Deng}}(2020)}]{Deng20}%
  \BibitemOpen
  \bibfield  {author} {\bibinfo {author} {\bibfnamefont {X.-M.}\ \bibnamefont {{Deng}}},\ }\href {\doibase 10.1016/j.dark.2020.100629} {\bibfield  {journal} {\bibinfo  {journal} {Phys. Dark Universe}\ }\textbf {\bibinfo {volume} {30}},\ \bibinfo {eid} {100629} (\bibinfo {year} {2020})}\BibitemShut {NoStop}%
\bibitem [{\citenamefont {Yang}\ \emph {et~al.}(2024)\citenamefont {Yang}, \citenamefont {Zhang}, \citenamefont {Zhu}, \citenamefont {Zhao},\ and\ \citenamefont {Liu}}]{yang2024}%
  \BibitemOpen
  \bibfield  {author} {\bibinfo {author} {\bibfnamefont {S.}~\bibnamefont {Yang}}, \bibinfo {author} {\bibfnamefont {Y.-P.}\ \bibnamefont {Zhang}}, \bibinfo {author} {\bibfnamefont {T.}~\bibnamefont {Zhu}}, \bibinfo {author} {\bibfnamefont {L.}~\bibnamefont {Zhao}}, \ and\ \bibinfo {author} {\bibfnamefont {Y.-X.}\ \bibnamefont {Liu}},\ }\href {https://arxiv.org/abs/2407.00283} {\enquote {\bibinfo {title} {Gravitational waveforms from periodic orbits around a quantum-corrected black hole},}\ } (\bibinfo {year} {2024}),\ \Eprint {http://arxiv.org/abs/2407.00283} {arXiv:2407.00283 [gr-qc]} \BibitemShut {NoStop}%
\bibitem [{\citenamefont {Liu}\ \emph {et~al.}(2019)\citenamefont {Liu}, \citenamefont {Ding},\ and\ \citenamefont {Jing}}]{Liu:2018vea}%
  \BibitemOpen
  \bibfield  {author} {\bibinfo {author} {\bibfnamefont {C.}~\bibnamefont {Liu}}, \bibinfo {author} {\bibfnamefont {C.}~\bibnamefont {Ding}}, \ and\ \bibinfo {author} {\bibfnamefont {J.}~\bibnamefont {Jing}},\ }\href {\doibase 10.1088/0253-6102/71/12/1461} {\bibfield  {journal} {\bibinfo  {journal} {Commun. Theor. Phys.}\ }\textbf {\bibinfo {volume} {71}},\ \bibinfo {pages} {1461} (\bibinfo {year} {2019})},\ \Eprint {http://arxiv.org/abs/1804.05883} {arXiv:1804.05883 [gr-qc]} \BibitemShut {NoStop}%
\bibitem [{\citenamefont {Lin}\ and\ \citenamefont {Deng}(2023{\natexlab{a}})}]{Lin:2023rmo}%
  \BibitemOpen
  \bibfield  {author} {\bibinfo {author} {\bibfnamefont {H.-Y.}\ \bibnamefont {Lin}}\ and\ \bibinfo {author} {\bibfnamefont {X.-M.}\ \bibnamefont {Deng}},\ }\href {\doibase 10.1140/epjc/s10052-023-11487-x} {\bibfield  {journal} {\bibinfo  {journal} {Eur. Phys. J. C}\ }\textbf {\bibinfo {volume} {83}},\ \bibinfo {pages} {311} (\bibinfo {year} {2023}{\natexlab{a}})}\BibitemShut {NoStop}%
\bibitem [{\citenamefont {Yao}\ and\ \citenamefont {Li}(2023)}]{Yao:2023ziq}%
  \BibitemOpen
  \bibfield  {author} {\bibinfo {author} {\bibfnamefont {J.-T.}\ \bibnamefont {Yao}}\ and\ \bibinfo {author} {\bibfnamefont {X.}~\bibnamefont {Li}},\ }\href {\doibase 10.1103/PhysRevD.108.084067} {\bibfield  {journal} {\bibinfo  {journal} {Phys. Rev. D}\ }\textbf {\bibinfo {volume} {108}},\ \bibinfo {pages} {084067} (\bibinfo {year} {2023})}\BibitemShut {NoStop}%
\bibitem [{\citenamefont {Lin}\ and\ \citenamefont {Deng}(2022{\natexlab{a}})}]{Lin:2022llz}%
  \BibitemOpen
  \bibfield  {author} {\bibinfo {author} {\bibfnamefont {H.-Y.}\ \bibnamefont {Lin}}\ and\ \bibinfo {author} {\bibfnamefont {X.-M.}\ \bibnamefont {Deng}},\ }\href {\doibase 10.3390/universe8050278} {\bibfield  {journal} {\bibinfo  {journal} {Universe}\ }\textbf {\bibinfo {volume} {8}},\ \bibinfo {pages} {278} (\bibinfo {year} {2022}{\natexlab{a}})}\BibitemShut {NoStop}%
\bibitem [{\citenamefont {Chan}\ and\ \citenamefont {Lim}(2025)}]{Chan:2025ocy}%
  \BibitemOpen
  \bibfield  {author} {\bibinfo {author} {\bibfnamefont {Z.~C.~S.}\ \bibnamefont {Chan}}\ and\ \bibinfo {author} {\bibfnamefont {Y.-K.}\ \bibnamefont {Lim}},\ }\href {\doibase 10.1007/s10714-025-03368-3} {\bibfield  {journal} {\bibinfo  {journal} {Gen. Rel. Grav.}\ }\textbf {\bibinfo {volume} {57}},\ \bibinfo {pages} {35} (\bibinfo {year} {2025})},\ \Eprint {http://arxiv.org/abs/2502.03082} {arXiv:2502.03082 [gr-qc]} \BibitemShut {NoStop}%
\bibitem [{\citenamefont {Wang}\ \emph {et~al.}(2022)\citenamefont {Wang}, \citenamefont {Gao},\ and\ \citenamefont {Chen}}]{Wang:2022tfo}%
  \BibitemOpen
  \bibfield  {author} {\bibinfo {author} {\bibfnamefont {R.}~\bibnamefont {Wang}}, \bibinfo {author} {\bibfnamefont {F.}~\bibnamefont {Gao}}, \ and\ \bibinfo {author} {\bibfnamefont {H.}~\bibnamefont {Chen}},\ }\href {\doibase 10.1016/j.aop.2022.169167} {\bibfield  {journal} {\bibinfo  {journal} {Annals Phys.}\ }\textbf {\bibinfo {volume} {447}},\ \bibinfo {pages} {169167} (\bibinfo {year} {2022})}\BibitemShut {NoStop}%
\bibitem [{\citenamefont {Lin}\ and\ \citenamefont {Deng}(2023{\natexlab{b}})}]{Lin:2023eyd}%
  \BibitemOpen
  \bibfield  {author} {\bibinfo {author} {\bibfnamefont {H.-Y.}\ \bibnamefont {Lin}}\ and\ \bibinfo {author} {\bibfnamefont {X.-M.}\ \bibnamefont {Deng}},\ }\href {\doibase 10.1016/j.aop.2023.169360} {\bibfield  {journal} {\bibinfo  {journal} {Annals Phys.}\ }\textbf {\bibinfo {volume} {455}},\ \bibinfo {pages} {169360} (\bibinfo {year} {2023}{\natexlab{b}})}\BibitemShut {NoStop}%
\bibitem [{\citenamefont {Haroon}\ and\ \citenamefont {Zhu}(2025)}]{Haroon:2025rzx}%
  \BibitemOpen
  \bibfield  {author} {\bibinfo {author} {\bibfnamefont {S.}~\bibnamefont {Haroon}}\ and\ \bibinfo {author} {\bibfnamefont {T.}~\bibnamefont {Zhu}},\ }\href {\doibase 10.1103/ckdt-wtsl} {\bibfield  {journal} {\bibinfo  {journal} {Phys. Rev. D}\ }\textbf {\bibinfo {volume} {112}},\ \bibinfo {pages} {044046} (\bibinfo {year} {2025})},\ \Eprint {http://arxiv.org/abs/2502.09171} {arXiv:2502.09171 [gr-qc]} \BibitemShut {NoStop}%
\bibitem [{\citenamefont {Habibina}\ and\ \citenamefont {Ramadhan}(2023)}]{Habibina:2022ztd}%
  \BibitemOpen
  \bibfield  {author} {\bibinfo {author} {\bibfnamefont {A.~S.}\ \bibnamefont {Habibina}}\ and\ \bibinfo {author} {\bibfnamefont {H.~S.}\ \bibnamefont {Ramadhan}},\ }\href {\doibase 10.1016/j.aop.2022.169169} {\bibfield  {journal} {\bibinfo  {journal} {Annals Phys.}\ }\textbf {\bibinfo {volume} {448}},\ \bibinfo {pages} {169169} (\bibinfo {year} {2023})},\ \Eprint {http://arxiv.org/abs/2205.14635} {arXiv:2205.14635 [gr-qc]} \BibitemShut {NoStop}%
\bibitem [{\citenamefont {Zhang}\ and\ \citenamefont {Xie}(2022{\natexlab{a}})}]{Zhang:2022psr}%
  \BibitemOpen
  \bibfield  {author} {\bibinfo {author} {\bibfnamefont {J.}~\bibnamefont {Zhang}}\ and\ \bibinfo {author} {\bibfnamefont {Y.}~\bibnamefont {Xie}},\ }\href {\doibase 10.1007/s10509-022-04046-5} {\bibfield  {journal} {\bibinfo  {journal} {Astrophys. Space Sci.}\ }\textbf {\bibinfo {volume} {367}},\ \bibinfo {pages} {17} (\bibinfo {year} {2022}{\natexlab{a}})}\BibitemShut {NoStop}%
\bibitem [{\citenamefont {Lin}\ and\ \citenamefont {Deng}(2022{\natexlab{b}})}]{Lin:2022wda}%
  \BibitemOpen
  \bibfield  {author} {\bibinfo {author} {\bibfnamefont {H.-Y.}\ \bibnamefont {Lin}}\ and\ \bibinfo {author} {\bibfnamefont {X.-M.}\ \bibnamefont {Deng}},\ }\href {\doibase 10.1140/epjp/s13360-022-02391-6} {\bibfield  {journal} {\bibinfo  {journal} {Eur. Phys. J. Plus}\ }\textbf {\bibinfo {volume} {137}},\ \bibinfo {pages} {176} (\bibinfo {year} {2022}{\natexlab{b}})}\BibitemShut {NoStop}%
\bibitem [{\citenamefont {Gao}\ and\ \citenamefont {Deng}(2021)}]{Gao:2021arw}%
  \BibitemOpen
  \bibfield  {author} {\bibinfo {author} {\bibfnamefont {B.}~\bibnamefont {Gao}}\ and\ \bibinfo {author} {\bibfnamefont {X.-M.}\ \bibnamefont {Deng}},\ }\href {\doibase 10.1142/S0217732321502370} {\bibfield  {journal} {\bibinfo  {journal} {Mod. Phys. Lett. A}\ }\textbf {\bibinfo {volume} {36}},\ \bibinfo {pages} {2150237} (\bibinfo {year} {2021})}\BibitemShut {NoStop}%
\bibitem [{\citenamefont {Lin}\ and\ \citenamefont {Deng}(2021)}]{Lin:2021noq}%
  \BibitemOpen
  \bibfield  {author} {\bibinfo {author} {\bibfnamefont {H.-Y.}\ \bibnamefont {Lin}}\ and\ \bibinfo {author} {\bibfnamefont {X.-M.}\ \bibnamefont {Deng}},\ }\href {\doibase 10.1016/j.dark.2020.100745} {\bibfield  {journal} {\bibinfo  {journal} {Phys. Dark Univ.}\ }\textbf {\bibinfo {volume} {31}},\ \bibinfo {pages} {100745} (\bibinfo {year} {2021})}\BibitemShut {NoStop}%
\bibitem [{\citenamefont {Gao}\ and\ \citenamefont {Deng}(2020)}]{Gao:2020wjz}%
  \BibitemOpen
  \bibfield  {author} {\bibinfo {author} {\bibfnamefont {B.}~\bibnamefont {Gao}}\ and\ \bibinfo {author} {\bibfnamefont {X.-M.}\ \bibnamefont {Deng}},\ }\href {\doibase 10.1016/j.aop.2020.168194} {\bibfield  {journal} {\bibinfo  {journal} {Annals Phys.}\ }\textbf {\bibinfo {volume} {418}},\ \bibinfo {pages} {168194} (\bibinfo {year} {2020})}\BibitemShut {NoStop}%
\bibitem [{\citenamefont {Deng}(2020)}]{Deng:2020hxw}%
  \BibitemOpen
  \bibfield  {author} {\bibinfo {author} {\bibfnamefont {X.-M.}\ \bibnamefont {Deng}},\ }\href {\doibase 10.1140/epjc/s10052-020-8067-7} {\bibfield  {journal} {\bibinfo  {journal} {Eur. Phys. J. C}\ }\textbf {\bibinfo {volume} {80}},\ \bibinfo {pages} {489} (\bibinfo {year} {2020})}\BibitemShut {NoStop}%
\bibitem [{\citenamefont {Azreg-A{\"\i}nou}\ \emph {et~al.}(2020)\citenamefont {Azreg-A{\"\i}nou}, \citenamefont {Chen}, \citenamefont {Deng}, \citenamefont {Jamil}, \citenamefont {Zhu}, \citenamefont {Wu},\ and\ \citenamefont {Lim}}]{Azreg-Ainou:2020bfl}%
  \BibitemOpen
  \bibfield  {author} {\bibinfo {author} {\bibfnamefont {M.}~\bibnamefont {Azreg-A{\"\i}nou}}, \bibinfo {author} {\bibfnamefont {Z.}~\bibnamefont {Chen}}, \bibinfo {author} {\bibfnamefont {B.}~\bibnamefont {Deng}}, \bibinfo {author} {\bibfnamefont {M.}~\bibnamefont {Jamil}}, \bibinfo {author} {\bibfnamefont {T.}~\bibnamefont {Zhu}}, \bibinfo {author} {\bibfnamefont {Q.}~\bibnamefont {Wu}}, \ and\ \bibinfo {author} {\bibfnamefont {Y.-K.}\ \bibnamefont {Lim}},\ }\href {\doibase 10.1103/PhysRevD.102.044028} {\bibfield  {journal} {\bibinfo  {journal} {Phys. Rev. D}\ }\textbf {\bibinfo {volume} {102}},\ \bibinfo {pages} {044028} (\bibinfo {year} {2020})},\ \Eprint {http://arxiv.org/abs/2004.02602} {arXiv:2004.02602 [gr-qc]} \BibitemShut {NoStop}%
\bibitem [{\citenamefont {Wei}\ \emph {et~al.}(2019{\natexlab{b}})\citenamefont {Wei}, \citenamefont {Yang},\ and\ \citenamefont {Liu}}]{Wei:2019zdf}%
  \BibitemOpen
  \bibfield  {author} {\bibinfo {author} {\bibfnamefont {S.-W.}\ \bibnamefont {Wei}}, \bibinfo {author} {\bibfnamefont {J.}~\bibnamefont {Yang}}, \ and\ \bibinfo {author} {\bibfnamefont {Y.-X.}\ \bibnamefont {Liu}},\ }\href {\doibase 10.1103/PhysRevD.99.104016} {\bibfield  {journal} {\bibinfo  {journal} {Phys. Rev. D}\ }\textbf {\bibinfo {volume} {99}},\ \bibinfo {pages} {104016} (\bibinfo {year} {2019}{\natexlab{b}})},\ \Eprint {http://arxiv.org/abs/1904.03129} {arXiv:1904.03129 [gr-qc]} \BibitemShut {NoStop}%
\bibitem [{\citenamefont {Pugliese}\ \emph {et~al.}(2017)\citenamefont {Pugliese}, \citenamefont {Quevedo},\ and\ \citenamefont {Ruffini}}]{Pugliese:2013xfa}%
  \BibitemOpen
  \bibfield  {author} {\bibinfo {author} {\bibfnamefont {D.}~\bibnamefont {Pugliese}}, \bibinfo {author} {\bibfnamefont {H.}~\bibnamefont {Quevedo}}, \ and\ \bibinfo {author} {\bibfnamefont {R.}~\bibnamefont {Ruffini}},\ }\href {\doibase 10.1140/epjc/s10052-017-4769-x} {\bibfield  {journal} {\bibinfo  {journal} {Eur. Phys. J. C}\ }\textbf {\bibinfo {volume} {77}},\ \bibinfo {pages} {206} (\bibinfo {year} {2017})},\ \Eprint {http://arxiv.org/abs/1304.2940} {arXiv:1304.2940 [gr-qc]} \BibitemShut {NoStop}%
\bibitem [{\citenamefont {Zhang}\ and\ \citenamefont {Xie}(2022{\natexlab{b}})}]{Zhang:2022zox}%
  \BibitemOpen
  \bibfield  {author} {\bibinfo {author} {\bibfnamefont {J.}~\bibnamefont {Zhang}}\ and\ \bibinfo {author} {\bibfnamefont {Y.}~\bibnamefont {Xie}},\ }\href {\doibase 10.1140/epjc/s10052-022-10846-4} {\bibfield  {journal} {\bibinfo  {journal} {Eur. Phys. J. C}\ }\textbf {\bibinfo {volume} {82}},\ \bibinfo {pages} {854} (\bibinfo {year} {2022}{\natexlab{b}})}\BibitemShut {NoStop}%
\bibitem [{\citenamefont {Healy}\ \emph {et~al.}(2009)\citenamefont {Healy}, \citenamefont {Levin},\ and\ \citenamefont {Shoemaker}}]{Healy:2009zm}%
  \BibitemOpen
  \bibfield  {author} {\bibinfo {author} {\bibfnamefont {J.}~\bibnamefont {Healy}}, \bibinfo {author} {\bibfnamefont {J.}~\bibnamefont {Levin}}, \ and\ \bibinfo {author} {\bibfnamefont {D.}~\bibnamefont {Shoemaker}},\ }\href {\doibase 10.1103/PhysRevLett.103.131101} {\bibfield  {journal} {\bibinfo  {journal} {Phys. Rev. Lett.}\ }\textbf {\bibinfo {volume} {103}},\ \bibinfo {pages} {131101} (\bibinfo {year} {2009})},\ \Eprint {http://arxiv.org/abs/0907.0671} {arXiv:0907.0671 [gr-qc]} \BibitemShut {NoStop}%
\bibitem [{\citenamefont {Wang}\ \emph {et~al.}(2025{\natexlab{a}})\citenamefont {Wang}, \citenamefont {Zhang}, \citenamefont {Zhu},\ and\ \citenamefont {Wei}}]{Wang:2025wob}%
  \BibitemOpen
  \bibfield  {author} {\bibinfo {author} {\bibfnamefont {C.-H.}\ \bibnamefont {Wang}}, \bibinfo {author} {\bibfnamefont {Y.-P.}\ \bibnamefont {Zhang}}, \bibinfo {author} {\bibfnamefont {T.}~\bibnamefont {Zhu}}, \ and\ \bibinfo {author} {\bibfnamefont {S.-W.}\ \bibnamefont {Wei}},\ }\href {https://arxiv.org/abs/2508.20558} {\enquote {\bibinfo {title} {A new type of multi-branch periodic orbits in dyonic black holes},}\ } (\bibinfo {year} {2025}{\natexlab{a}}),\ \Eprint {http://arxiv.org/abs/2508.20558} {arXiv:2508.20558 [gr-qc]} \BibitemShut {NoStop}%
\bibitem [{\citenamefont {Alloqulov}\ \emph {et~al.}(2026{\natexlab{a}})\citenamefont {Alloqulov}, \citenamefont {Shaymatov}, \citenamefont {Ahmedov},\ and\ \citenamefont {Zhu}}]{Alloqulov:2025bxh}%
  \BibitemOpen
  \bibfield  {author} {\bibinfo {author} {\bibfnamefont {M.}~\bibnamefont {Alloqulov}}, \bibinfo {author} {\bibfnamefont {S.}~\bibnamefont {Shaymatov}}, \bibinfo {author} {\bibfnamefont {B.}~\bibnamefont {Ahmedov}}, \ and\ \bibinfo {author} {\bibfnamefont {T.}~\bibnamefont {Zhu}},\ }\href {\doibase 10.1140/epjc/s10052-025-15251-1} {\bibfield  {journal} {\bibinfo  {journal} {Eur. Phys. J. C}\ }\textbf {\bibinfo {volume} {86}},\ \bibinfo {pages} {117} (\bibinfo {year} {2026}{\natexlab{a}})}\BibitemShut {NoStop}%
\bibitem [{\citenamefont {Wei}\ \emph {et~al.}(2025)\citenamefont {Wei}, \citenamefont {Zhang}, \citenamefont {Xie},\ and\ \citenamefont {Yin}}]{Wei:2025qlh}%
  \BibitemOpen
  \bibfield  {author} {\bibinfo {author} {\bibfnamefont {Z.-L.}\ \bibnamefont {Wei}}, \bibinfo {author} {\bibfnamefont {J.}~\bibnamefont {Zhang}}, \bibinfo {author} {\bibfnamefont {Y.}~\bibnamefont {Xie}}, \ and\ \bibinfo {author} {\bibfnamefont {P.-L.}\ \bibnamefont {Yin}},\ }\href {\doibase 10.1140/epjc/s10052-025-14437-x} {\bibfield  {journal} {\bibinfo  {journal} {Eur. Phys. J. C}\ }\textbf {\bibinfo {volume} {85}},\ \bibinfo {pages} {698} (\bibinfo {year} {2025})}\BibitemShut {NoStop}%
\bibitem [{\citenamefont {Yang}\ \emph {et~al.}(2025{\natexlab{a}})\citenamefont {Yang}, \citenamefont {Zhang}, \citenamefont {Zhu}, \citenamefont {Zhao},\ and\ \citenamefont {Liu}}]{Yang:2024lmj}%
  \BibitemOpen
  \bibfield  {author} {\bibinfo {author} {\bibfnamefont {S.}~\bibnamefont {Yang}}, \bibinfo {author} {\bibfnamefont {Y.-P.}\ \bibnamefont {Zhang}}, \bibinfo {author} {\bibfnamefont {T.}~\bibnamefont {Zhu}}, \bibinfo {author} {\bibfnamefont {L.}~\bibnamefont {Zhao}}, \ and\ \bibinfo {author} {\bibfnamefont {Y.-X.}\ \bibnamefont {Liu}},\ }\href {\doibase 10.1088/1475-7516/2025/01/091} {\bibfield  {journal} {\bibinfo  {journal} {JCAP}\ }\textbf {\bibinfo {volume} {01}},\ \bibinfo {pages} {091} (\bibinfo {year} {2025}{\natexlab{a}})},\ \Eprint {http://arxiv.org/abs/2407.00283} {arXiv:2407.00283 [gr-qc]} \BibitemShut {NoStop}%
\bibitem [{\citenamefont {Shabbir}\ \emph {et~al.}(2025)\citenamefont {Shabbir}, \citenamefont {Jamil},\ and\ \citenamefont {Azreg-A{\"\i}nou}}]{Shabbir:2025kqh}%
  \BibitemOpen
  \bibfield  {author} {\bibinfo {author} {\bibfnamefont {O.}~\bibnamefont {Shabbir}}, \bibinfo {author} {\bibfnamefont {M.}~\bibnamefont {Jamil}}, \ and\ \bibinfo {author} {\bibfnamefont {M.}~\bibnamefont {Azreg-A{\"\i}nou}},\ }\href {\doibase 10.1016/j.dark.2025.101816} {\bibfield  {journal} {\bibinfo  {journal} {Phys. Dark Univ.}\ }\textbf {\bibinfo {volume} {47}},\ \bibinfo {pages} {101816} (\bibinfo {year} {2025})},\ \Eprint {http://arxiv.org/abs/2501.04367} {arXiv:2501.04367 [gr-qc]} \BibitemShut {NoStop}%
\bibitem [{\citenamefont {Junior}\ \emph {et~al.}(2025)\citenamefont {Junior}, \citenamefont {Junior}, \citenamefont {Lobo}, \citenamefont {Rodrigues}, \citenamefont {Rubiera-Garcia}, \citenamefont {da~Silva},\ and\ \citenamefont {Vieira}}]{Junior:2024tmi}%
  \BibitemOpen
  \bibfield  {author} {\bibinfo {author} {\bibfnamefont {E.~L.~B.}\ \bibnamefont {Junior}}, \bibinfo {author} {\bibfnamefont {J.~T. S.~S.}\ \bibnamefont {Junior}}, \bibinfo {author} {\bibfnamefont {F.~S.~N.}\ \bibnamefont {Lobo}}, \bibinfo {author} {\bibfnamefont {M.~E.}\ \bibnamefont {Rodrigues}}, \bibinfo {author} {\bibfnamefont {D.}~\bibnamefont {Rubiera-Garcia}}, \bibinfo {author} {\bibfnamefont {L.~F.~D.}\ \bibnamefont {da~Silva}}, \ and\ \bibinfo {author} {\bibfnamefont {H.~A.}\ \bibnamefont {Vieira}},\ }\href {\doibase 10.1140/epjc/s10052-025-14299-3} {\bibfield  {journal} {\bibinfo  {journal} {Eur. Phys. J. C}\ }\textbf {\bibinfo {volume} {85}},\ \bibinfo {pages} {557} (\bibinfo {year} {2025})},\ \Eprint {http://arxiv.org/abs/2412.00769} {arXiv:2412.00769 [gr-qc]} \BibitemShut {NoStop}%
\bibitem [{\citenamefont {Jiang}\ \emph {et~al.}(2024)\citenamefont {Jiang}, \citenamefont {Alloqulov}, \citenamefont {Wu}, \citenamefont {Shaymatov},\ and\ \citenamefont {Zhu}}]{Jiang:2024cpe}%
  \BibitemOpen
  \bibfield  {author} {\bibinfo {author} {\bibfnamefont {H.}~\bibnamefont {Jiang}}, \bibinfo {author} {\bibfnamefont {M.}~\bibnamefont {Alloqulov}}, \bibinfo {author} {\bibfnamefont {Q.}~\bibnamefont {Wu}}, \bibinfo {author} {\bibfnamefont {S.}~\bibnamefont {Shaymatov}}, \ and\ \bibinfo {author} {\bibfnamefont {T.}~\bibnamefont {Zhu}},\ }\href {\doibase 10.1016/j.dark.2024.101627} {\bibfield  {journal} {\bibinfo  {journal} {Phys. Dark Univ.}\ }\textbf {\bibinfo {volume} {46}},\ \bibinfo {pages} {101627} (\bibinfo {year} {2024})}\BibitemShut {NoStop}%
\bibitem [{\citenamefont {Yang}\ \emph {et~al.}(2025{\natexlab{b}})\citenamefont {Yang}, \citenamefont {Zhang}, \citenamefont {Zhu}, \citenamefont {Zhao},\ and\ \citenamefont {Liu}}]{Yang:2024cnd}%
  \BibitemOpen
  \bibfield  {author} {\bibinfo {author} {\bibfnamefont {S.}~\bibnamefont {Yang}}, \bibinfo {author} {\bibfnamefont {Y.-P.}\ \bibnamefont {Zhang}}, \bibinfo {author} {\bibfnamefont {T.}~\bibnamefont {Zhu}}, \bibinfo {author} {\bibfnamefont {L.}~\bibnamefont {Zhao}}, \ and\ \bibinfo {author} {\bibfnamefont {Y.-X.}\ \bibnamefont {Liu}},\ }\href {\doibase 10.1088/1674-1137/adef1a} {\bibfield  {journal} {\bibinfo  {journal} {Chin. Phys.}\ }\textbf {\bibinfo {volume} {49}},\ \bibinfo {pages} {115107} (\bibinfo {year} {2025}{\natexlab{b}})},\ \Eprint {http://arxiv.org/abs/2412.04302} {arXiv:2412.04302 [gr-qc]} \BibitemShut {NoStop}%
\bibitem [{\citenamefont {Qi}\ \emph {et~al.}(2024)\citenamefont {Qi}, \citenamefont {Kuang}, \citenamefont {Li},\ and\ \citenamefont {Sang}}]{QiQi:2024dwc}%
  \BibitemOpen
  \bibfield  {author} {\bibinfo {author} {\bibfnamefont {Q.}~\bibnamefont {Qi}}, \bibinfo {author} {\bibfnamefont {X.-M.}\ \bibnamefont {Kuang}}, \bibinfo {author} {\bibfnamefont {Y.-Z.}\ \bibnamefont {Li}}, \ and\ \bibinfo {author} {\bibfnamefont {Y.}~\bibnamefont {Sang}},\ }\href {\doibase 10.1140/epjc/s10052-024-12989-y} {\bibfield  {journal} {\bibinfo  {journal} {Eur. Phys. J. C}\ }\textbf {\bibinfo {volume} {84}},\ \bibinfo {pages} {645} (\bibinfo {year} {2024})},\ \Eprint {http://arxiv.org/abs/2407.01958} {arXiv:2407.01958 [gr-qc]} \BibitemShut {NoStop}%
\bibitem [{\citenamefont {Alloqulov}\ \emph {et~al.}(2025{\natexlab{b}})\citenamefont {Alloqulov}, \citenamefont {Xamidov}, \citenamefont {Shaymatov},\ and\ \citenamefont {Ahmedov}}]{Alloqulov:2025ucf}%
  \BibitemOpen
  \bibfield  {author} {\bibinfo {author} {\bibfnamefont {M.}~\bibnamefont {Alloqulov}}, \bibinfo {author} {\bibfnamefont {T.}~\bibnamefont {Xamidov}}, \bibinfo {author} {\bibfnamefont {S.}~\bibnamefont {Shaymatov}}, \ and\ \bibinfo {author} {\bibfnamefont {B.}~\bibnamefont {Ahmedov}},\ }\href {\doibase 10.1140/epjc/s10052-025-14529-8} {\bibfield  {journal} {\bibinfo  {journal} {Eur. Phys. J. C}\ }\textbf {\bibinfo {volume} {85}},\ \bibinfo {pages} {798} (\bibinfo {year} {2025}{\natexlab{b}})},\ \Eprint {http://arxiv.org/abs/2504.05236} {arXiv:2504.05236 [gr-qc]} \BibitemShut {NoStop}%
\bibitem [{\citenamefont {Wang}\ \emph {et~al.}(2025{\natexlab{b}})\citenamefont {Wang}, \citenamefont {Meng}, \citenamefont {Zhang}, \citenamefont {Zhu},\ and\ \citenamefont {Wei}}]{Wang:2025hla}%
  \BibitemOpen
  \bibfield  {author} {\bibinfo {author} {\bibfnamefont {C.-H.}\ \bibnamefont {Wang}}, \bibinfo {author} {\bibfnamefont {X.-C.}\ \bibnamefont {Meng}}, \bibinfo {author} {\bibfnamefont {Y.-P.}\ \bibnamefont {Zhang}}, \bibinfo {author} {\bibfnamefont {T.}~\bibnamefont {Zhu}}, \ and\ \bibinfo {author} {\bibfnamefont {S.-W.}\ \bibnamefont {Wei}},\ }\href {\doibase 10.1088/1475-7516/2025/07/021} {\  (\bibinfo {year} {2025}{\natexlab{b}}),\ 10.1088/1475-7516/2025/07/021},\ \Eprint {http://arxiv.org/abs/2502.08994} {arXiv:2502.08994 [gr-qc]} \BibitemShut {NoStop}%
\bibitem [{\citenamefont {Lu}\ and\ \citenamefont {Zhu}(2025)}]{Lu:2025cxx}%
  \BibitemOpen
  \bibfield  {author} {\bibinfo {author} {\bibfnamefont {S.}~\bibnamefont {Lu}}\ and\ \bibinfo {author} {\bibfnamefont {T.}~\bibnamefont {Zhu}},\ }\href {\doibase 10.1016/j.dark.2025.102141} {\bibfield  {journal} {\bibinfo  {journal} {Phys. Dark Univ.}\ }\textbf {\bibinfo {volume} {50}},\ \bibinfo {pages} {102141} (\bibinfo {year} {2025})},\ \Eprint {http://arxiv.org/abs/2505.00294} {arXiv:2505.00294 [gr-qc]} \BibitemShut {NoStop}%
\bibitem [{\citenamefont {Zare}\ \emph {et~al.}(2025)\citenamefont {Zare}, \citenamefont {Zhu}, \citenamefont {Nieto}, \citenamefont {Lu},\ and\ \citenamefont {Hassanabadi}}]{Zare:2025aek}%
  \BibitemOpen
  \bibfield  {author} {\bibinfo {author} {\bibfnamefont {S.}~\bibnamefont {Zare}}, \bibinfo {author} {\bibfnamefont {T.}~\bibnamefont {Zhu}}, \bibinfo {author} {\bibfnamefont {L.~M.}\ \bibnamefont {Nieto}}, \bibinfo {author} {\bibfnamefont {S.}~\bibnamefont {Lu}}, \ and\ \bibinfo {author} {\bibfnamefont {H.}~\bibnamefont {Hassanabadi}},\ }\href@noop {} {\  (\bibinfo {year} {2025})},\ \Eprint {http://arxiv.org/abs/2510.05166} {arXiv:2510.05166 [gr-qc]} \BibitemShut {NoStop}%
\bibitem [{\citenamefont {Gong}\ \emph {et~al.}(2025)\citenamefont {Gong}, \citenamefont {Long}, \citenamefont {Wang}, \citenamefont {Xia}, \citenamefont {Wu},\ and\ \citenamefont {Pan}}]{Gong:2025mne}%
  \BibitemOpen
  \bibfield  {author} {\bibinfo {author} {\bibfnamefont {H.}~\bibnamefont {Gong}}, \bibinfo {author} {\bibfnamefont {S.}~\bibnamefont {Long}}, \bibinfo {author} {\bibfnamefont {X.-J.}\ \bibnamefont {Wang}}, \bibinfo {author} {\bibfnamefont {Z.}~\bibnamefont {Xia}}, \bibinfo {author} {\bibfnamefont {J.-P.}\ \bibnamefont {Wu}}, \ and\ \bibinfo {author} {\bibfnamefont {Q.}~\bibnamefont {Pan}},\ }\href@noop {} {\  (\bibinfo {year} {2025})},\ \Eprint {http://arxiv.org/abs/2509.23318} {arXiv:2509.23318 [gr-qc]} \BibitemShut {NoStop}%
\bibitem [{\citenamefont {Li}\ and\ \citenamefont {Kuang}(2025)}]{Li:2025sfe}%
  \BibitemOpen
  \bibfield  {author} {\bibinfo {author} {\bibfnamefont {Y.-Z.}\ \bibnamefont {Li}}\ and\ \bibinfo {author} {\bibfnamefont {X.-M.}\ \bibnamefont {Kuang}},\ }\href@noop {} {\  (\bibinfo {year} {2025})},\ \Eprint {http://arxiv.org/abs/2509.07333} {arXiv:2509.07333 [gr-qc]} \BibitemShut {NoStop}%
\bibitem [{\citenamefont {Choudhury}\ \emph {et~al.}(2025)\citenamefont {Choudhury}, \citenamefont {Hossain}, \citenamefont {Bauyrzhan},\ and\ \citenamefont {Yerzhanov}}]{Choudhury:2025qsh}%
  \BibitemOpen
  \bibfield  {author} {\bibinfo {author} {\bibfnamefont {S.}~\bibnamefont {Choudhury}}, \bibinfo {author} {\bibfnamefont {M.~K.}\ \bibnamefont {Hossain}}, \bibinfo {author} {\bibfnamefont {G.}~\bibnamefont {Bauyrzhan}}, \ and\ \bibinfo {author} {\bibfnamefont {K.}~\bibnamefont {Yerzhanov}},\ }\href@noop {} {\  (\bibinfo {year} {2025})},\ \Eprint {http://arxiv.org/abs/2507.00904} {arXiv:2507.00904 [gr-qc]} \BibitemShut {NoStop}%
\bibitem [{\citenamefont {Chen}\ and\ \citenamefont {Yang}(2025)}]{Chen:2025aqh}%
  \BibitemOpen
  \bibfield  {author} {\bibinfo {author} {\bibfnamefont {J.}~\bibnamefont {Chen}}\ and\ \bibinfo {author} {\bibfnamefont {J.}~\bibnamefont {Yang}},\ }\href {\doibase 10.1140/epjc/s10052-025-14457-7} {\bibfield  {journal} {\bibinfo  {journal} {Eur. Phys. J. C}\ }\textbf {\bibinfo {volume} {85}},\ \bibinfo {pages} {726} (\bibinfo {year} {2025})},\ \Eprint {http://arxiv.org/abs/2505.02660} {arXiv:2505.02660 [gr-qc]} \BibitemShut {NoStop}%
\bibitem [{\citenamefont {Deng}\ \emph {et~al.}(2025)\citenamefont {Deng}, \citenamefont {Long}, \citenamefont {Tan},\ and\ \citenamefont {Jing}}]{Deng:2025wzz}%
  \BibitemOpen
  \bibfield  {author} {\bibinfo {author} {\bibfnamefont {W.}~\bibnamefont {Deng}}, \bibinfo {author} {\bibfnamefont {S.}~\bibnamefont {Long}}, \bibinfo {author} {\bibfnamefont {Q.}~\bibnamefont {Tan}}, \ and\ \bibinfo {author} {\bibfnamefont {J.}~\bibnamefont {Jing}},\ }\href@noop {} {\  (\bibinfo {year} {2025})},\ \Eprint {http://arxiv.org/abs/2510.24468} {arXiv:2510.24468 [gr-qc]} \BibitemShut {NoStop}%
\bibitem [{\citenamefont {Li}\ \emph {et~al.}(2025)\citenamefont {Li}, \citenamefont {Qiao},\ and\ \citenamefont {Tao}}]{Li:2025eln}%
  \BibitemOpen
  \bibfield  {author} {\bibinfo {author} {\bibfnamefont {G.-H.}\ \bibnamefont {Li}}, \bibinfo {author} {\bibfnamefont {C.-K.}\ \bibnamefont {Qiao}}, \ and\ \bibinfo {author} {\bibfnamefont {J.}~\bibnamefont {Tao}},\ }\href@noop {} {\  (\bibinfo {year} {2025})},\ \Eprint {http://arxiv.org/abs/2510.24989} {arXiv:2510.24989 [gr-qc]} \BibitemShut {NoStop}%
\bibitem [{\citenamefont {{Sharipov}}\ \emph {et~al.}(2026)\citenamefont {{Sharipov}}, \citenamefont {{Xamidov}}, \citenamefont {{Wu}}, \citenamefont {{Shaymatov}},\ and\ \citenamefont {{Zhu}}}]{Sharipov:2026CPC}%
  \BibitemOpen
  \bibfield  {author} {\bibinfo {author} {\bibfnamefont {J.}~\bibnamefont {{Sharipov}}}, \bibinfo {author} {\bibfnamefont {T.}~\bibnamefont {{Xamidov}}}, \bibinfo {author} {\bibfnamefont {Q.}~\bibnamefont {{Wu}}}, \bibinfo {author} {\bibfnamefont {S.}~\bibnamefont {{Shaymatov}}}, \ and\ \bibinfo {author} {\bibfnamefont {T.}~\bibnamefont {{Zhu}}},\ }\href {\doibase 10.1088/1674-1137/ae6b2f} {\bibfield  {journal} {\bibinfo  {journal} {Chin. Phys. C.}\ }\textbf {\bibinfo {volume} {50}},\ \bibinfo {eid} {085105} (\bibinfo {year} {2026})},\ \Eprint {http://arxiv.org/abs/2511.10043} {arXiv:2511.10043 [gr-qc]} \BibitemShut {NoStop}%
\bibitem [{\citenamefont {Ahmed}\ \emph {et~al.}(2025)\citenamefont {Ahmed}, \citenamefont {Wu}, \citenamefont {Ghosh},\ and\ \citenamefont {Zhu}}]{Ahmed:2025azu}%
  \BibitemOpen
  \bibfield  {author} {\bibinfo {author} {\bibfnamefont {F.}~\bibnamefont {Ahmed}}, \bibinfo {author} {\bibfnamefont {Q.}~\bibnamefont {Wu}}, \bibinfo {author} {\bibfnamefont {S.~G.}\ \bibnamefont {Ghosh}}, \ and\ \bibinfo {author} {\bibfnamefont {T.}~\bibnamefont {Zhu}},\ }\href@noop {} {\  (\bibinfo {year} {2025})},\ \Eprint {http://arxiv.org/abs/2511.08456} {arXiv:2511.08456 [gr-qc]} \BibitemShut {NoStop}%
\bibitem [{\citenamefont {Glampedakis}\ and\ \citenamefont {Kennefick}(2002)}]{Glampedakis2002}%
  \BibitemOpen
  \bibfield  {author} {\bibinfo {author} {\bibfnamefont {K.}~\bibnamefont {Glampedakis}}\ and\ \bibinfo {author} {\bibfnamefont {D.}~\bibnamefont {Kennefick}},\ }\href {\doibase 10.1103/PhysRevD.66.044002} {\bibfield  {journal} {\bibinfo  {journal} {Physical Review D}\ }\textbf {\bibinfo {volume} {66}},\ \bibinfo {pages} {044002} (\bibinfo {year} {2002})}\BibitemShut {NoStop}%
\bibitem [{\citenamefont {Alloqulov}\ \emph {et~al.}(2026{\natexlab{b}})\citenamefont {Alloqulov}, \citenamefont {Shaymatov}, \citenamefont {Ahmedov},\ and\ \citenamefont {Zhu}}]{Alloqulovmod}%
  \BibitemOpen
  \bibfield  {author} {\bibinfo {author} {\bibfnamefont {M.}~\bibnamefont {Alloqulov}}, \bibinfo {author} {\bibfnamefont {S.}~\bibnamefont {Shaymatov}}, \bibinfo {author} {\bibfnamefont {B.}~\bibnamefont {Ahmedov}}, \ and\ \bibinfo {author} {\bibfnamefont {T.}~\bibnamefont {Zhu}},\ }\href {\doibase 10.1140/epjc/s10052-026-15469-7} {\bibfield  {journal} {\bibinfo  {journal} {Eur. Phys. J. C}\ }\textbf {\bibinfo {volume} {86}},\ \bibinfo {pages} {259} (\bibinfo {year} {2026}{\natexlab{b}})}\BibitemShut {NoStop}%
\bibitem [{\citenamefont {Tan}\ \emph {et~al.}(2026)\citenamefont {Tan}, \citenamefont {Jiang}, \citenamefont {Li}, \citenamefont {Hu}, \citenamefont {Deng},\ and\ \citenamefont {Lin}}]{Tan_2026}%
  \BibitemOpen
  \bibfield  {author} {\bibinfo {author} {\bibfnamefont {S.}~\bibnamefont {Tan}}, \bibinfo {author} {\bibfnamefont {C.}~\bibnamefont {Jiang}}, \bibinfo {author} {\bibfnamefont {D.}~\bibnamefont {Li}}, \bibinfo {author} {\bibfnamefont {S.}~\bibnamefont {Hu}}, \bibinfo {author} {\bibfnamefont {C.}~\bibnamefont {Deng}}, \ and\ \bibinfo {author} {\bibfnamefont {W.}~\bibnamefont {Lin}},\ }\href {\doibase 10.1016/j.jheap.2026.100685} {\bibfield  {journal} {\bibinfo  {journal} {Journal of High Energy Astrophysics}\ }\textbf {\bibinfo {volume} {54}},\ \bibinfo {pages} {100685} (\bibinfo {year} {2026})}\BibitemShut {NoStop}%
\bibitem [{\citenamefont {Shi}\ \emph {et~al.}(2026)\citenamefont {Shi}, \citenamefont {Zhang},\ and\ \citenamefont {Liu}}]{Shi_2026}%
  \BibitemOpen
  \bibfield  {author} {\bibinfo {author} {\bibfnamefont {Z.}~\bibnamefont {Shi}}, \bibinfo {author} {\bibfnamefont {X.}~\bibnamefont {Zhang}}, \ and\ \bibinfo {author} {\bibfnamefont {Y.}~\bibnamefont {Liu}},\ }\href {\doibase 10.1016/j.dark.2026.102349} {\bibfield  {journal} {\bibinfo  {journal} {Physics of the Dark Universe}\ }\textbf {\bibinfo {volume} {52}},\ \bibinfo {pages} {102349} (\bibinfo {year} {2026})}\BibitemShut {NoStop}%
\bibitem [{\citenamefont {Huang}\ \emph {et~al.}(2026)\citenamefont {Huang}, \citenamefont {Qiao},\ and\ \citenamefont {Tao}}]{Huang:2026tul}%
  \BibitemOpen
  \bibfield  {author} {\bibinfo {author} {\bibfnamefont {J.-L.}\ \bibnamefont {Huang}}, \bibinfo {author} {\bibfnamefont {C.-K.}\ \bibnamefont {Qiao}}, \ and\ \bibinfo {author} {\bibfnamefont {J.}~\bibnamefont {Tao}},\ }\href@noop {} {\  (\bibinfo {year} {2026})},\ \Eprint {http://arxiv.org/abs/2608.30049} {arXiv:2608.30049 [gr-qc]} \BibitemShut {NoStop}%
\bibitem [{\citenamefont {Wang}\ \emph {et~al.}(2026)\citenamefont {Wang}, \citenamefont {Zhang}, \citenamefont {Zhu},\ and\ \citenamefont {Wei}}]{wang2026newtypemultibranchperiodic}%
  \BibitemOpen
  \bibfield  {author} {\bibinfo {author} {\bibfnamefont {C.-H.}\ \bibnamefont {Wang}}, \bibinfo {author} {\bibfnamefont {Y.-P.}\ \bibnamefont {Zhang}}, \bibinfo {author} {\bibfnamefont {T.}~\bibnamefont {Zhu}}, \ and\ \bibinfo {author} {\bibfnamefont {S.-W.}\ \bibnamefont {Wei}},\ }\href {https://arxiv.org/abs/2508.20558} {\enquote {\bibinfo {title} {A new type of multi-branch periodic orbits in dyonic black holes},}\ } (\bibinfo {year} {2026}),\ \Eprint {http://arxiv.org/abs/2508.20558} {arXiv:2508.20558 [gr-qc]} \BibitemShut {NoStop}%
\bibitem [{\citenamefont {Gogoi}\ \emph {et~al.}(2026)\citenamefont {Gogoi}, \citenamefont {Bora},\ and\ \citenamefont {Övgün}}]{Gogoi_2026}%
  \BibitemOpen
  \bibfield  {author} {\bibinfo {author} {\bibfnamefont {D.~J.}\ \bibnamefont {Gogoi}}, \bibinfo {author} {\bibfnamefont {J.}~\bibnamefont {Bora}}, \ and\ \bibinfo {author} {\bibfnamefont {A.}~\bibnamefont {Övgün}},\ }\href {\doibase 10.1088/1475-7516/2026/07/087} {\bibfield  {journal} {\bibinfo  {journal} {Journal of Cosmology and Astroparticle Physics}\ }\textbf {\bibinfo {volume} {2026}},\ \bibinfo {pages} {087} (\bibinfo {year} {2026})}\BibitemShut {NoStop}%
\bibitem [{\citenamefont {Zhang}\ and\ \citenamefont {Zhu}(2026)}]{Zhang_2026}%
  \BibitemOpen
  \bibfield  {author} {\bibinfo {author} {\bibfnamefont {C.}~\bibnamefont {Zhang}}\ and\ \bibinfo {author} {\bibfnamefont {T.}~\bibnamefont {Zhu}},\ }\href {\doibase 10.1140/epjc/s10052-026-15606-2} {\bibfield  {journal} {\bibinfo  {journal} {Eur. Phys. J. C}\ }\textbf {\bibinfo {volume} {86}} (\bibinfo {year} {2026}),\ 10.1140/epjc/s10052-026-15606-2}\BibitemShut {NoStop}%
\bibitem [{\citenamefont {{Shokirov}}\ \emph {et~al.}(2026)\citenamefont {{Shokirov}}, \citenamefont {{Mirzakulov}}, \citenamefont {{Xamidov}},\ and\ \citenamefont {{Shaymatov}}}]{2026arXiv260700812S}%
  \BibitemOpen
  \bibfield  {author} {\bibinfo {author} {\bibfnamefont {B.}~\bibnamefont {{Shokirov}}}, \bibinfo {author} {\bibfnamefont {A.}~\bibnamefont {{Mirzakulov}}}, \bibinfo {author} {\bibfnamefont {T.}~\bibnamefont {{Xamidov}}}, \ and\ \bibinfo {author} {\bibfnamefont {S.}~\bibnamefont {{Shaymatov}}},\ }\href {\doibase 10.48550/arXiv.2607.00812} {\bibfield  {journal} {\bibinfo  {journal} {arXiv e-prints}\ ,\ \bibinfo {eid} {arXiv:2607.00812}} (\bibinfo {year} {2026})},\ \Eprint {http://arxiv.org/abs/2607.00812} {arXiv:2607.00812 [gr-qc]} \BibitemShut {NoStop}%
\bibitem [{\citenamefont {Lu}\ \emph {et~al.}(2026)\citenamefont {Lu}, \citenamefont {Lin}, \citenamefont {Zhu}, \citenamefont {Liu},\ and\ \citenamefont {Zhang}}]{Lu_2026}%
  \BibitemOpen
  \bibfield  {author} {\bibinfo {author} {\bibfnamefont {S.}~\bibnamefont {Lu}}, \bibinfo {author} {\bibfnamefont {H.-J.}\ \bibnamefont {Lin}}, \bibinfo {author} {\bibfnamefont {T.}~\bibnamefont {Zhu}}, \bibinfo {author} {\bibfnamefont {Y.-X.}\ \bibnamefont {Liu}}, \ and\ \bibinfo {author} {\bibfnamefont {X.}~\bibnamefont {Zhang}},\ }\href {\doibase 10.1140/epjc/s10052-026-15466-w} {\bibfield  {journal} {\bibinfo  {journal} {Eur. Phys. J. C}\ }\textbf {\bibinfo {volume} {86}} (\bibinfo {year} {2026}),\ 10.1140/epjc/s10052-026-15466-w}\BibitemShut {NoStop}%
\bibitem [{\citenamefont {Das}\ \emph {et~al.}(2026)\citenamefont {Das}, \citenamefont {Dalui}, \citenamefont {Lee},\ and\ \citenamefont {Cai}}]{Das_2026}%
  \BibitemOpen
  \bibfield  {author} {\bibinfo {author} {\bibfnamefont {S.}~\bibnamefont {Das}}, \bibinfo {author} {\bibfnamefont {S.}~\bibnamefont {Dalui}}, \bibinfo {author} {\bibfnamefont {B.-H.}\ \bibnamefont {Lee}}, \ and\ \bibinfo {author} {\bibfnamefont {Y.-F.}\ \bibnamefont {Cai}},\ }\href {\doibase 10.1007/s11433-026-3011-6} {\bibfield  {journal} {\bibinfo  {journal} {Science China Physics, Mechanics {\&} Astronomy}\ }\textbf {\bibinfo {volume} {69}} (\bibinfo {year} {2026}),\ 10.1007/s11433-026-3011-6}\BibitemShut {NoStop}%
\bibitem [{\citenamefont {Hua}\ \emph {et~al.}(2026)\citenamefont {Hua}, \citenamefont {He}, \citenamefont {Lai}, \citenamefont {Jiao},\ and\ \citenamefont {Tian}}]{Hua_2026}%
  \BibitemOpen
  \bibfield  {author} {\bibinfo {author} {\bibfnamefont {Z.}~\bibnamefont {Hua}}, \bibinfo {author} {\bibfnamefont {Z.-T.}\ \bibnamefont {He}}, \bibinfo {author} {\bibfnamefont {J.-Q.}\ \bibnamefont {Lai}}, \bibinfo {author} {\bibfnamefont {J.}~\bibnamefont {Jiao}}, \ and\ \bibinfo {author} {\bibfnamefont {Y.}~\bibnamefont {Tian}},\ }\href {\doibase 10.1016/j.physletb.2026.140402} {\bibfield  {journal} {\bibinfo  {journal} {Physics Letters B}\ }\textbf {\bibinfo {volume} {876}},\ \bibinfo {pages} {140402} (\bibinfo {year} {2026})}\BibitemShut {NoStop}%
\bibitem [{\citenamefont {{Ahmed}}\ \emph {et~al.}(2026)\citenamefont {{Ahmed}}, \citenamefont {{Shaymatov}}, \citenamefont {{Yuan}},\ and\ \citenamefont {{Nasri}}}]{2026arXiv260724154A}%
  \BibitemOpen
  \bibfield  {author} {\bibinfo {author} {\bibfnamefont {F.}~\bibnamefont {{Ahmed}}}, \bibinfo {author} {\bibfnamefont {S.}~\bibnamefont {{Shaymatov}}}, \bibinfo {author} {\bibfnamefont {C.}~\bibnamefont {{Yuan}}}, \ and\ \bibinfo {author} {\bibfnamefont {S.}~\bibnamefont {{Nasri}}},\ }\href {\doibase 10.48550/arXiv.2607.24154} {\bibfield  {journal} {\bibinfo  {journal} {arXiv e-prints}\ ,\ \bibinfo {eid} {arXiv:2607.24154}} (\bibinfo {year} {2026})},\ \Eprint {http://arxiv.org/abs/2607.24154} {arXiv:2607.24154 [gr-qc]} \BibitemShut {NoStop}%
\bibitem [{\citenamefont {Bravo-Gaete}\ \emph {et~al.}(2026)\citenamefont {Bravo-Gaete}, \citenamefont {Lin}, \citenamefont {Liu},\ and\ \citenamefont {Zhang}}]{bravogaete2026}%
  \BibitemOpen
  \bibfield  {author} {\bibinfo {author} {\bibfnamefont {M.}~\bibnamefont {Bravo-Gaete}}, \bibinfo {author} {\bibfnamefont {J.}~\bibnamefont {Lin}}, \bibinfo {author} {\bibfnamefont {Y.}~\bibnamefont {Liu}}, \ and\ \bibinfo {author} {\bibfnamefont {X.}~\bibnamefont {Zhang}},\ }\href {https://arxiv.org/abs/2602.15609} {\enquote {\bibinfo {title} {Periodic orbits and gravitational waveforms of spinning particles in nonlocal gravity},}\ } (\bibinfo {year} {2026}),\ \Eprint {http://arxiv.org/abs/2602.15609} {arXiv:2602.15609 [gr-qc]} \BibitemShut {NoStop}%
\bibitem [{\citenamefont {{Xamidov}}\ \emph {et~al.}(2026)\citenamefont {{Xamidov}}, \citenamefont {{Shaymatov}}, \citenamefont {{Wu}},\ and\ \citenamefont {{Zhu}}}]{2026arXiv260209453X}%
  \BibitemOpen
  \bibfield  {author} {\bibinfo {author} {\bibfnamefont {T.}~\bibnamefont {{Xamidov}}}, \bibinfo {author} {\bibfnamefont {S.}~\bibnamefont {{Shaymatov}}}, \bibinfo {author} {\bibfnamefont {Q.}~\bibnamefont {{Wu}}}, \ and\ \bibinfo {author} {\bibfnamefont {T.}~\bibnamefont {{Zhu}}},\ }\href {\doibase 10.48550/arXiv.2602.09453} {\bibfield  {journal} {\bibinfo  {journal} {arXiv e-prints}\ ,\ \bibinfo {eid} {arXiv:2602.09453}} (\bibinfo {year} {2026})},\ \Eprint {http://arxiv.org/abs/2602.09453} {arXiv:2602.09453 [gr-qc]} \BibitemShut {NoStop}%
\bibitem [{\citenamefont {Chandrasekhar}(1984)}]{1983mtbh.book.....C}%
  \BibitemOpen
  \bibfield  {author} {\bibinfo {author} {\bibfnamefont {S.}~\bibnamefont {Chandrasekhar}},\ }\href {https://doi.org/10.1007/978-94-009-6469-3_2} {\emph {\bibinfo {title} {General Relativity and Gravitation: Invited Papers and Discussion Reports of the 10th International Conference on General Relativity and Gravitation, Padua, July 3--8, 1983}}},\ edited by\ \bibinfo {editor} {\bibfnamefont {B.}~\bibnamefont {Bertotti}}, \bibinfo {editor} {\bibfnamefont {F.}~\bibnamefont {de~Felice}}, \ and\ \bibinfo {editor} {\bibfnamefont {A.}~\bibnamefont {Pascolini}}\ (\bibinfo  {publisher} {Springer Netherlands},\ \bibinfo {address} {Dordrecht},\ \bibinfo {year} {1984})\ pp.\ \bibinfo {pages} {5--26}\BibitemShut {NoStop}%
\bibitem [{\citenamefont {{Dadhich}}\ and\ \citenamefont {{Shaymatov}}(2022)}]{Dadhich22a}%
  \BibitemOpen
  \bibfield  {author} {\bibinfo {author} {\bibfnamefont {N.}~\bibnamefont {{Dadhich}}}\ and\ \bibinfo {author} {\bibfnamefont {S.}~\bibnamefont {{Shaymatov}}},\ }\href {\doibase 10.1016/j.dark.2022.100986} {\bibfield  {journal} {\bibinfo  {journal} {Phys. Dark Universe}\ }\textbf {\bibinfo {volume} {35}},\ \bibinfo {eid} {100986} (\bibinfo {year} {2022})},\ \Eprint {http://arxiv.org/abs/2104.00427} {arXiv:2104.00427 [gr-qc]} \BibitemShut {NoStop}%
\bibitem [{\citenamefont {Thorne}(1980)}]{RevMod}%
  \BibitemOpen
  \bibfield  {author} {\bibinfo {author} {\bibfnamefont {K.~S.}\ \bibnamefont {Thorne}},\ }\href {\doibase 10.1103/RevModPhys.52.299} {\bibfield  {journal} {\bibinfo  {journal} {Rev. Mod. Phys.}\ }\textbf {\bibinfo {volume} {52}},\ \bibinfo {pages} {299} (\bibinfo {year} {1980})}\BibitemShut {NoStop}%
\bibitem [{\citenamefont {Poisson}\ and\ \citenamefont {Will}(2014)}]{Poisson_Will_2014}%
  \BibitemOpen
  \bibfield  {author} {\bibinfo {author} {\bibfnamefont {E.}~\bibnamefont {Poisson}}\ and\ \bibinfo {author} {\bibfnamefont {C.~M.}\ \bibnamefont {Will}},\ }\href@noop {} {\emph {\bibinfo {title} {Gravity: Newtonian, Post-Newtonian, Relativistic}}}\ (\bibinfo  {publisher} {Cambridge University Press},\ \bibinfo {year} {2014})\BibitemShut {NoStop}%
\bibitem [{\citenamefont {{Meng}}\ \emph {et~al.}(2024)\citenamefont {{Meng}}, \citenamefont {{Xu}},\ and\ \citenamefont {{Tang}}}]{2024arXiv241101858M}%
  \BibitemOpen
  \bibfield  {author} {\bibinfo {author} {\bibfnamefont {L.}~\bibnamefont {{Meng}}}, \bibinfo {author} {\bibfnamefont {Z.}~\bibnamefont {{Xu}}}, \ and\ \bibinfo {author} {\bibfnamefont {M.}~\bibnamefont {{Tang}}},\ }\href {\doibase 10.48550/arXiv.2411.01858} {\bibfield  {journal} {\bibinfo  {journal} {arXiv e-prints}\ ,\ \bibinfo {eid} {arXiv:2411.01858}} (\bibinfo {year} {2024})},\ \Eprint {http://arxiv.org/abs/2411.01858} {arXiv:2411.01858 [gr-qc]} \BibitemShut {NoStop}%
\end{thebibliography}%

\end{document}